\documentclass[12pt,preprint]{aastex}

\usepackage{natbib}

\begin{document}
\title{Near-Ultraviolet Properties of a Large Sample of Type Ia Supernovae 
as Observed with the {\it Swift} UVOT}
\author{Peter~A.~Milne\altaffilmark{1},
Peter~J.~Brown\altaffilmark{2},
Peter~W. A.~Roming\altaffilmark{2}, \\
Stephen T. Holland\altaffilmark{3,4}, 
Stefan Immler\altaffilmark{3,4},
Alexei V. Filippenko\altaffilmark{5},
Mohan Ganeshalingam\altaffilmark{5},
Weidong Li\altaffilmark{5},
Maximilian Stritzinger\altaffilmark{6},
Mark M. Phillips\altaffilmark{6},
Malcolm Hicken\altaffilmark{7},
Robert P. Kirshner\altaffilmark{7}, 
Peter J. Challis\altaffilmark{7}, 
Paolo Mazzali\altaffilmark{8,9},
Brian P. Schmidt\altaffilmark{10},
Filamena Bufano\altaffilmark{3,11},
Neil Gehrels\altaffilmark{3}, and
Daniel Vanden Berk\altaffilmark{2,12}
}
\altaffiltext{1}{University of Arizona, Steward Observatory, 
                    933 N. Cherry Ave., Tucson, AZ 85719.}
\altaffiltext{2}{Pennsylvania State University,
                 Department of Astronomy \& Astrophysics,
                 University Park, PA 16802.}
\altaffiltext{3}{NASA-Goddard Space Flight Center, Astrophysics Science 
Division, Codes 660.1 \& 662, Greenbelt, MD 20771.}
\altaffiltext{4}{Universities Space Research Association, 
Columbia, MD 21044.}
\altaffiltext{5}{Department of Astronomy, University of California, 
   Berkeley, CA  94720-3411.}
\altaffiltext{6}{Las Campanas Observatory, Carnegie Observatories,
  Casilla 601, La Serena, Chile.}
\altaffiltext{7}{Harvard-Smithsonian Center for Astrophysics, 
60 Garden St., Cambridge, MA 02138.}
\altaffiltext{8}{INAF, Observatorio Astronomico di Padova, 
Vicolo dell'Osservatorio 5, I-35122 Padova, Italy}
\altaffiltext{9}{Max-Planck-Institut f\"ur Astrophysik, 
Karl-Schwartzschild-Str. 1, D-85748 Garching, Germany.}
\altaffiltext{10}{Australian National University, Mt. Stromlo Observatory, 
The Research School of Astronomy and Astrophysics,
Weston Creek, ACT 2611 Australia.}
\altaffiltext{11}{Universita' degli Studi di Padova, Dipartmento di 
Astronomia, Padova, Italy 35122}
\altaffiltext{12}{St. Vincent College, 300 Fraser Purchase Rd., 
Latrobe, PA 1560-2690.}

\begin{abstract}
We present ultraviolet (UV) and optical photometry of 26 Type Ia supernovae 
(SNe~Ia) observed from March 2005 to March 2008 with the NASA {\it Swift}
Ultraviolet and Optical Telescope (UVOT). 
The dataset consists of 2133 individual 
observations, making it by far the most complete study of the 
UV emission from SNe~Ia to date. Grouping the SNe into three  
subclasses as derived from optical observations, 
we investigate the evolution of the colors of 
these SNe, finding a high degree of homogeneity within the 
normal subclass, but dramatic differences between that group 
and the subluminous and SN 2002cx-like groups. For the normal 
events, the redder UV filters on UVOT ($u$, $uvw1$) show more homogeneity 
than do the bluer UV filters ($uvm2$, $uvw2$). Searching for 
purely UV characteristics to determine existing optically based 
groupings, we find the peak width to be a poor discriminant, but we do 
see a variation in the time delay between peak emission 
and the late, flat phase of the light curves. The UV light curves 
peak a few days before the $B$ band for most subclasses (as was 
previously reported by Jha et al. 2006a),
although the SN 2002cx-like objects peak at a very early epoch in the 
UV. That group also features the bluest emission observed among
SNe~Ia. As the observational campaign is ongoing, we discuss 
the critical times to observe, as determined by this study, 
in order to maximize the scientific output of future observations.  
\end{abstract}


\section{Introduction}

A Type Ia supernova (SN~Ia), driven by a thermonuclear explosion of
a white dwarf generally near the Chandrasekhar limit, 
is an exceedingly luminous event ($L \approx 4 \times 10^{9}$~L$_{\odot}$), 
among the most powerful phenomena in the universe. This characteristic, 
combined 
with a relatively modest observed dispersion in the peak luminosity, 
has led to considerable interest in utilizing Type Ia supernovae (SNe~Ia) 
to study the expansion history of the universe; see \citep{Fil05} for
a review. The factor of $\sim$10 variation in peak luminosity 
has been shown to correlate with observable features in the SN emission, 
making SNe~Ia standardizable candles at the level of $\sim$8\% 
in distance 
\citep[see, e.g.,][and references therein]{Jha_etal_2007}.
They have been used to measure the Hubble constant 
\citep[e.g.,][]{Freedman_etal_2001,Riess_etal_2005,Riess_etal_2009}, and were
the first objects to clearly reveal the current acceleration
of the universe \citep{Riess_etal_1998,Perlmutter_etal_1999}.
There is also the desire to understand the progenitor systems, 
nucleosynthetic products, explosion mechanisms, and remnants
of SNe~Ia. 

Based largely on the nature of the optical emission, 
SNe~Ia have been categorized according to the width of the peak 
of their optical light curves \citep{Phillips_1993,
Hamuy_etal_1996shapes,Riess_96MLCS,Phillips_etal_1999, wang03}, 
their overall optical spectra \citep[e.g.,][]{Fil_etal_1992a,
Fil_etal_1992b,Phillips_etal_1992,Leibundgut_etal_1993},
the velocity gradient of the Si~II $\lambda$6355 absorption line
\citep{Benetti_etal_2005}, the nature of ``high velocity" 
absorption features in the early-epoch spectra \citep{mazz05}, 
the width of the nebular emission lines \citep{Mazzali_etal_1998}, 
and so on. Decades of observational 
and theoretical investigations have led to the emerging understanding 
that there may be multiple paths for a white dwarf in a binary 
star system to reach the critical mass for
a thermonuclear runaway, and that advancing our understanding 
of these events depends, in part, on broadening the wavelength
range of observations.

Observations at near-infrared (NIR) and ultraviolet (UV) wavelengths 
have led to an improved understanding of SNe~Ia in a variety of ways. 
There are spectral features in the NIR and/or UV that either provide 
unique information or are less ambiguous than optical features alone. 
An appreciable fraction of the bolometric luminosity  ($\leq$~20\%) 
is emitted outside the optical wavelength range
\citep[e.g.,][]{Wang_etal_2009a}, making NIR/UV observations 
important to minimize the fraction of the bolometric luminosity that 
is extrapolated from the observed wavelength range.\footnote{This was 
well illustrated by observations of the peculiar SN 2000cx, where the 
$\gtrsim$400 day NIR emission 
exhibited a plateau, while the optical emission followed a linear 
decline \citep{soll04}.}
The availability of ground-based NIR instruments has led to 
considerable progress in our understanding of SNe~Ia in this 
wavelength range. 
Several of the major observational programs for SNe~Ia include NIR 
observations of the nearest SNe~Ia, notably the CfA Supernova 
Program which employs PAIRITEL (Bloom et al. 2006) 
and the Carnegie Supernova Project (CSP; Hamuy et al. 2006). 
NIR light curves are being studied to determine
if the entire range of SN~Ia events can be considered standard
candles (Krisciunas et al. 2004; Wood-Vasey et al. 2008).
Observations are also being expanded toward the bluest emission 
observable from the ground. Jha et al. (2006a) presented 
``CfA2," a $UBVRI$ study of 44 SNe~Ia collected during the years 
1997--2000, 
with an emphasis on the completeness of the $U$-band survey. 
Hicken et al. (2009) presented ``CfA3," a $UBVRI$ study of an
additional 185 SNe~Ia.
current, ongoing campaigns include $U$-band or $u$-band photometry, 
with the promise of dramatically increasing the number of SNe~Ia 
with well-sampled $U$ or $u$ light curves.

Recognizing the importance of UV observations, SNe~Ia have been observed 
with space-based UV satellites for over 30 years. The {\it International 
Ultraviolet Explorer (IUE)} observed 12 
SNe~Ia over 13 years (see Panagia 2007 for a summary), 
with the most extensive campaigns being 7 observations 
of SN 1990N (Jeffery et al. 1992) and 7 observations of 
SN 1992A (Kirshner et al. 1993; also see 
\citep{branch86}). The {\it Hubble Space Telescope (HST)}
has obtained UV spectra of 10 SNe~Ia, with notable campaigns on 
SNe 1992A \citep{Kirshner_etal_1993}, 
2001eh and 2001ep \citep{Sauer_etal_2008}, and 2005cf \citep{Wang_etal_2009a}.
These observations have afforded insight into the nature of the 
UV emission from SNe~Ia, and have been the 
basis for characterizing the UV portion of the bolometric luminosity.
A theoretical understanding of the UV flux is slowly emerging, as 
radiative-transfer models are tested against these observations in an 
effort to reproduce the fundamental characteristics of the UV spectrum 
(Pauldrach et al. 1996; Fisher et al. 1997; H\"{o}flich et al. 1996; 
Mazzali et al. 2001; Lentz et al. 2001; Sauer et al. 2008). 
Foley et al. (2008b) present a collected study of {\it IUE} and {\it HST} 
observations, in particular finding a tentative correlation between 
light-curve peak width and 
a flux ratio of two UV wavelength bands, possibly providing a means of 
classifying SNe~Ia based only on UV emission. Despite these 
successes, however, it is clear that our understanding of the UV emission 
in SNe~Ia has been rather limited. 

The Ultraviolet and Optical Telescope (UVOT; Roming et al. 2005) 
onboard the NASA {\it Swift} satellite \citep{Gehrels_etal_2004} 
possesses a number of characteristics that make it useful for SN~Ia 
studies. First, the telescope is one 
of only a handful capable of detecting UV emission from nearby 
SNe~Ia.\footnote{GALEX is another UV telescope that has observed 
SNe during the period of these UVOT 
observations (Gezari et al. 2008; Gal-Yam et al 2008). 
See Roming et al. (2009) for a review of  
4 current UV-capable missions.} 
For the very nearest subset, UV spectra have been obtained and will be 
presented by Bufano et al. (2009).
Second, the primary science goal of the {\it Swift} mission, 
prompt follow-up observations of gamma-ray bursts (GRBs), 
permits a large fraction of the total time for secondary targets. 
UVOT can monitor a SN~Ia daily for weeks, obtaining unprecedented 
temporal sampling of a variety  of SNe~Ia. Third, scheduling 
of targets can be arranged in less than 24 hours for important 
transients, including SNe. Indeed, SN 2006bp, a Type~IIP 
SN, was observed within 24 hours of discovery, which was within 
2.5 days after the explosion \citep{immler07}. 
This prompt scheduling led to the discovery of short-lived 
X-ray emission from SN 2006bp, a discovery that would not be possible 
with a less agile observatory, where all schedule changes require a much 
longer delay.  

Between March 2005 and March 2008, 
UVOT observed 28 SNe~Ia. In this paper, we work with the 
photometry of 25 of these events. In \S 2, we introduce 
the UVOT instrument and the analysis methods employed to accurately 
determine the UV emission from the observed SNe~Ia. The SN~Ia sample 
is detailed in \S 3, including the categorizations possible 
from ground-based optical observations. 
In \S 4, we fit the light curves of the well-observed events, 
comparing their UV behavior to parameters that have been defined in the 
past based upon the optical emission. Section 5 presents the 
UV to optical, as well as UV to UV, colors of all the SNe~Ia 
included in this study. We discuss in \S 6 how these 
observations have improved the understanding of SNe~Ia. The absolute 
UV magnitudes of our SNe~Ia at peak brightness are not given in this 
paper; instead, the peak apparent magnitudes derived here
contribute to a companion paper (Brown et al. 2010, hereafter B10) 
that explores the UV emission from SNe~Ia as a potential standard 
candle.

\section{UVOT Analysis and Photometry}

\begin{figure}
\epsscale{1.0} \plotone{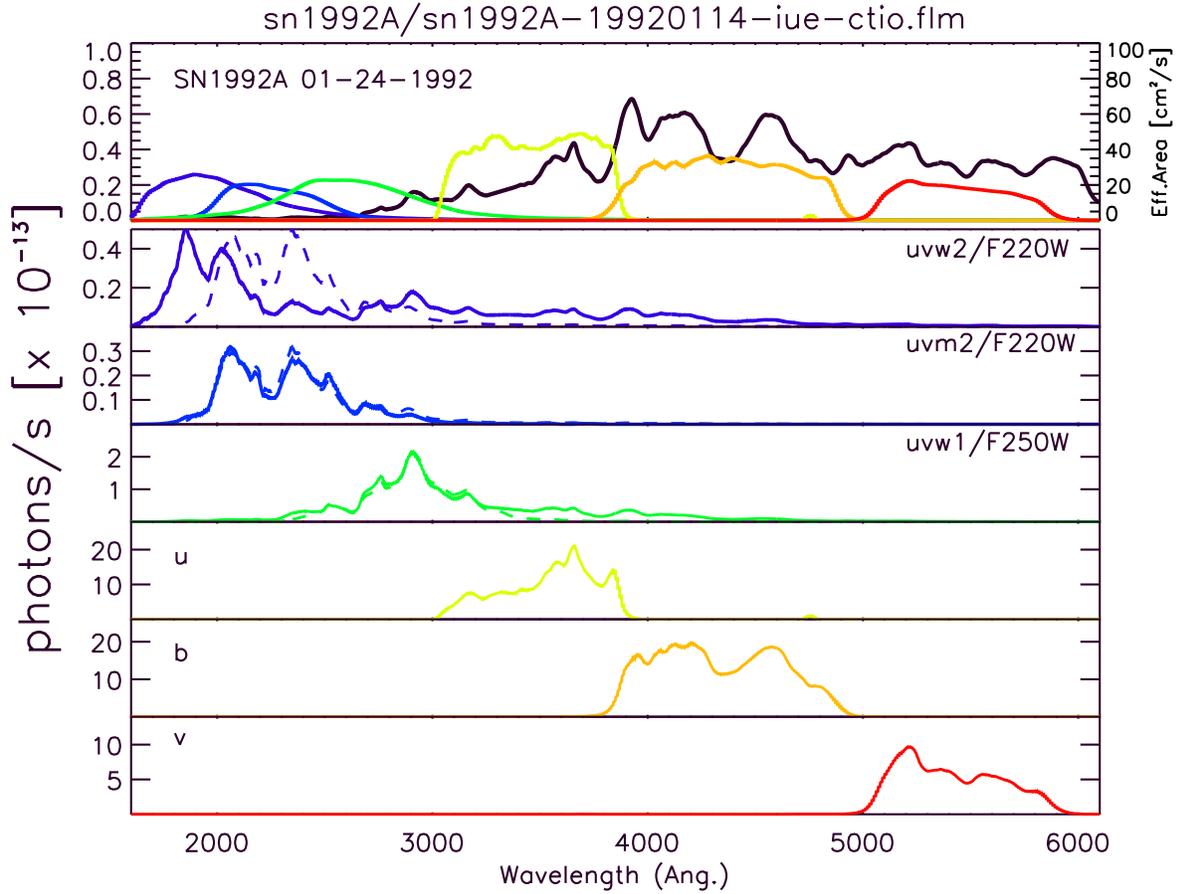}
\caption{The six UVOT filter transmission curves folded through the 
{\it HST}/CTIO spectrum of SN 1992A (Kirshner et al. 1993). In the top 
panel, the spectrum is shown in black, compared to the transmission 
curves for the six UVOT filters. The next six panels show the 
SN spectrum folded through the UVOT transmission curves (solid lines), 
as well as the {\it HST} $F220W$ and $F250W$ filters (dashed lines).} 
\label{transcurves}
\end{figure}

The UVOT has an aperture of 30-cm, and is equipped with a photon-counting 
detector sensitive to emission in the 1600--8000~\AA\ wavelength range. 
The effective UVOT field of view is $17' \times 17'$, with a 
point-spread function (PSF) of $\sim 2''$ full width at half-maximum
intensity (FWHM). UVOT features six filters, with three in the 
UV wavelength range ($uvw2$, $\lambda_c \approx 1928$~\AA; 
$uvm2$, $\lambda_c \approx 2246$~\AA; $uvw1$, $\lambda_c \approx
2600$~\AA) and three in the optical ($u$, $b$, $v$). 
The filter transmission curves are shown 
in Figure \ref{transcurves}, compared to the {\it HST} $F220W$ and $F250W$ 
filters. 

The $uvw1$ filter transmission peaks at a wavelength similar 
to that of the {\it HST} $F250W$ and $F275W$ filters;
however, the $uvw1$ filter has greater sensitivity to redder-wavelength 
emission. For the {\it HST} SN 1992A spectrum, this $uvw1$ ``red leak" 
leads to 52\% of the emission emanating from wavelengths longer than 
3000~\AA\ compared to 36\% and 46\% for the $F250W$ and $F275W$ filters,
respectively. The $uvw2$ filter transmission peaks at 
$\sim$1900~\AA, but also has sensitivity to redder-wavelength 
emission. Using the {\it HST} SN 1992A spectrum, the $uvw2$ red leak 
leads to 44\% of the emission emanating from wavelengths longer than 
3000~\AA\ compared to 10\% for the {\it HST} $F220W$ filter. 
Due to the observed tendency of SN emission
to peak toward the red edge of each filter, the filter center is not
the ideal wavelength for characterizing each filter. As extinction further
reddens the SN emission, highly extinguished SNe~Ia suffer the most from 
this effect. For our study, we treat the highly extinguished SNe~Ia
separately in order to minimize this effect. UV photometry of SN 2005cf 
shows encouraging agreement between UVOT and {\it HST} filters 
($F250W/uvw1$, $F220W/uvm2$; Wang et al. 2009a). 

In our companion paper (B10), a spectrum of SN 1992A (Fig. 1) 
is employed to approximately remove the red leak from the $uvw1$ and 
$uvw2$ filters, creating $uvw1_{rc}$ and $uvw2_{rc}$ pseudofilters. That 
paper concentrates on the absolute near-UV magnitudes of SNe~Ia, which 
justifies the additional effort to create pseudofilters. Here we 
present the original photometry with a note of caution that,  
especially for the $uvw2$ filter, the light collected is not 
entirely near-UV emission. 

The $uvm2$ filter is a medium-bandpass filter optimized for the 
2000--2500~\AA\ wavelength range. This filter has the 
advantage of having a sharp red cutoff, so it has the
largest fraction (99\%) of its total emission blueward of 
3000~\AA\ for the {\it HST} SN 1992A spectrum.  
The disadvantage of this filter is that the narrow bandpass 
leads to lower signal-to-noise ratios for our SN sample. 
As the $uvw2$ and $uvm2$ filters feature a high degree of overlap, we 
will emphasize the comparison and contrast between the 
features seen in each filter. 

The UVOT $u$-band filter detects bluer emission than 
ground-based $U$-band filters. Although the instrumental filter transmission 
curves are similar for ground-based systems with optimal blue sensitivity, 
the atmospheric absorption of light toward the blue edge of the ground-based 
$U$ band results in the UVOT $u$ band being more sensitive 
to light in the $\sim$3000--3500~\AA\ range. As 
atmospheric conditions change by site and by night, comparisons between 
ground-based $U$-band systems are particularly difficult \cite{Suntzeff_1996}. 
One goal for future investigations is to attempt such comparisons by 
determining whether the differences between spectral responses can 
account for the differences between UVOT $u$-band photometry 
and photometry in the $U$ band from various ground-based systems, 
employing S-corrections (see Stritzinger et al. 2002).

In this work, we include ground-based $U$-band photometry for a handful of 
SNe~Ia to give the reader the ability to compare our findings with 
previous studies. In situations where the same SN was observed with 
both the ground-based $U$ and UVOT $u$ systems, we briefly discuss the 
level of agreement between the two photometric sets, and we use this
to estimate how much UVOT $u$ and ground-based $U$ photometry can 
differ.    

The UVOT $b$ and $v$ filters are largely comparable to ground-based 
$B$ and $V$ systems. As this paper concentrates on the UV filters, 
we will present S-corrections between UVOT filters and ground-based 
systems elsewhere. 

A large fraction of the observed SNe occurred in regions where 
emission from the host galaxy was an important consideration. 
The methodology of host-galaxy removal is different 
for the UVOT detector than for standard CCD photometry, due to the
photon-counting nature of the UVOT detector. After obtaining a 
post-SN image, rather than subtracting that image from the science
images, the 3$\arcsec$ aperture is placed at the location of the SN 
in the post-SN image and the counts attributed to the host galaxy are
subtracted from the SN counts. This is dictated by coincidence loss 
(similar to pile-up in X-ray instruments), which makes it 
critical to account for the full number of counts in each pixel in 
every image. See Brown et al. (2009) for a complete description 
of the method employed. 

The majority of our SN photometry comes from Brown et al. (2009). 
UVOT photometry of SN~2007ax is from Kasliwal et al. (2008), while 
UVOT photometry of SN~2008ha is from Foley et al. (2009) and 
UVOT photometry of SN~2007gi is from Wang et al. (in preparation). 
In Appendix A, we present new UVOT photometry for SNe 2005mz, 2007on, 2007sr, 
2008A, 2008Q, and 2008ec following the same photometric procedure and 
galaxy subtraction as in Brown et al. (2009).

\section{UVOT SN Ia Sample}

Between March 2005 and March 2008, {\it Swift} UVOT observed 28 SNe~Ia,
and we have chosen to include 25 of those 28 in this study. 
SNe 2005bc and 2005gj were detected at
a single epoch in the UV filters, and are not included here.
SN 2007bm suffered from high extinction, with the additional 
difficulty of the host-galaxy count rate being very high in the 
$b$ and $v$ bands. It is excluded from this work, but 
UVOT photometry will be combined with ground-based photometry for  
inclusion in a future study that uses SNe~Ia to probe extinction 
(see \S 5.1.4).  

\subsection{SN Ia Subclasses}

It has long been recognized that there exists diversity in both
the light curves and optical spectra among the 
events categorized as SNe~Ia (see Filippenko 1997 for a review).
Subsequent to the recognition of this diversity, there was 
the development of a 
single-parameter categorization that placed individual events 
along a continuum of SNe~Ia. This single parameter has been 
based upon the light-curve shape 
($\Delta m_{15}(B)$: \citep{Phillips_1993, Hamuy_etal_1996shapes}; 
$\Delta$: \citep{Riess_96MLCS,Jha_etal_2006b}; 
stretch: \citep{Goldhaber_etal_2001}); $C_{12}$: \citep{wang03}),
spectral features \citep{Nugent_etal_1995, hach08}, and colors at
peak ($B-V$: \citep{Phillips_etal_1999,Garnavich_etal_2004}; $U-B$: 
\citep{Jha_etal_2006}), with a high level of correlation between 
the various techniques. The large majority of SNe~Ia 
have been suitably categorized by these parameters, but there 
have been notable outliers from single-parameter categorization
\citep[e.g.,][]{Li_etal_2001,Li_etal_2003}.

Of the SNe~Ia that are suitably categorized by a single parameter, 
we will present the events according to their $B$-band decline in the
first 15 days after maximum brightness, the $\Delta m_{15}(B)$ 
value. This parameter has the benefit of simplicity,
and our categorization does not require high precision.
We separate normal and superluminous events from subluminous 
events at $\Delta m_{15}(B) \approx 1.6$ mag; the 
fast-declining SNe~Ia are typically subluminous (e.g., SN 1991bg,
Filippenko et al. 1992b; Leibundgut et al. 1993). We do not 
discriminate between normal and superluminous events (e.g., SN 1991T,
Filippenko et al. 1992a; Phillips et al. 1992). 
None of the SNe in this study are of the peculiar ``overluminous" 
group (possibly super-Chandrasekhar mass; Howell et al. 2006; 
Hicken et al. 2007); however, SN 2009dc (Yamanaka et al. 2009) is a
member of that group and it will be included in our follow-up studies. 
We note that a number of observed characteristics of SNe~Ia 
show that the  normal and/or superluminous events vary relatively 
little, whereas subluminous events can differ 
appreciably. Examples of this are the peak 
$B-V$ colors \citep{Garnavich_etal_2004}, 
the late-time optical light curves (Milne, The, \& Leising 2001), and the  
peak $U-B$ colors \citep{Jha_etal_2006}. 
The subluminous events are redder than normal events, and feature 
lower expansion velocities, lower $^{56}$Ni yields, stronger Ti~II lines, 
lower ionization, and 
possibly higher intrinsic polarization 
\citep{Fil_etal_1992b,Leibundgut_etal_1993,howell01,Garnavich_etal_2004}.

There are indications of variations within the group of ``normal" 
SNe~Ia. It has been suggested that the velocity of the absorption 
lines of intermediate-mass elements in the spectra near peak brightness
might permit subdivision of the normal subclass 
\citep{Benetti_etal_2004,Benetti_etal_2005,Wang_etal_2006,Wang_etal_2008}. 
This subdivision distinguishes high-velocity (HV) normal SNe~Ia 
from normal-velocity normal SNe~Ia, with the inference that the HV group 
might have different progenitor properties and/or the presence of 
circumstellar matter (CSM) surrounding the SN event. Wang et al. (2009b) 
applied this method to a large sample of normal SNe~Ia, including a
majority of the normal events in this study. 

As mentioned above, some SNe~Ia are not suitably categorized 
by these single parameters. Their anomalous nature can reveal itself in 
light curves explainable by none of the range of templates,
or by spectral characteristics that do not match expectations based 
upon the light-curve shapes. In particular, SN 2002cx 
\citep{Fil03,Li_etal_2003} has emerged as 
the namesake event for an emerging subclass of SNe~Ia. 
SN 2002cx-like events are characterized by high-ionization iron 
lines in the premaximum spectra, but low expansion velocities 
and low peak luminosities 
\citep{Jha_etal_2006b,Phillips_etal_2007,Vale09,fole09}. 
Furthermore, the $R$ and $I$-band light curves did not feature the
normal secondary 
maxima. It has been suggested that this subclass of SNe~Ia 
may be due to a deflagration, rather than a delayed detonation 
\citep{Phillips_etal_2007}. Alternatively, perhaps
SN 2002cx-like events are not of thermonuclear origin, but rather a variety 
of stripped-envelope, core-collapse events (Valenti et al. 2009, but see 
Foley et al. 2009). 

\subsection{SN Ia Sample}

All of the SNe presented in this study were also observed with 
ground-based telescopes. Those observations have been used with the 
UVOT data to categorize the SNe in this sample. 
The Carnegie Supernova Project (CSP) observed SNe 2005am, 2005hk, 2005ke, 
2007co, 2007cq, 2007cv, 2007gi, and 2007on.  
The Harvard-Smithsonian  
Center for Astrophysics supernova program (CfA) observed 
SNe 2005am, 2005cf, 2005hk, 2005ke, 2005mz, 2006ej, 2007S, 2007af, 
2007ax, 2007co, 2007cq, 2007cv, and 2008A. Using the 0.76-m
Katzman Automatic Imaging Telescope (KAIT; Filippenko et al. 2001;
Filippenko 2003) at Lick Observatory, the U.~C. Berkeley SN group observed
SNe 2005am, 2005cf, 2006dm, 2006ej, 2007co, 2007cq, and 2008ha. 
The Australian National University observed SN 2005df; that photometry 
is presented in Appendix B.

Tables \ref{snparams_0506} and \ref{snparams_0708} 
show the basic optical parameters for each SN~Ia 
included in this study. When multiple $\Delta m_{15}(B)$ values 
were available, we use the average of the values in Tables 1 and 2. 
Fifteen of the SNe have $\Delta m_{15}(B)$ values
(hereafter referred to as ``decline rates'') 
less than 1.6 mag, by far the largest group. Two more SNe~Ia do not have 
known $\Delta m_{15}(B)$ values, one was spectroscopically categorized as
a member of the normal subclass (2006E), the other as a member of the 
SN 2002cx-like subclass (2008A). Five have faster decline 
rates, with a few of the SNe with undetermined decline rates being 
likely members of this subclass based upon spectroscopic properties. 
The optically brightest events, in terms of apparent magnitude, were 
SNe 2005cf, 2005df, and 2007af; they are three of the best-studied 
SNe~Ia in the sample. The high-velocity normal SNe~Ia are listed in 
column 4 as ``NHV," the normal-velocity normal SNe~Ia as ``NNN," and 
the normal SNe~Ia with undetermined absorption velocity features as ``N."

\section{Fitting Light Curves}

For the best-sampled subset of the UVOT SN~Ia dataset, the fitting of 
functions to 
the multi-band light curves permits the direct study of individual 
events. We explore specifically the UV light curves compared to the 
optical light-curve shapes for the purpose of probing the SN~Ia 
events. The derived peak magnitudes contribute 
to the absolute-magnitude study presented by B10. We also 
emphasize considerations related to the utilization of SNe~Ia 
as very high-redshift distance indicators. Observations of SNe~Ia at 
very high redshifts sample little or none of the rest-frame optical 
wavelength range, so individual events must be categorized according
to their rest-frame UV emission. In this section, 
we explore possible indicators of SN~Ia subclass from 
UV emission. 

\subsection{Fitting with Functions}

Fifteen SNe~Ia were determined to be adequately observed in 
the UVOT $uvw1$ filter to warrant fitting the light curves with 
functions. Seven SNe~Ia were well observed in the $uvw2$ filter, 
ten in the $u$ and $b$ filters, and twelve in the $v$ filter. 

\begin{figure}
\epsscale{1.0} \plotone{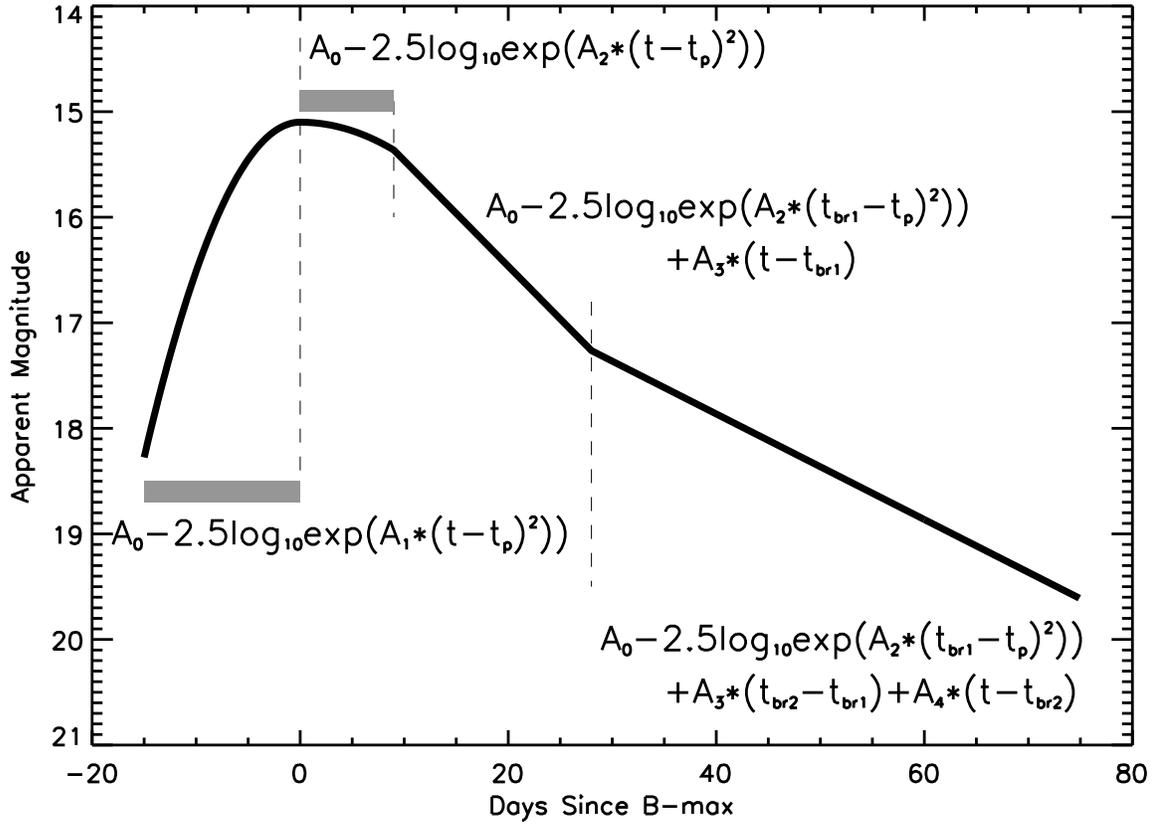}
\caption{A schematic of the function fit to the SN light curves.
The function is characterized by eight parameters: $A_{0}-A_{4}$, $t_{p}$, 
$t_{br1}$, and $t_{br2}$.}
\label{lc_schematic}
\end{figure}

The complete function chosen contains four terms: a 
Gaussian rise to peak, an initial Gaussian decline,  
a middle-epoch linear decline, and a late-epoch 
linear decline (Fig. \ref{lc_schematic}). 
In total, eight parameters are used for the 
complete function (``cmplt"), as follows:

$m =  \left\{
\begin{array}{lr}
A_{0} -2.5\,{\rm log}_{10}({\rm exp}(A_{1}(t-t_{p})^{2})) \hspace{8mm} 
({\rm before~peak})
                         &  t \leq t_{p}, \\
A_{0} -2.5\,{\rm log}_{10}({\rm exp}(A_{2}(t-t_{p})^{2})) \hspace{5mm} ({\rm Gaussian~decline})
       & t_{p} \leq t \leq t_{br1} \\
A_{0} -2.5\,{\rm log}_{10}({\rm exp}(A_{2}(t_{br1}-t_{p})^{2}))
      + A_{3}(t-t_{br1}) & \\
\hspace{2in} ({\rm steep~linear~decline}) & t_{br1} \leq t \leq t_{br2}  \\
A_{0} -2.5\,{\rm log}_{10}({\rm exp}(A_{2}(t_{br1}-t_{p})^{2}))
      + A_{3}(t_{br2}-t_{br1}) 
    &  \\
\hspace{0.8in} + A_{4}*(t - t_{br2}) \hspace{6mm} ({\rm late~linear~decline}) &  t \geq t_{br2}.
\end{array}
\right.
$



\begin{figure}
\plotone{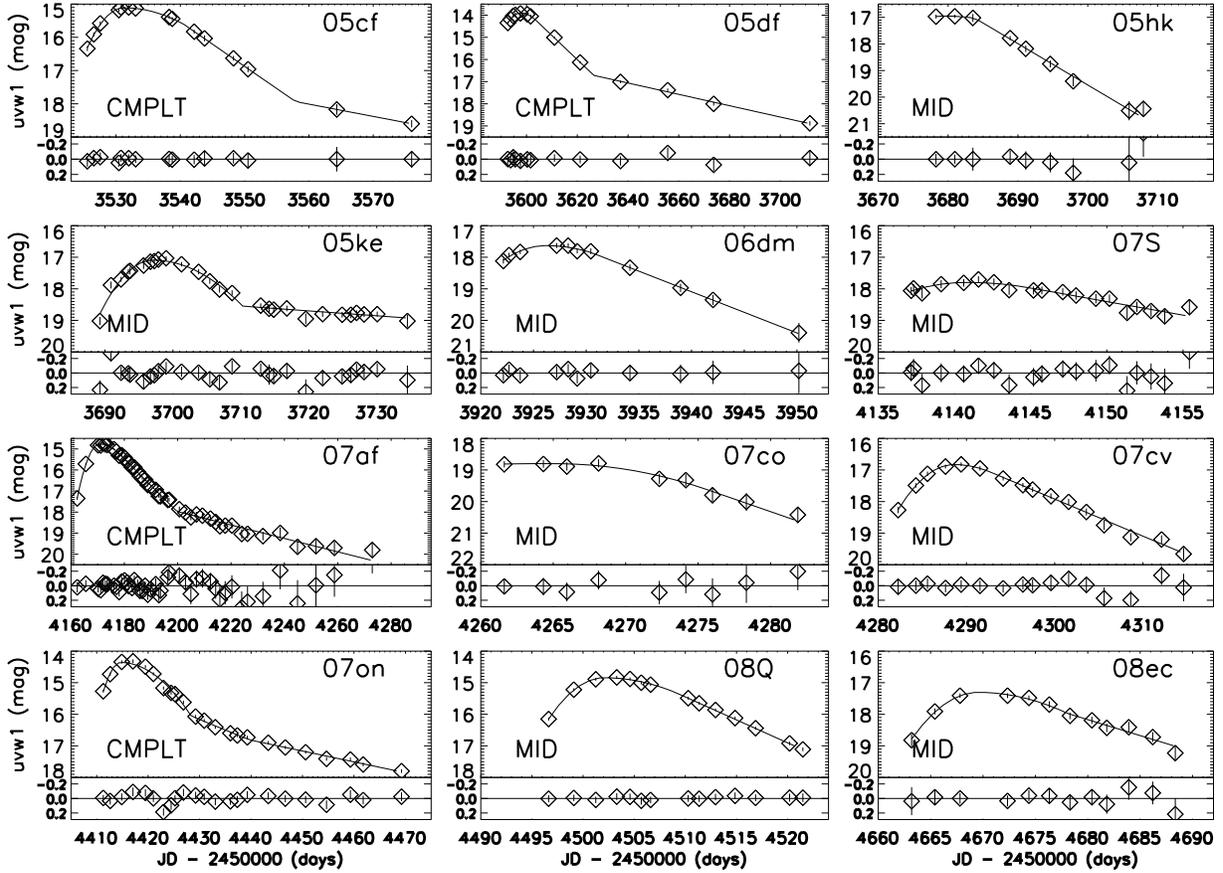}
\caption{$uvw1$-band light curves of twelve SNe~Ia observed near peak, 
fitted with a multi-component function, with residuals. 
The dataset for SN 2005ke was trunctated to allow fitting. 
See Fig. \ref{late_break_lc} for later-epoch data.} 
\label{fit_w1}
\end{figure}

\begin{figure}
\epsscale{1.0} \plotone{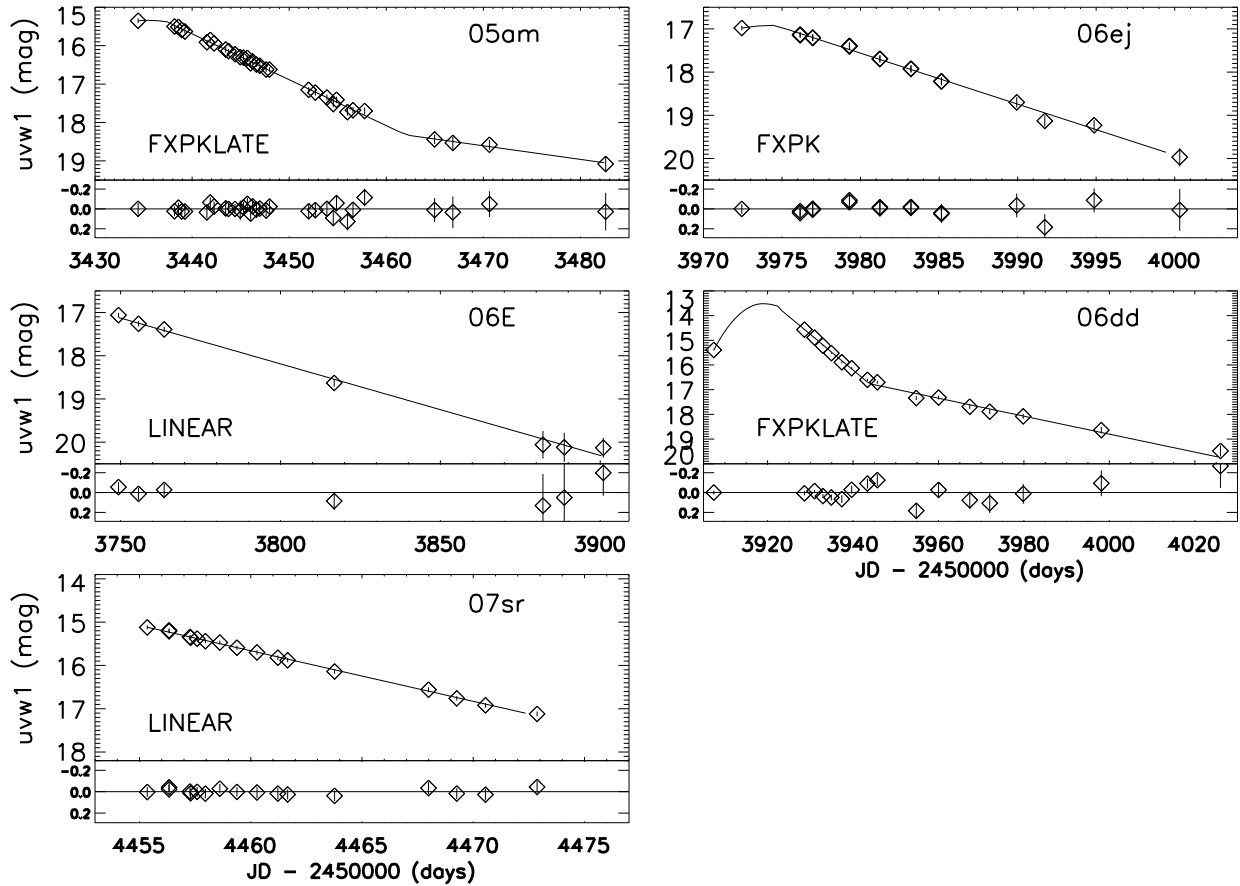}
\caption{$uvw1$-band light curves of five SNe~Ia not observed during the 
peak phase. Light curves were fitted with a model that fixed the 
Gaussian rise and decline constants.}
\label{fix_peak}
\end{figure}

\begin{figure}
\epsscale{1.0} \plotone{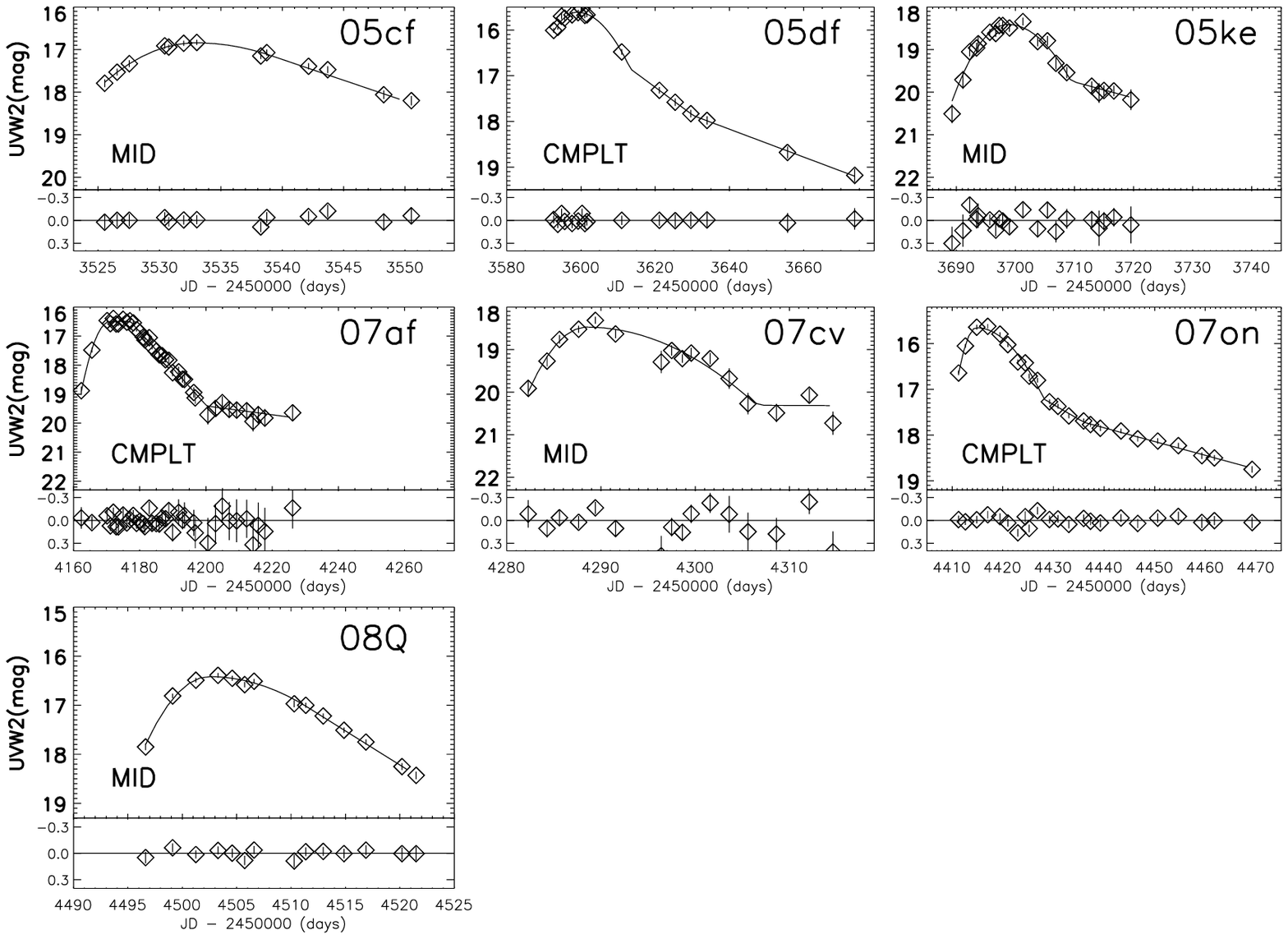}
\caption{$uvw2$-band light curves of seven SNe~Ia, fitted with a 
multi-component function, with residuals. The dataset for SN 2005ke was 
truncated, as in Fig. \ref{fit_w1}.}
\label{fit_w2}
\end{figure}

\begin{figure}
\epsscale{1.0} \plotone{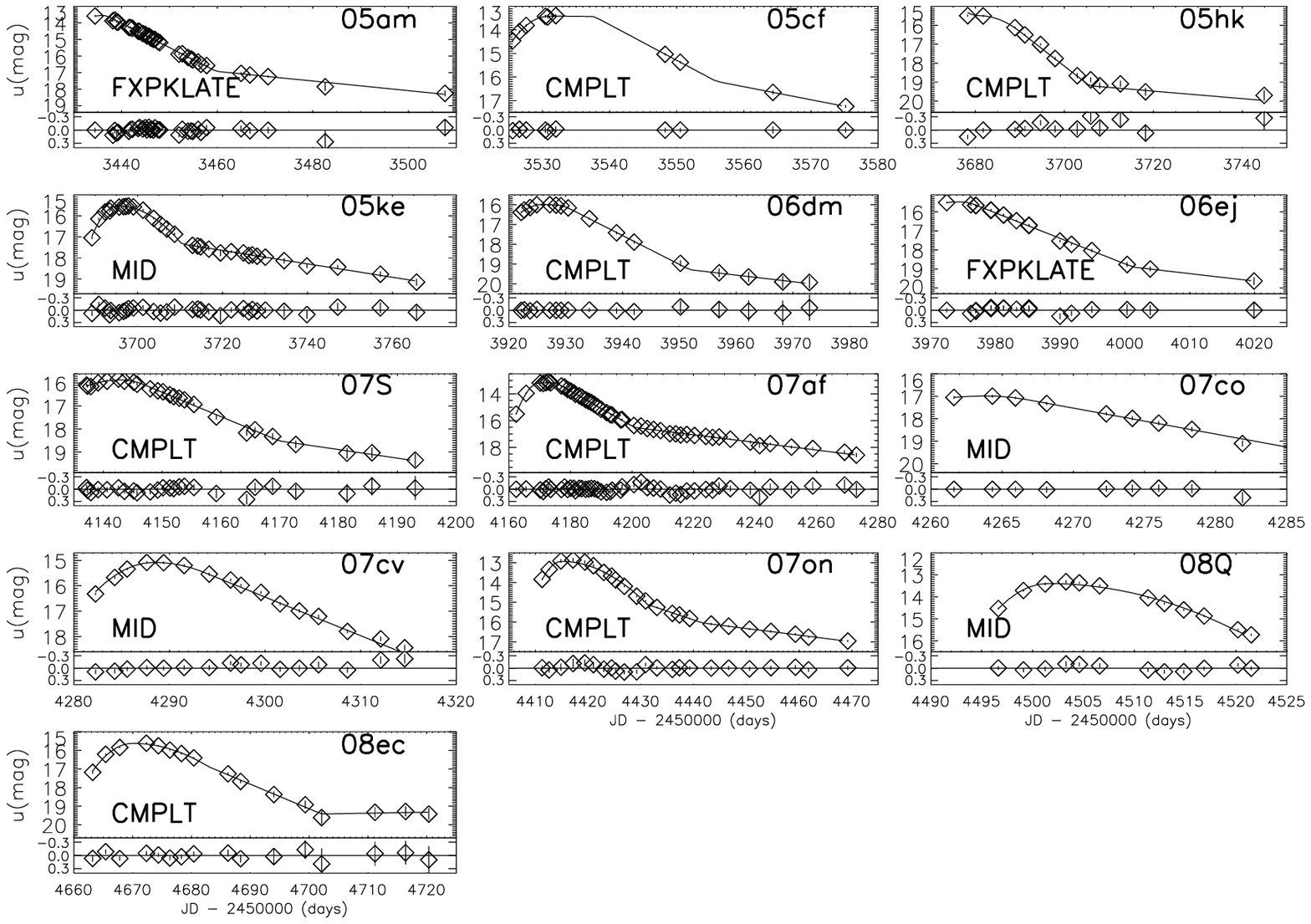}
\caption{$u$-band light curves of ten SNe Ia, fitted with a multi-component
function, with residuals. The dataset for SN 2005ke was truncated, as in
Fig. \ref{fit_w1}.} 
\label{fit_u}
\end{figure}

\begin{figure}
\epsscale{1.0} \plotone{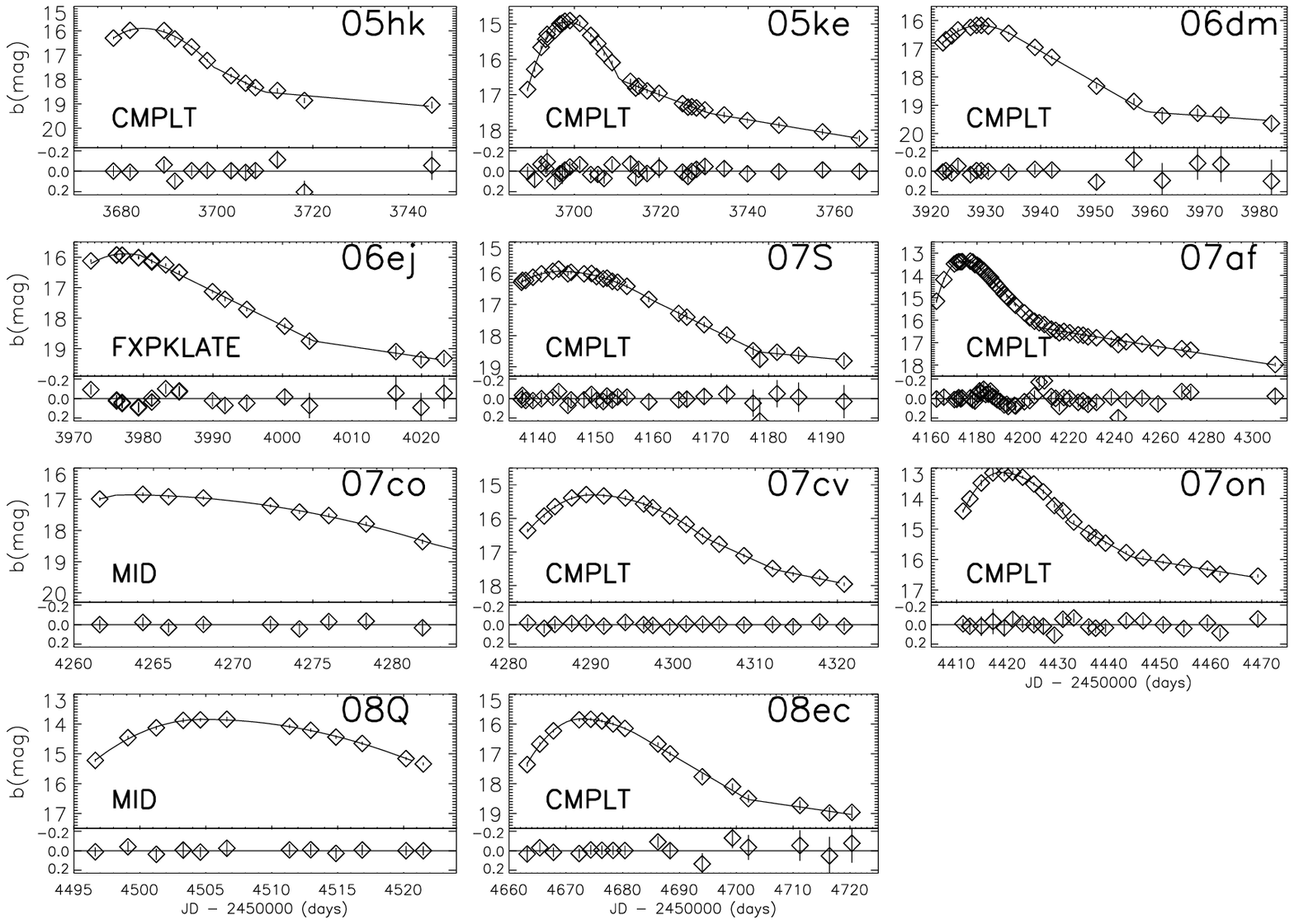}
\caption{$b$-band light curves of ten SNe~Ia, fitted with a multi-component
function, with residuals. The dataset for SN 2005ke was truncated, as in
Fig. \ref{fit_w1}.}
\label{fit_b}
\end{figure}

\begin{figure}
\epsscale{1.0} \plotone{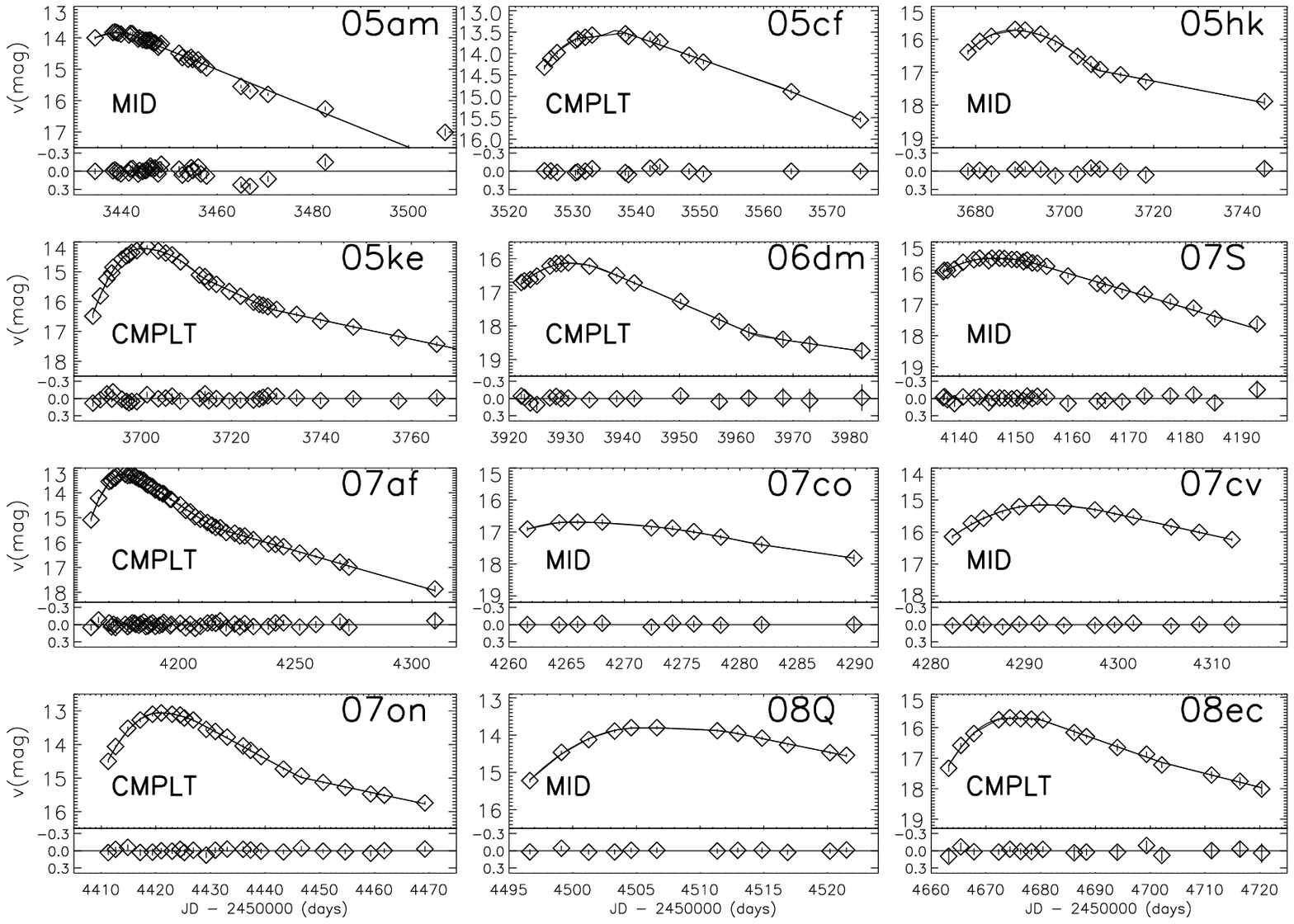}
\caption{$v$-band light curves of twelve SNe~Ia, fitted with a multi-component
function, with residuals. The dataset for SN 2005ke was truncated, as in
Fig. \ref{fit_w1}.}
\label{fit_v}
\end{figure}

\noindent
The fitting function is not physically motivated; it was chosen with 
the goal of approximating the light curves with the minimum number 
of parameters. We follow the methodology of 
Contardo, Leibundgut, \& Vacca (2000), who note that 
Pinto \& Eastman (2000) predict a Gaussian shape for the peak in 
models with constant opacity and $^{56}$Ni buried well within the ejecta.   
The fitted light curves are shown in Figures \ref{fit_w1}--\ref{fit_v}. 
Tables \ref{fit_params_uv}, \ref{fit_params_opt}, and \ref{fit_params_late} 
show parameters derived from those fits. 

For many of the SNe, the campaigns were incomplete,  being 
sparsely sampled in one of the four epochs. For 
these SNe, the adopted model either eliminated the late-epoch 
linear decline (``mid": $A_{4}$ term) or 
both linear declines (``peak": $A_{3}$ and $A_{4}$ terms). 
SNe poorly sampled at peak but well observed post-peak 
were fitted with ``fix-peak" models, where the $A_{1}$ and 
$A_{2}$ terms were fixed to be 0.013 and 0.008, respectively 
(``FXPKLATE": fit $A_{0}$, $A_{3}$, and $A_{4}$; ``FXPK": fit 
$A_{0}$ and $A_{3}$). SNe only observed at late epochs were 
fit with a straight line (``LINEAR").    
For all models, the fitted parameters were not constrained, 
but SNe~Ia that were best fitted with models in which the peak, 
first time break, and second time break were not positively 
separated in time were fit instead with ``fix-peak" models. 
In all cases, we attribute 
the fitting problems to be a result of sparse sampling, and we 
treat those SNe in \S 4.2. 
Removing a decline epoch results in elimination of two parameters. 
The model used for each SN is listed in Table \ref{fit_params_uv}. 
We note that for SN~2005ke, the dataset was truncated about 
25 days after the late break for fitting, as the light-curve shape differs 
appreciably from the function shown above.

The $uvw1$, $uvw2$, and $u$ bands all peak days earlier than the $b$ band. 
This can be seen by comparing (in Tables 3 and 4) columns 2, 7, and 
13 with column 18. 
\citep{Jha_etal_2006} report that the $U$ band peaks $2.3 \pm 0.4$ days 
before the $b$ band for their sample of 44 SNe~Ia. We find similar 
early peaking for a smaller sample. One SN~Ia stands out as 
peaking much earlier (5.6 days), SN~2005hk. The UV peaks of SN~2005hk 
were not well constrained at early epochs, but were definitely 
earlier than the normal SNe~Ia. We will show in $\S$5 that this is consistent 
with the UV color curves, where the bluest emission was seen in the first 
observations. 

Using the information obtained from fitting the normal SNe~Ia 
(Fig. \ref{fit_w1}), we fixed the peak parameters to 
fit the $uvw1$ light curves of four normal SNe~Ia either observed 
only post-peak (SNe 2005am, 2006ej, 2006E) or observed before 
the peak phase, but not within 5 days of the peak (SN~2006dd).  
The resulting fitted light 
curves are shown in Figure \ref{fix_peak}. The light curves 
either exhibit a late break, or constrain such a break to being 
later than 20 days past maximum brightness.

\subsubsection{The $\Delta m_{15}(i)$ Parameter}

\begin{figure}
\epsscale{1.0} \plotone{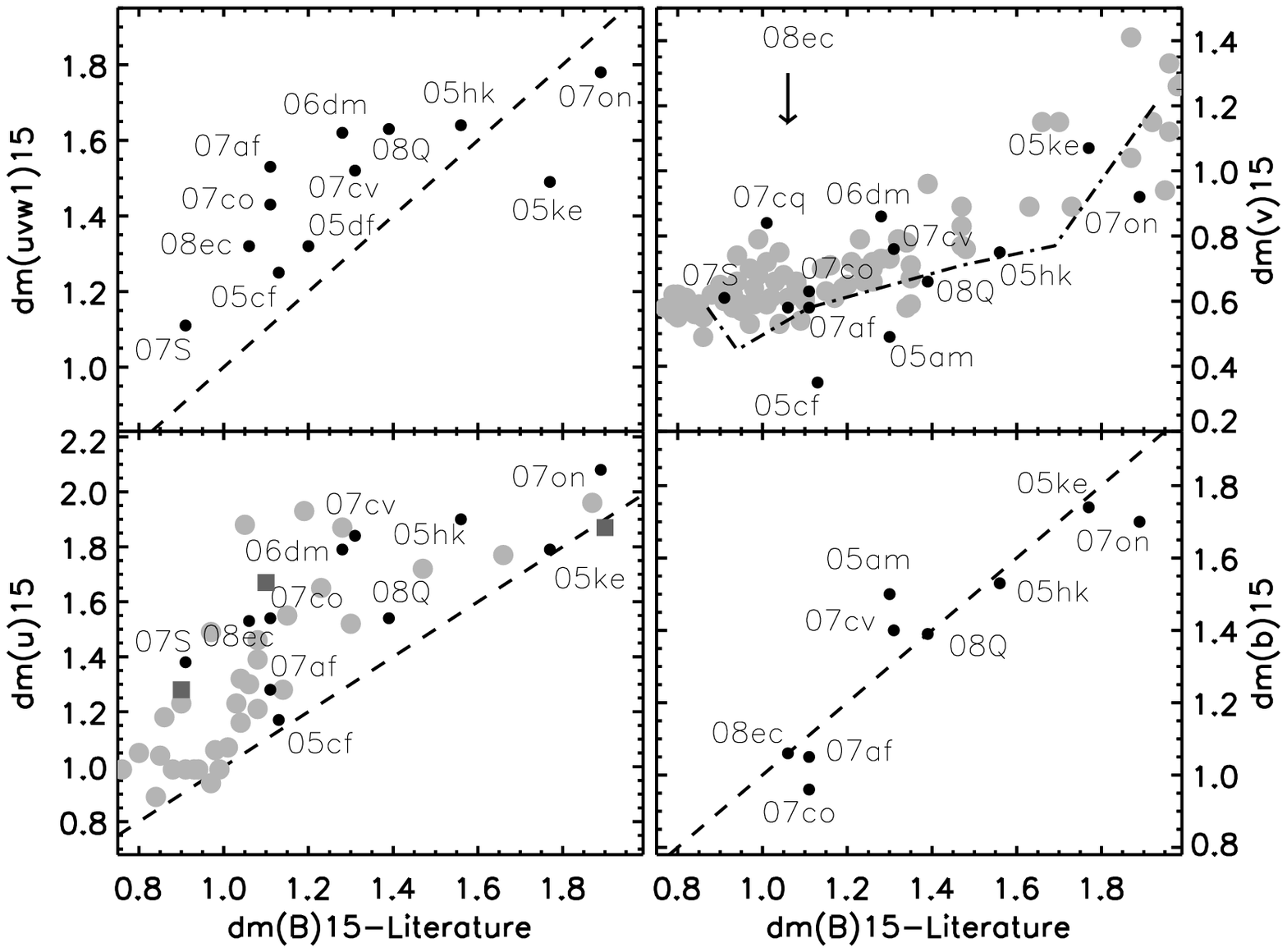}
\caption{15-day decline rates for UVOT data. The $uvw1$-band decline 
rates are compared to the $B$-band rates (upper-left panel), the  
$u$-band rates are compared to the $B$ band (lower-left panel), and the 
$v$-band rates are compared to the $B$ band (upper-right panel). UVOT 
$b$-band rates are compared to ground-based $B$-band values for the 
same SNe (lower-right panel); the dashed line indicates equality.
The dot-dashed line shows values for SN~Ia templates
(Hamuy et al. 1996). Values of $\Delta m_{15}(U)$, $\Delta m_{15}(B)$, 
and $\Delta m_{15}(V)$ from 101 SNe~Ia in the CfA3 study are shown as 
grey circles. Values of $\Delta m_{15}(U)$ versus $\Delta m_{15}(B)$ 
for the overluminous, normal 
and subluminous templates from Nugent are shown as grey squares.} 
\label{decline_rates}
\end{figure}

As a check of the fitting outputs for UVOT data, in 
the lower-right panel of Figure~\ref{decline_rates} we compare the 
$\Delta m_{15}(b)$ values from fitting the UVOT data to the
$\Delta m_{15}(B)$ values from 
ground-based telescopes (see Tables 1 and 2 for references). 
 There is reasonable agreement between 
the two methods, as the mean $\Delta m_{15}(b)$ - $\Delta m_{15}(B)$ 
(i.e., UVOT minus literature) value of $-0.02 \pm 0.12$ mag is 
consistent with zero. 

Extending the comparisons of the UVOT study to other ground-based 
studies, the UVOT $v$-band values in the  $\Delta m_{15}(b)$ versus 
$\Delta m_{15}(v)$ plane (Fig. \ref{decline_rates}, upper-right panel) 
agree well with the \citep{Hamuy_etal_1996shapes} templates and 
with 101 SNe~Ia in the CfA3 study. Comparing $\Delta m_{15}(b)$ versus 
$\Delta m_{15}(u)$ (Fig. \ref{decline_rates}, lower-left panel), 
we see that the UVOT $u$-band data largely agree with the 53 CfA3 
SNe~Ia and with the P. Nugent (2009, private communication) 
$U$-band templates for superluminous, 
normal, and subluminous SNe~Ia. The CfA3 study suggests a correlation 
between $\Delta m_{15}(B)$ and $\Delta m_{15}(U)$. As the UVOT dataset 
generally agrees with the CfA3 dataset in terms of $\Delta m_{15}(B)$ 
and $\Delta$m$_{15}(U)$, that suggestion is largely supported, 
albeit with increased scatter for the fast-declining events. 

The comparison between $\Delta m_{15}(B)$ and $\Delta m_{15}(uvw1)$ 
can only be performed with UVOT data (Fig. \ref{decline_rates}, 
upper-left panel). This comparison is particularly 
interesting, as the peak width has been the dominant method used 
to determine SN~Ia subclass for high-redshift SN searches. The 
trends are generally the same as seen for $\Delta m_{15}(B)$ versus  
$\Delta m_{15}(u)$, although for this comparison, 
SN~2005ke stands out with a relatively slow $uvw1$ decline rate. 
The decline rates are not as well distributed in $\Delta m_{15}(uvw1)$, 
appearing to clump at 1.3 and 1.6 mag.

For the goal of characterizing the SN subclass via purely
UV observables, $\Delta m_{15}(uvw1)$ would
be inadequate. This is of particular importance for the cosmological
utilization of SNe~Ia at high redshifts, where rest-frame UV emission
would be shifted into optical or even NIR wavelength ranges.
Although the peak width has been the dominant method used to determine 
SN~Ia subclass for high-redshift SN searches for rest-frame optical 
emission, these light curves suggest that single-band UV light-curve peak 
widths would be inadequate to categorize precisely a given SN~Ia. 

\begin{figure}
\epsscale{1.0} \plotone{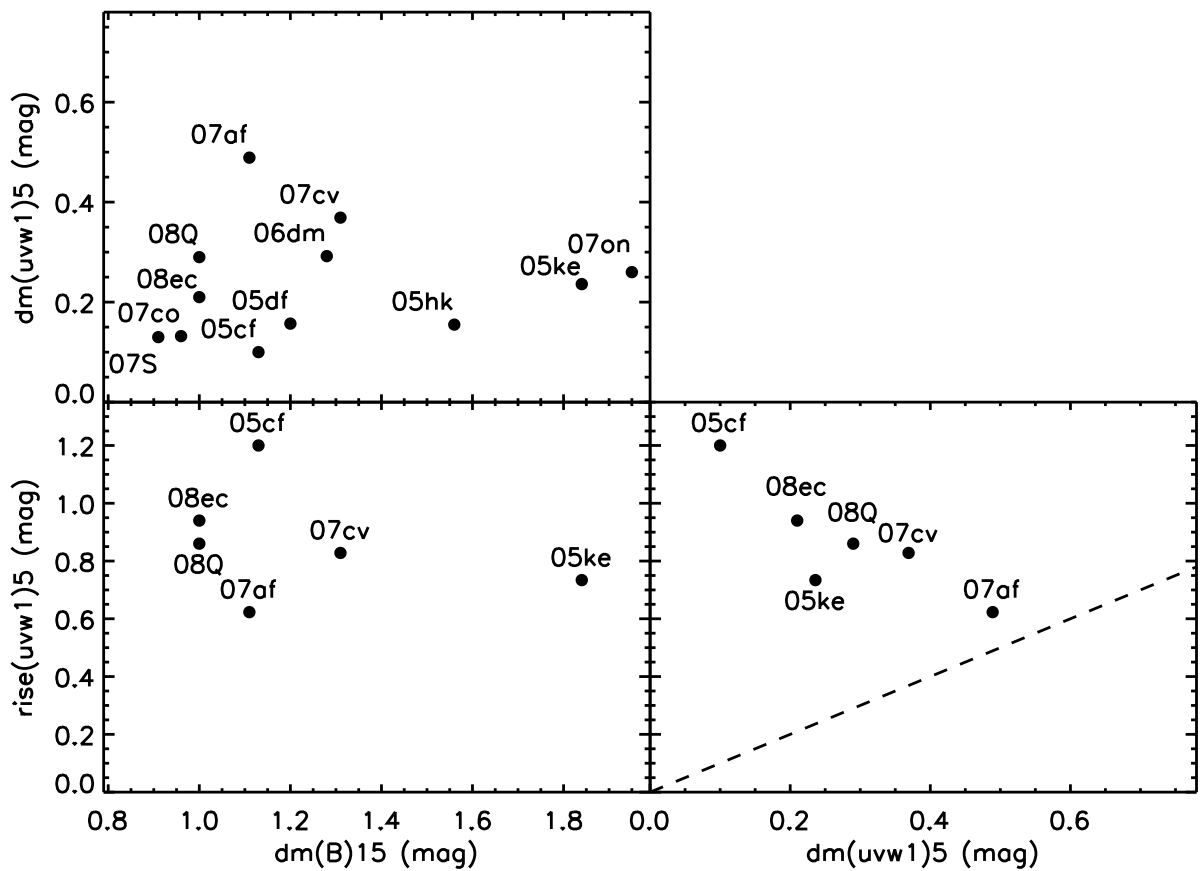}
\caption{The 5-day rise (lower left) and 5-day declines (upper left) 
of the $uvw1$ emission from twelve SNe~Ia. Six SNe~Ia had measurements 
for both parameters, which are plotted against each other in the 
lower-right panel. The light curves rise to the peak more steeply than 
they decline from the peak. Values of $\Delta m_{15}(B)$ are taken from 
Tables 1 and 2.}  
\label{rise_vs_dec}
\end{figure}

The 5-day rise to peak and decline from peak are shown in Figure 
\ref{rise_vs_dec}; clearly, within 5 days of peak brightness
the light curves rise faster than they fall. 
There are no other obvious tendencies in the rise and decline rates 
relative to the optical decline rates, as  
correlation with the $\Delta m_{15}(B)$ 
parameter is not apparent to the statistical precision of these
data (Fig.~\ref{rise_vs_dec}, upper-left and lower-left panels). 
There is the appearance of an inverse correlation between the 
5-day rise and the 5-day decline (Fig. \ref{rise_vs_dec}, lower-right panel), 
but many more SNe~Ia need to be observed to further explore that possibility. 

\subsubsection{Late Break to a Shallow Decline}

\begin{figure}
\plotone{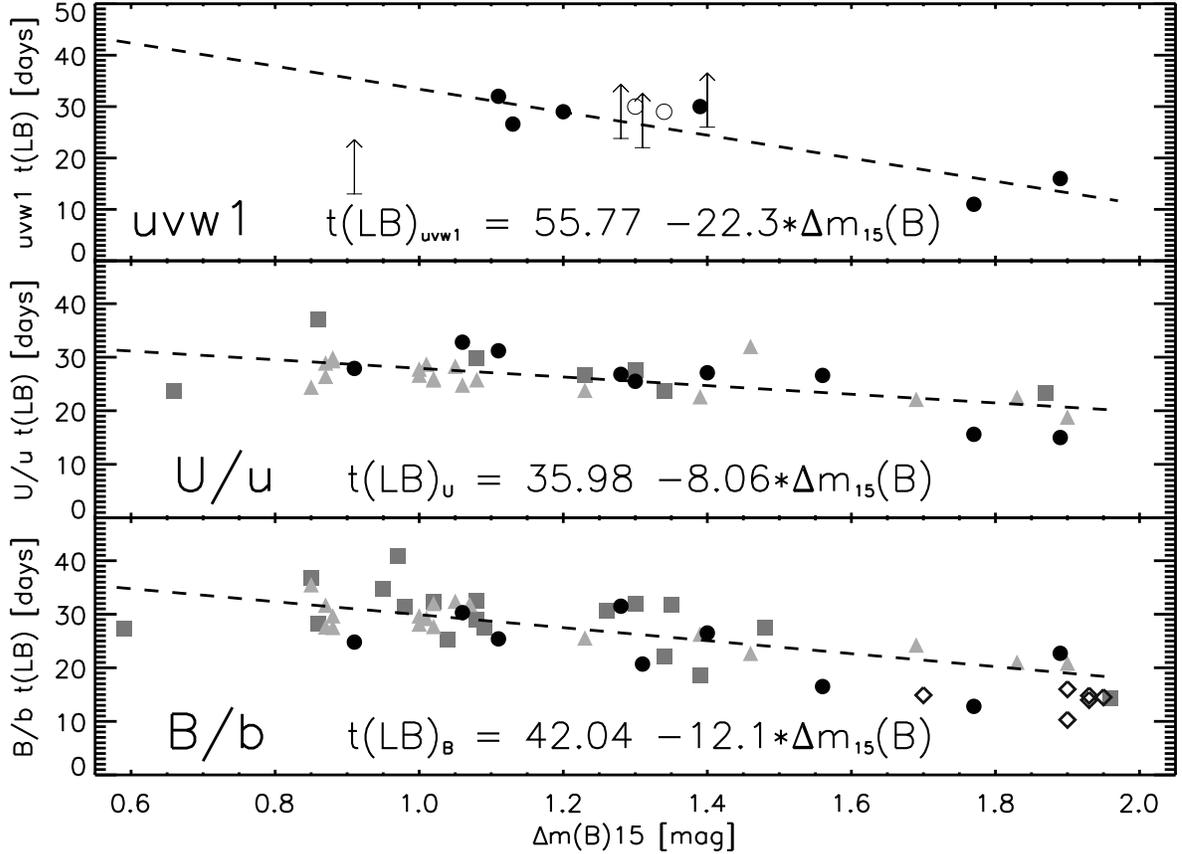}
\caption{Late-time light-curve breaks in the $uvw1$ band (upper panel),
$U$ and $u$ bands (middle panel), and $B$ and $b$ bands (lower panel) for 53
SNe~Ia shown as a function of \protect$\Delta m_{15}(B)$.
UVOT SNe with well-defined peak dates are shown
as filled circles, poorly defined peak dates are
shown as open circles, lower limits as arrows. CfA2 SNe~Ia are
shown as grey triangles, and CfA3 SNe~Ia are shown as grey squares.
Subluminous SNe~Ia from Kasliwal et al. (2008) are shown as
open diamonds. The subluminous
SNe~Ia break sooner than do the normal SNe~Ia. Each filter is
fitted with a dashed line, following the function listed.}
\label{late_break_times}
\end{figure}

\begin{figure}
\plotone{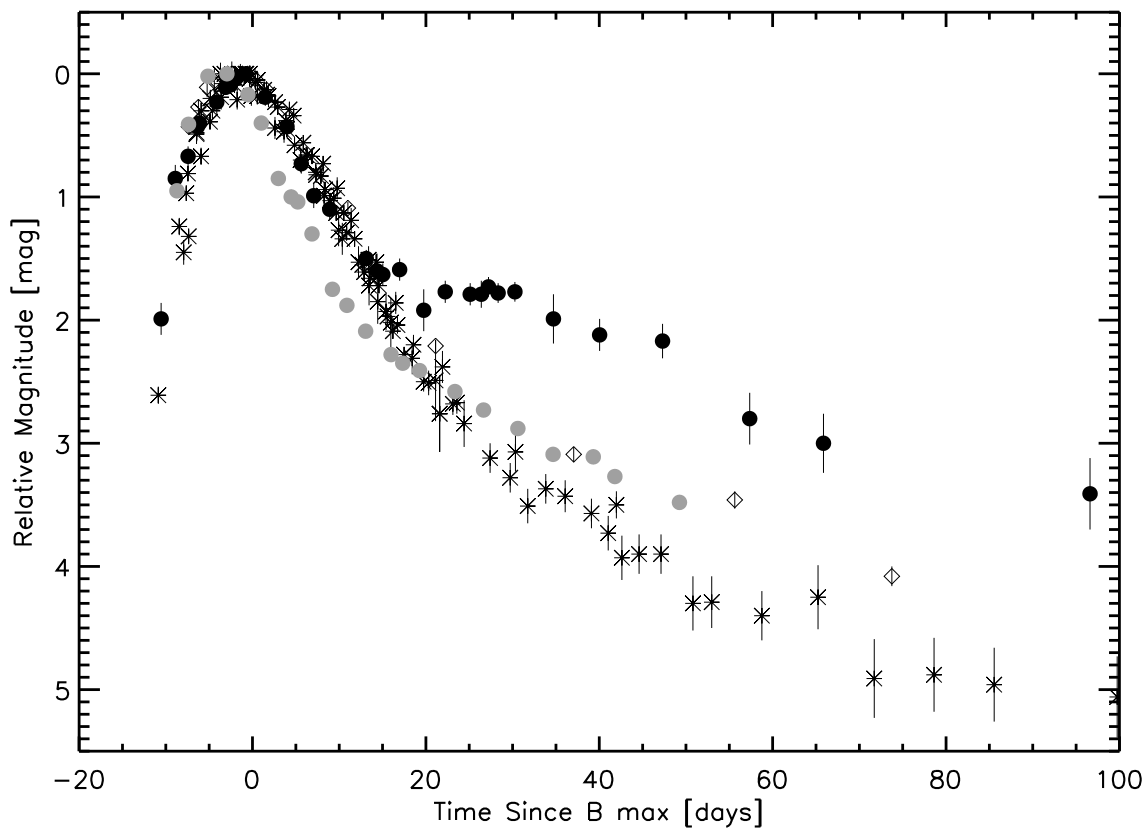}
\caption{$uvw1$-band light curves of SNe~Ia aligned by peak magnitude. 
Seven normal SNe~Ia 
(asterisks/open diamonds) are shown compared to two subluminous SNe~Ia 
(SNe 2005ke and 2007on). 
SN 2005ke (dark filled circles) breaks to a bright plateau, while SN 2007on 
(grey filled circles) 
breaks at an early epoch, but to a flat decline fainter than that of 
SN 2005ke. SN 2005df, a normal-luminosity SN~Ia (open diamonds), 
features an early-time light curve similar to that of the 
other normal SNe~Ia, but breaks 
at a late epoch to a very shallow decline.}
\label{late_break_lc}
\end{figure}

The delays between the peak in the $uvw1$ filter and the 
late break (at which time the light curves flatten) are shown 
in Figure \ref{late_break_times}. The normal SNe~Ia have delays that 
cluster around $\sim$28 days, whereas the subluminous 
SNe 2005ke and 2007on  
and the SN~2002cx-like SN~2005hk break at far earlier epochs. 
This characteristic could be used to distinguish between normal 
SNe~Ia and these other two subclasses if only rest-frame UV emission 
can be observed. Such an approach has been used in the optical wavelength 
range before \citep{psko84,Hamuy_etal_1996shapes};
it was even employed on one SN in our dataset, SN~2007ax, 
for the $B$ band \citep{Kasliwal_etal_2008}. 
Figure \ref{late_break_lc} shows $uvw1$-band light curves of 
SNe 2005ke and 2007on compared with those of a selection of normal 
SNe~Ia. All SNe have been normalized to the same magnitude at 
$B_{\rm peak}$. The two subluminous SNe~Ia have different light-curve 
shapes, but both break at earlier epochs than do the normal 
SNe~Ia. SN~2005ke has a light-curve shape similar to that of the 
normal SNe~Ia before 
the break, while SN~2007on has a narrower light curve peak before the 
break. SN~2005df is interesting in featuring a late break, but breaking to 
a shallow slope similar to that of SN~2007on.

The initial and late linear slopes, parameterized as $\beta$ and $\gamma$  
respectively (Pskovskii 1984), 
have only been measured for a few SNe~Ia (Table \ref{fit_params_late}). 
A detailed study of the late-time light curves will be performed in a 
future work. It is worth noting that to date, no SN has been observed 
in the UV sufficiently well to determine 
$\gamma_{1}$ (50--150 days) and $\gamma_{2}$ (later than 150 days), a 
splitting of the late-time decline shown in the Salvo et al. (2001) 
study of SN 1996X.    

One unfortunate aspect of utilizing the late-break 
delay to determine SN~Ia subclass is the fact that the SN has 
faded from peak by as much as 4 mag before the existence 
of the late break is apparent. For magnitude-limited surveys, it might 
not be possible to observe a SN at ``peak + 4" mag. 
This problem is similar to the problems that arise by trying to 
estimate the extinction via the Lira (1995) relation, where the SNe must 
be observed 60 or more days after the peak. Nonetheless, there is a 
clear difference between the subclasses in this observed parameter, 
making it useful if no other observable can be found to perform the 
same task. The ``$R_{\rm UV}$" ratio suggested by Foley et al. (2008b) is 
another potential observable. A worthwhile project would be to obtain 
both parameters, as well as optical information, for a collection of 
SNe Ia to test the accuracy of each as a SN subclass predictor. 
The appreciable scatter evident in Figure \ref{late_break_times} 
suggests that the timing of the late break might only be useful 
to eliminate subluminous events from a sample. 

\subsubsection{Limits of Stretch Fitting}

The stretch method has been employed in the optical
bands to show that the light curves of 
a large range of SNe~Ia can be explained as
a function of only the timescale of the light-curve evolution.
\citep{Jha_etal_2006} used stretch to fit $U$-band light curves of a
collection of SNe~Ia from $-10$ to +40 days relative to $B_{\rm peak}$ 
in the SN rest frame, 
following the formalism of Goldhaber et al. (2001). 
We have shown that the peak widths and late-break 
times of these SNe in the $uvw1$ filter do not support the idea that the 
entire range of normal to subluminous events can be explained with the 
stretch of a single, characteristic light-curve shape during that entire 
epoch. This is supported by the  
color evolution shown in Figure \ref{col_N_4panel}, which 
suggests that stretch can be suitably employed on the normal subclass
(as they have similar color evolution and stretch has been employed
with the $V$ band). However, the other subclasses have very different
color evolution, so fitting with stretch would be at best approximate. 
Hicken et al. (2009) arrived at a similar conclusion with the $U$-band 
dataset of CfA3.
The fact that the Nugent 
$U$-band templates cross, due to the flatness of the late-time 
light curves for the subluminous subclass, is further grounds for
rejecting the use of
stretch for the entire range of SN~Ia events in the $U$ band or UV. 
That being said, subluminous
SNe~Ia are not large outliers in relations that compare the $U$-band
stretch with other values as presented by \citep{Jha_etal_2006}.
We do not attempt to characterize exact ranges of SN subclass and 
epoch where stretch would be appropriate, as
this will be better addressed at a later date with a larger sample.

\subsection{Fitting Normal SNe Ia to a Mean Template}

The similarity of the UV light curves of the normal SNe~Ia suggests that 
a single, mean template could be generated for each filter. 
There are a number of advantages to creating a 
mean template, and comparing that against individual SNe~Ia and 
light-curve templates in other filters: (1) the combined dataset better 
samples the range of epochs than any individual SN, (2) the 
templates can be analyzed for structure beyond that revealed in 
 the function fitting performed earlier in this section, 
(3) individual SNe can be compared to search for subtle variations 
from the mean behavior, and (4) the mean template can be used to estimate 
the peak magnitude and peak date for SNe with gaps in their 
sampling. Determining peak magnitudes and dates via template fitting has 
the drawback that it assumes the individual SN evolves exactly as 
the mean template, an assumption that is not fully 
supported by the available light curves. 
The subluminous and SN~2002cx-like subclasses are not 
included, as there is adequate evidence of differences from the light-curve 
fitting, and from the color curves that we will present in the 
next section.    
  
\begin{figure}
\epsscale{0.9} \plotone{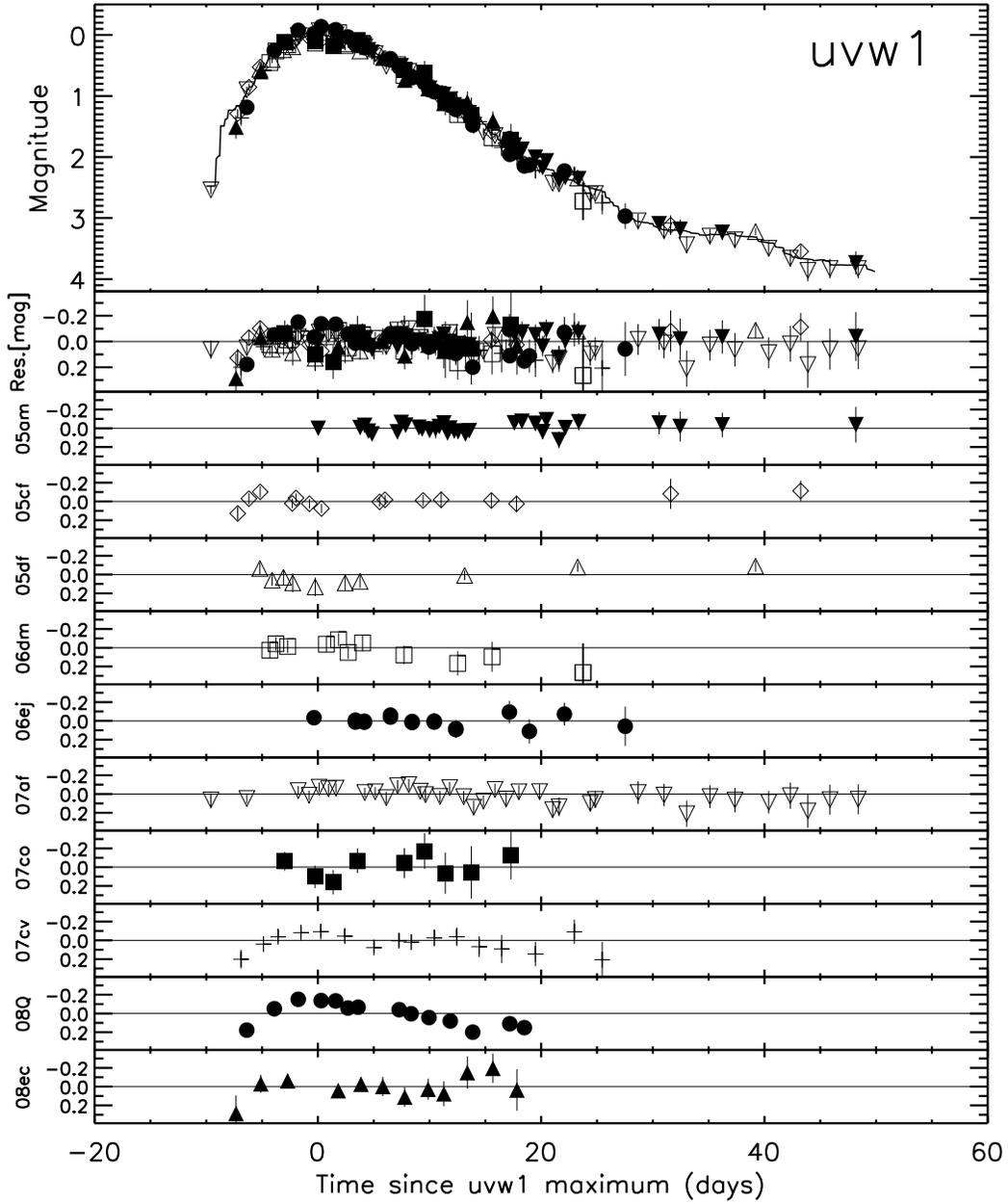}
\caption{Mean template of eight SNe~Ia in the $uvw1$ filter. The 
light curves were shifted in time and magnitude to produce the 
mean template (upper panel). The residuals of individual SNe 
relative to the mean template are shown in the lower panels. 
The template was calculated from $-12$ to +50 days relative to
$uvw1$ maximum.}
\label{mean_temp_W1}
\end{figure}

The $uvw1$-band mean template is shown in Figure \ref{mean_temp_W1}. 
It was calculated from $-12$ days to +50 days relative to $uvw1$ maximum, 
by including data 
within 2.5 days of each time point and weighting the data with inverse 
proportion to the time separation and the photometric error bar. 
The mean template was created 
iteratively by allowing individual SNe to move in time and 
magnitude to afford the best fit to the mean shape.  
The mean template features a steep rise to peak, and a long period of 
linear (in magnitudes) decline (from a few days after peak to roughly 
30 days after peak). There 
are suggestions that the later decline is not uniform in all events, 
but overall the mean template represents the subclass very well. 

We further explore the shape of the $uvw1$ mean template by fitting 
it to the Nugent $U$-band template for normal SNe~Ia
(Fig. \ref{meantemp_vs_utemp}, top panel).
The rise to peak is steeper for the $uvw1$ filter, but the decline 
from peak is fairly similar. There are a few bumps and dips in the 
$uvw1$ template not present in the Nugent $U$-band template; we will 
continue to observe SNe~Ia at these epochs to determine whether such 
features are characteristic of normal SNe~Ia. The narrower peak of 
the $uvw1$ template compared to the Nugent $U$-band template also 
manifests itself in the $uvw1$-$u$ color curves shown in the next 
section. This feature might be the net effect of iron-peak elements 
in the emitting region.

\begin{figure}
\epsscale{0.9} \plotone{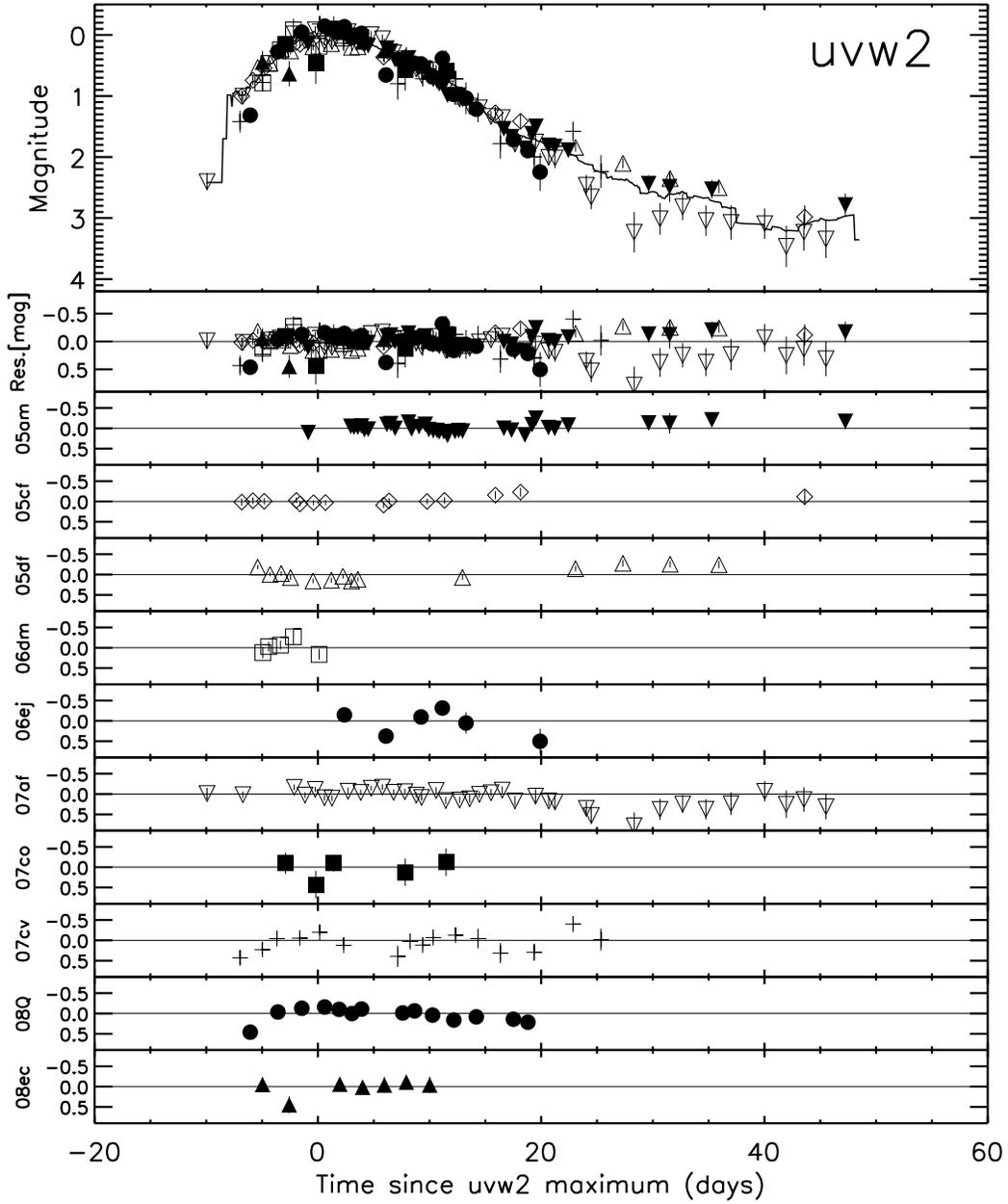}
\caption{Mean template of eight SNe~Ia in the $uvw2$ filter. 
The mean template is shown in the upper panel, while the 
residuals of individual SNe
relative to the mean template are shown in the lower panels.
The template was calculated from $-12$ to +50 days
relative to $uvw2$ maximum.}
\label{mean_temp_W2}
\end{figure}

\begin{figure}
\epsscale{0.85} \plotone{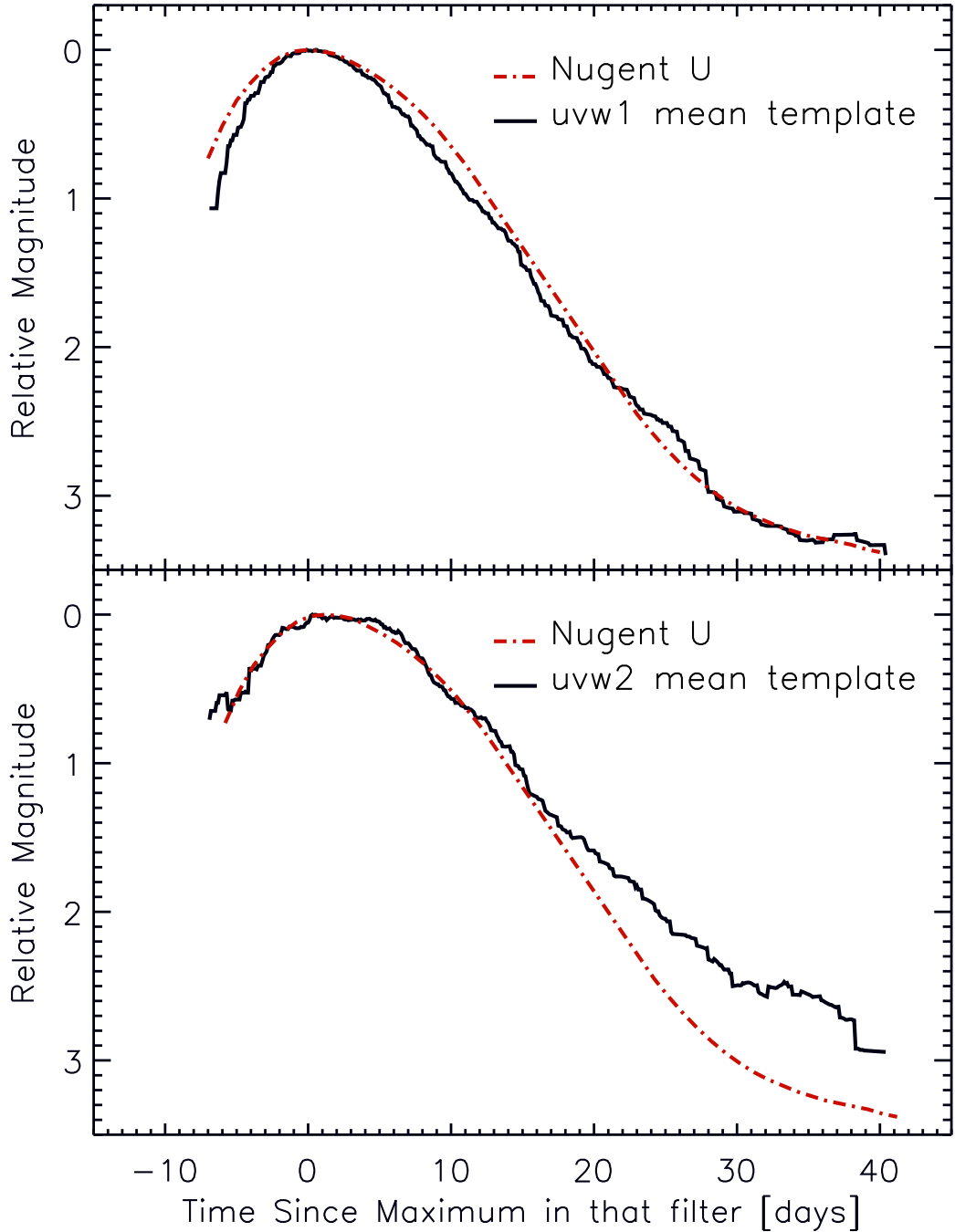}
\caption{$uvw1$ and $uvw2$ mean templates compared to the normal 
$U$-band template from P. Nugent (2009, private communication). 
The $uvw1$ template (upper panel) rises more steeply 
to peak, and is somewhat faster than the $U$-band template 
in the initial decline, coming to better agreement at late times. 
The $uvw2$ template (lower panel) roughly matches the 
$U$-band template until +15 days, remaining brighter at later epochs.}
\label{meantemp_vs_utemp}
\end{figure}

\begin{figure}
\epsscale{1.0} \plotone{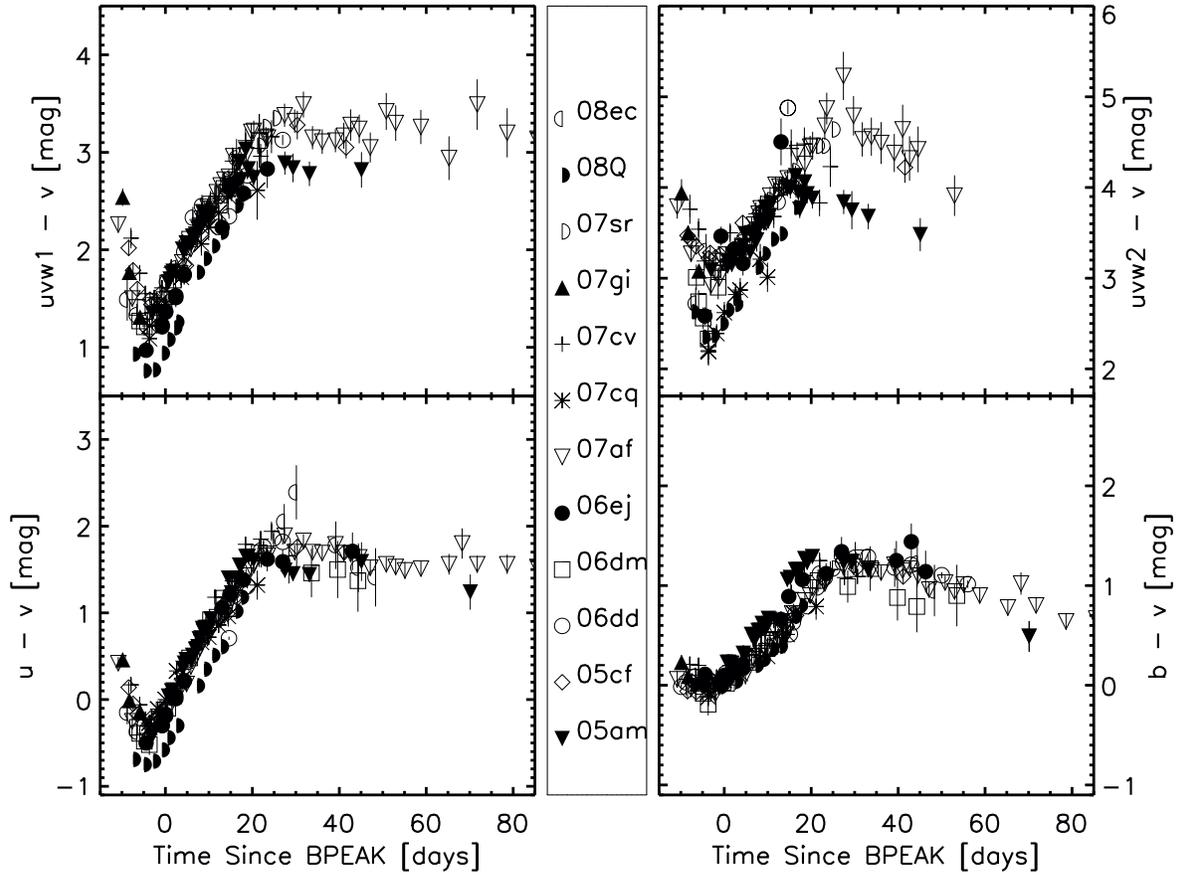}
\caption{Colors of twelve normal SNe~Ia relative to the $v$ band. 
There is a high degree of homogeneity in these SNe out to 
+25 days.}
\label{col_N_4panel}  
\end{figure}

The $uvw2$ mean template is shown in Figure \ref{mean_temp_W2}. 
It was calculated with an algorithm similar to that used to generate 
the $uvw1$ mean template. The dominant features are the same as for 
the $uvw1$ template. Beyond +20 days, the scatter is appreciable, 
as SNe~Ia are difficult to detect in that filter when the SN is 
faint. The shape of the $uvw2$ mean template is similar to 
the Nugent $U$-band normal SN~Ia template until +15 days. This might 
be symptomatic of the higher degree of red leak in the $uvw2$ filter, 
causing $u$-band emission to drive the basic shape of the light curve. 

Table \ref{mean_temp_table} shows peak magnitudes and dates 
derived from fitting these SNe to the mean templates. 
The $uvw1$ peak dates are compared with the $B$-band peak dates, 
revealing that the $B$ band peaks more than 1.5 days after the 
$uvw1$ band. The mean delay (ignoring SN~2005hk) is roughly equal to 
the $2.3 \pm 0.4$ day $U$-peak to $B$-peak delay found by Jha et al. (2006), 
as was also seen from functional fitting in \S 4.1. 
SN~2005hk has a delay in excess of the other SNe~Ia. It appears that 
SN~2002cx-like SNe~Ia peak many days earlier in the 
UV than in the $B$ band. This is potentially a useful diagnostic for 
theoretical simulations of that SN explosion type. 
We will explore this further when we look at the colors of SN~2008A 
compared to SN~2005hk.

\section{Colors}

The {\it Swift} SN~Ia sample was not observed with a uniform cadence;
consequently, the light curves are very heterogeneous
in terms of the epochs of observation. 
To study the color curves of the SNe~Ia in our 
sample, we require an estimate of the $B$-band peak date for each SN; 
these values are listed in Tables \ref{snparams_0506} and 
\ref{snparams_0708}.
Extinction can be important, so our initial comparisons 
concentrate on low-extinction SNe. 
SN~2005df is included only in the $uvw2-uvw1$ colors, as 
UVOT obtained no $ubv$ data for that SN.

\subsection{Colors Relative to the $v$ Band}

First, we show comparisons relative to the {\it Swift} $v$ band. 
This was motivated 
by the desire to allow comparisons and contrasts with the commonly used 
$B-V$ color index. In addition, comparisons 
with the $v$ band permit the maximal separation between the UV filters and 
the optical emission observable with UVOT. 

\subsubsection{Colors Relative to $v$: Normal SNe~Ia} 

The $uvw1-v$, $uvw2-v$, $u-v$, and $b-v$ color curves for the 
twelve SNe~Ia considered to be normal with $E(B-V) < 0.20$ mag 
are shown in Figure \ref{col_N_4panel}. 
Most striking is the high degree of homogeneity among this 
large collection of SNe in the $uvw1-v$ and $u-v$  
color curves. All show an initial, dramatic shift toward bluer 
emission  until reaching a ``blue peak" roughly 5 days before the 
$b$-band peak. This peak 
is followed immediately by a shift toward redder emission until 
$\sim$20 days past $b$-band peak. This trend is also present in the 
$b-v$ color curve, but the individual color curves have a larger 
offset in this index. After +20 days, the $b-v$ color curve 
settles onto the slope that is the basis of the Lira (1995) relation,
while in the bluer filters the transition is to a fairly flat 
evolution. For context, the Mazzali et al. (2008) study of the 
normal SN~2004eo reports that at -11 days, the photospheric 
velocity is 12,000 km s$^{-1}$,  which corresponds to an enclosed mass of
$\sim$1.1 M$_{\cdot}$. They report abundant iron-peak elements, 
in particular providing line blocking in the UV. That study 
suggests that iron-peak elements are in the photosphere well 
before the blue peak.  

UV emission in SNe Ia at early times is the product of reverse fluorescence 
in metal lines \citep{mazz2000}. In regions near the photosphere
the UV opacity is so large that essentially no UV photons can escape, as they
are absorbed in lines of metals (mostly iron-group ions) and reemitted at
optical wavelengths. At layers well above the photosphere, the inverse
process can occur if metals are sufficiently abundant: red photons are
absorbed in metal lines and reemitted in the UV, since the UV radiation field
in those layers is negligible. Thus the initial shift toward bluer emission 
could mark the onset of the reverse fluorescence process, and it indicates 
the presence of metals at velocities well above the photosphere. The
subsequent shift toward redder emission is a cooling phase,
where the majority of emission tends to the red. The sharpness of the 
``blue peak" in the $u$ and $uvw1$ filters is a feature that invites 
theoretical investigation.

Concentrating on the near-peak epoch, the individual events do 
exhibit variations (Fig.~\ref{col_N_zoom}), with the $uvw2-v$  
color exhibiting more variations than the $uvw1-v$ or $u-v$ color 
curves. Perhaps this is a direct probe of variations in the 
distribution of iron-peak elements in the outer layers of the 
ejecta. We will explore this further in \S 5.1.4. 

\subsubsection{Colors Relative to $v$: Subluminous SNe~Ia} 

Five subluminous SNe~Ia were observed by UVOT, with SN~2005ke 
and 2007on being 
by far the best studied. Their color curves   
are shown in Figure \ref{col_N_sb}; the color
curves for the normal SNe~Ia have been replaced by a 
shaded region. Clearly, the evolution is far different for 
this subclass than for normal SNe~Ia. Whereas the normal SNe~Ia
are becoming bluer from $-10$ to $-5$ days relative to the $b$ peak, 
the subluminous SNe~Ia 
are becoming redder, reaching a ``red peak" at +5 to +10 days. 
For context, the Hachinger et al. (2009) study of the subluminous 
SN~2005bl report that at -6 days, the photospheric velocity 
is 8400 km s$^{-1}$, which corresponds to an enclosed mass of 
$\sim$0.7 M$_{\cdot}$. They report that oxygen dominates the zone, with no 
evidence of iron. Indeed, there is no evidence of iron in the  
-3 day spectrum for that event. The color evolution in these 
events as well as the tomography of these events differs from 
normal events.  

After +10 days, the color indices become bluer, either matching the colors of 
the normal SNe~Ia ($b-v$, $u-v$), or even becoming bluer at late epochs 
($uvw1-v$). SN~2005ke and 2007on evolve with similar patterns, with 
SN 2007on being bluer at all epochs. 
SNe 2005mz, 2006mr, and 2007ax add few data points, but 
they generally support the trends exhibited by 
SN~2005ke.\footnote{The $\sim$+30 day $uvw1-v$ color point for 
SN~2007ax was obtained by interpolating two Super-LOTIS $V$-band 
data points from \citep{Kasliwal_etal_2008}.} 
The $B-V$ evolution has been explored in previous studies. Indeed, 
other investigations contain SNe~Ia that bridge the gap between very 
subluminous events like SN~1991bg (Filippenko et al. 1992b; Leibundgut
et al. 1993) and transitionally subluminous 
events (SN~1986G is a well-studied example; Phillips et al. 1987). 
The earliest observations of both SNe~2005ke and 2007on 
hint at a ``blue peak" at roughly $-$9 days. It is fortunate that 
although only two subluminous SNe Ia have been well sampled with 
UVOT, both modes of the bimodal distribution claimed by 
Krisciunas et al. (2009) have one event. Additional
early-time observations are needed to test the suggestion that 
even very subluminous SNe~Ia also have a blue peak, 
but it occurs at such an early epoch that it is seldom detected.

The existence of a red peak is apparent in all four colors. 
In the $b-v$ color, the peak is slightly redder than an extrapolation of 
the Lira (1995) relation to earlier epochs. This is readily apparent in 
$B-V$ curves shown in 
previous studies \citep{Phillips_etal_1999,Garnavich_etal_2004}. 
The red-peaked nature is more dramatic in the UV$-v$ colors, 
making that wavelength range ideal for the study of the physics 
that drives that phenomenon. 

Of particular interest is the color evolution from +5 to +30 days 
relative to the $b$-band peak.
SN~2005ke was reported by {\it Swift} to exhibit a UV excess, the 
start of which (at $\sim$+16 days) was correlated in time with the 
weak detection of X-ray emission with XRT \citep{Immler_etal_2006}.
The SN~2005ke light curves remain brighter than those of normal 
SNe~Ia for more 
than 40 days, when normalized to the peak magnitude. The slopes of the  
color curves of SN~2005ke do not appear to change at +16 days, with
linear evolution from +5 to +25 days. Indeed, it is the 
normal SNe~Ia that appear to experience an inflection point between 
+15 and +20 days. If there is an excess of emission due to 
the SN shock interacting with CSM, 
it would need to produce emission such that the 
multi-band color evolution continues in a manner similar to that of 
the earlier-epoch intrinsic SN emission. These puzzling findings call
into question whether 
the differences between the $uvw1$, $uvw2$, and $u$-band 
light curves of SN~2005ke 
versus those of normal SNe~Ia are caused by shocked CSM emission. 
Alternatively, Kasliwal et al. (2008) 
present observations of SN~2007ax and  
suggest a photospheric origin for the flat late-time decline. 

\subsubsection{Colors Relative to $V$: SN~2002cx-Like SNe~Ia} 

Four SNe~Ia that are considered members of 
the ``SN~2002cx-like" subclass have 
been observed (see \citep{Jha_etal_2006b} for a description of this 
subclass and Foley et al. 2009 
for observations of the very subluminous, SN~2002cx-like  
SN~2008ha). The color curves are illustrated in 
Figure \ref{col_N_02cx}; as in Figure \ref{col_N_sb}, 
the normal SNe~Ia are shown as a 
shaded region, and external data for other members of this class 
are included. The color curves for three of the four SNe are notable by being 
bluer than those of normal SNe~Ia at the pre-peak epochs and by following a 
steeper slope during the reddening phase. The 
slope of the three SNe in $uvw1-v$ is consistent with a straight line from 
$-7$ to +10 days 
(Fig. \ref{slope_02cx}). The very linear evolution is a characteristic 
that must be reproduced by simulations of this subclass of SNe~Ia. 
The colors of SN~2008ha are much redder than those of 
the other three objects, a characteristic 
unlikely to be due to extinction, as Foley et al. (2009) 
measured only upper limits for interstellar Na~I~D absorption. 
They assumed zero host-galaxy extinction, and corrected for 
$A_{V} = 0.25$ mag of Milky Way extinction. 
The SN~2008ha optical light curves were 
reported to be well fit by those of the other SN~2002cx-like SNe 
when a stretch was applied to the 
timescale, but this is not an 
explanation for the red color seen in Figure \ref{col_N_02cx}. 
The UVOT observations were obtained near 
optical peak, and would not be affected appreciably 
by a stretch of $\sim$0.75 centered at peak light.

The SN~2002cx-like subclass is characterized by the early appearance of 
iron-peak elements in the pre-peak spectra, but with narrow lines, 
a suggested characteristic of a deflagration 
explosion (Phillips et al. 2007). 
It is possible that the very blue colors at early epochs 
for SNe~2005hk and 2008A 
are a consequence of reverse fluorescence from these iron-peak 
elements, but this is speculative until 
demonstrated by theoretical simulations.  

\subsubsection{Extinction Correction}

The homogeneity of the near-peak color evolution of the normal SNe~Ia 
in $uvw1-v$ and $u-v$ invites further investigation, as this 
feature could be used to determine rest-frame UV extinction. 
Specifically, we test two hypotheses: (1) the color evolution from 
$-12$ to +12 days relative to $B_{\rm peak}$
can be fitted with two lines, and (2) the color 
evolution is the same for all normal SNe~Ia in that color. 
We provide no basis for these hypotheses, presenting them {\it ad~hoc}, 
but these tests allow us to simultaneously study SN~Ia emission 
and the net wavelength dependence of extinction.   
SN~2008Q is excluded from this preliminary investigation, due to its 
very blue color and anomalous light-curve shapes. 
The follow-up study that will concentrate on using 
SNe~Ia as probes of UV extinction will address SN~2008Q in detail. 

In Figure \ref{ext_corr_w1}, we display the 
residuals of the low-extinction SNe (i.e., small values of $E(B-V)$) 
to a best-fit two-line evolution 
(upper two panels). The standard deviation about that fit is
$\sigma = 0.134$ mag. We then include higher-extinction SNe to the 
sample (third panel), and apply a reddening correction that allows 
the value of $R_{uvw1} - R_{V}$ to be a fitted parameter for 
each individual SN.\footnote{Note that 
$E(uvw1-V) = A_{uvw1} - A_{V} = (R_{uvw1} - R_{V}) \times E(B-V)$.} 
We bounded $R_{uvw1} - R_{V}$ such that 
$0.0 \leq R_{uvw1} - R_{V} \leq 5.0$. 
In Table \ref{ext_coeffs}, we show the optimal $R_{uvw1} - R_{V}$ 
values for the three SNe~Ia with $E(B-V) > 0.20$ mag. The $R_{uvw1} - 
R_{V}$ values range from 1.09 to 1.93, 
and result in the larger group (which includes the higher-extinction SNe)
having a standard deviation comparable to that of 
the low-extinction subset, 0.138 mag versus 0.134 mag. 

Figure \ref{ext_corr_u} displays the $u-v$ color curves following the 
same procedures as in Figure \ref{ext_corr_w1}. The standard deviation 
of the low-extinction SNe without reddening correction is 
0.098 mag. The correction decreases the standard deviation for 
the larger group that includes the higher-extinction SNe from 
0.675 mag to 0.104 mag. The $R_{u} - R_{V}$ values range from 
0.81 to 2.17 (Table \ref{ext_coeffs}). 

Whereas the variations in the $uvw1-v$ colors are suggestive of 
being dominated by the effects of extinction, the $uvw2-v$ light curves 
appear to vary in 
shape as well as vertical offset (Fig. \ref{col_N_zoom}). 
Applying the same approach as done for $uvw1-v$ and $u-v$, 
we investigate the 
idea that the differences are purely due to extinction 
(Fig. \ref{ext_corr_w2}). 
The residuals of most SNe appear to deviate little, or slightly 
toward the red relative to the two-line fit a few 
days before $B$-band peak. By contrast, some objects (SNe 2006dm and 
2007cq) appear to deviate toward the blue during the same epoch. 
Neither group represents the extremes of $E(B-V)$, so it seems 
unlikely that the differences are related to reddening. 
Rather, this appears to be an intrinsic difference among 
normal SNe~Ia. The $R_{uvw2} - R_{v}$ values range from 
1.20 to 1.63 (Table \ref{ext_coeffs}). The standard deviation
of the low-extinction SNe without reddening correction is
0.249 mag. The correction decreases the standard deviation for
the larger group that includes the higher-extinction SNe from
0.611 mag to 0.230 mag.

Collectively, this shows that the $uvw1-v$ and $u-v$ color curves 
can be well characterized by two lines that form a blue peak. 
For the $uvw2-v$ color curves, the characterization is more approximate. 
Application of a reddening correction optimized for each individual 
SN reduces the scatter of the total sample to the same order as the 
low-extinction subset. These preliminary findings show that using 
SN~Ia color curves to probe the UV extinction in other galaxies 
has promise. The limit to this technique will be the 
intrinsic variations in the UV$-v$ colors, and more work 
is required to further explore the possibilities. In 
particular, it is important to have multiple low-extinction SNe 
of the same variety as each high-extinction event to isolate 
extinction as the cause of color differences. We note that this sample 
is biased in that two of the highest-extinction events are toward the 
luminous (broad-peaked light curve) extreme of SNe Ia, with spectral 
features that are similar to those of SN 1991T. 
It will be important to observe multiple SN 1991T-like SNe~Ia that have 
low extinction. 

\subsubsection{$uvm2-v$ Colors}

In Figure \ref{col_m2_v} we show the $uvm2-v$ colors for 15 SNe~Ia. 
Twelve of these are normal SNe~Ia, but 
SNe 2005ke and 2007on are members of the subluminous subclass and 
SN~2005hk belongs to the SN~2002cx-like subclass. Among the normal
SNe~Ia, the $uvm2-v$ colors for SNe 2005am, 2005cf, 2007af, and 2007gi  
are initially red, in agreement with the $uvw2-v$ colors (dashed line). 
The $uvm2-v$ colors for SNe 2006ej, 2007cq, and 2008Q are initially
blue, transitioning to redder emission after the peak (dot-dashed line). 
Many of the red-colored group are normal-velocity normal SNe~Ia, while 
SN 2006ej is HV and in the blue group. However, SN 2007gi is red, but 
of the HV group, and a number of SNe~Ia have yet to be categorized as 
HV or NN. Whether there is a correlation between HV categorization and 
UV colors remains uncertain and will be explored in future papers.  

By contrast, the SN~2005ke and SN~2005hk $uvm2-v$ color curves appear to 
match the trends seen in the $u$, $uvw1$, and $uvw2$ filters. SN~2005ke 
reddens to reach a red peak 5--10 days after $B_{\rm peak}$, followed 
by a blueward shift to colors as blue as those of the normal SNe~Ia after 
+25 days. SN~2005hk is initially very blue, but quickly reddens to fall 
within the range of normal events. The fact that this evolution is 
observed for a range of color curves suggests that 
the SN~2002cx-like color curves are not dominated by specific absorption or 
emission features, but by emission over a wider wavelength range.  

\begin{figure}
\epsscale{1.0} \plotone{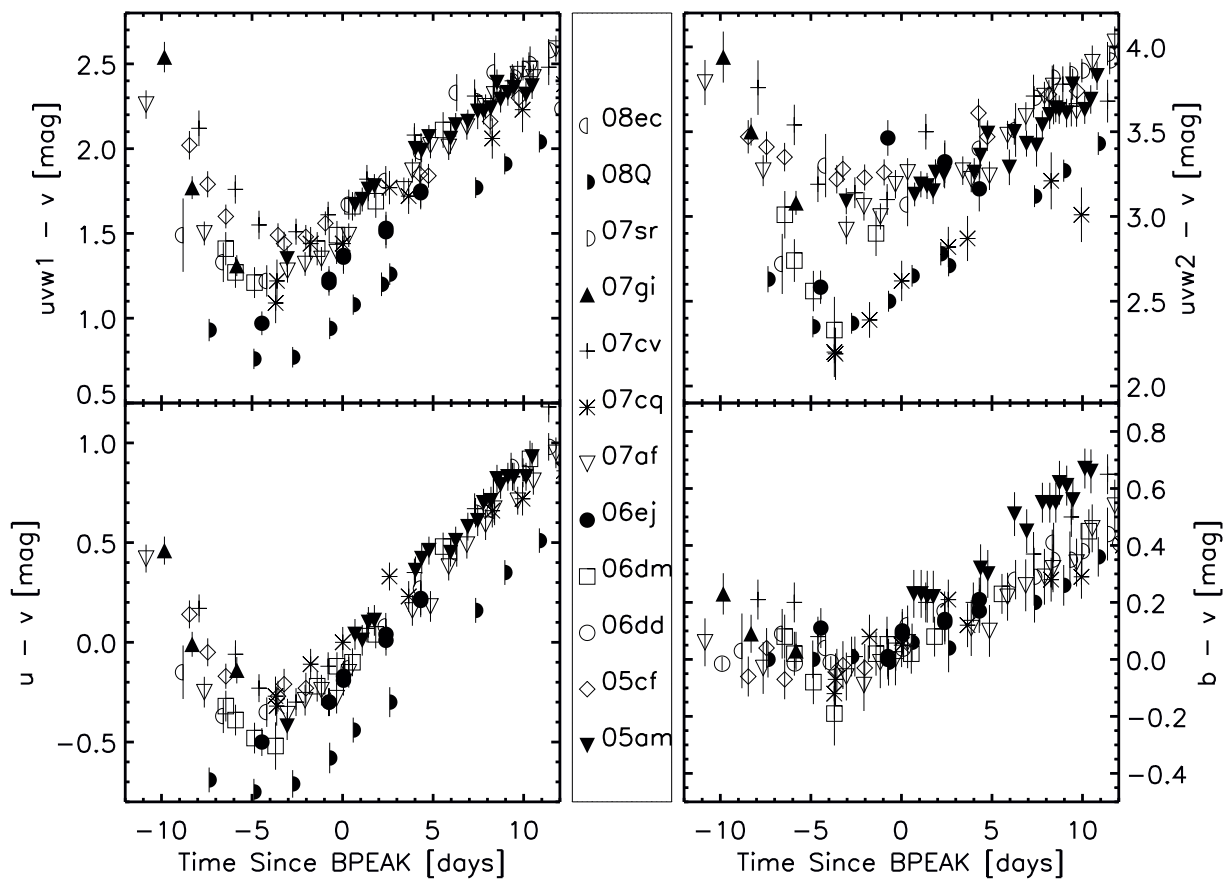}
\caption{Near-peak colors of twelve normal SNe~Ia compared to the $v$ band.
When concentrating upon the near-peak epoch, variations between 
individual events are apparent. 
The variations of the ``blue peak" are as small as $\sim$0.2 mag 
in $u-v$, but over 1.0 mag in $uvw2-v$.}
\label{col_N_zoom}
\end{figure}

\begin{figure}
\epsscale{1.0} \plotone{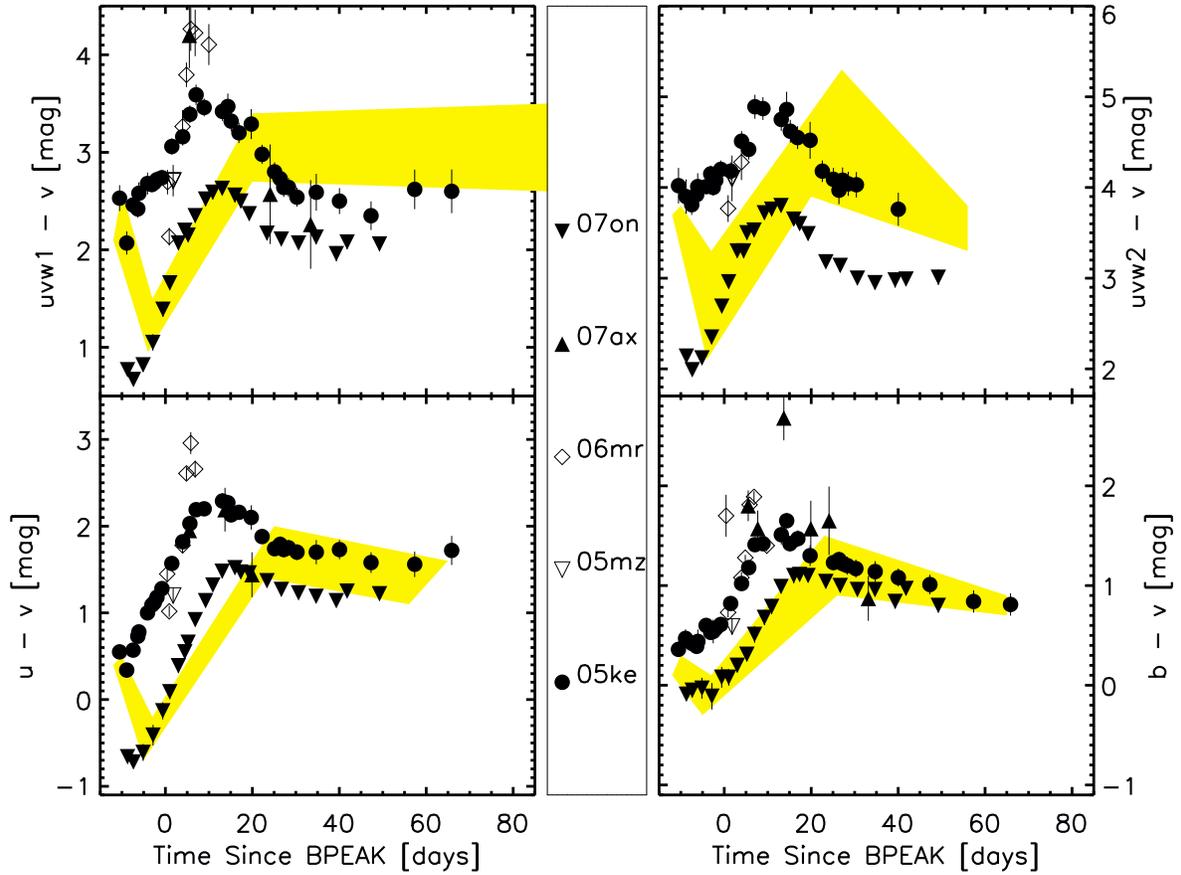}
\caption{Colors of four subluminous SNe~Ia compared to the $v$ band.
In each panel, the 
range of colors of the normal SNe~Ia is shown as the shaded 
region. The colors of the subluminous SNe~Ia differ appreciably 
from those of the normal SNe~Ia. }
\label{col_N_sb}
\end{figure}

\begin{figure}
\epsscale{1.0} \plotone{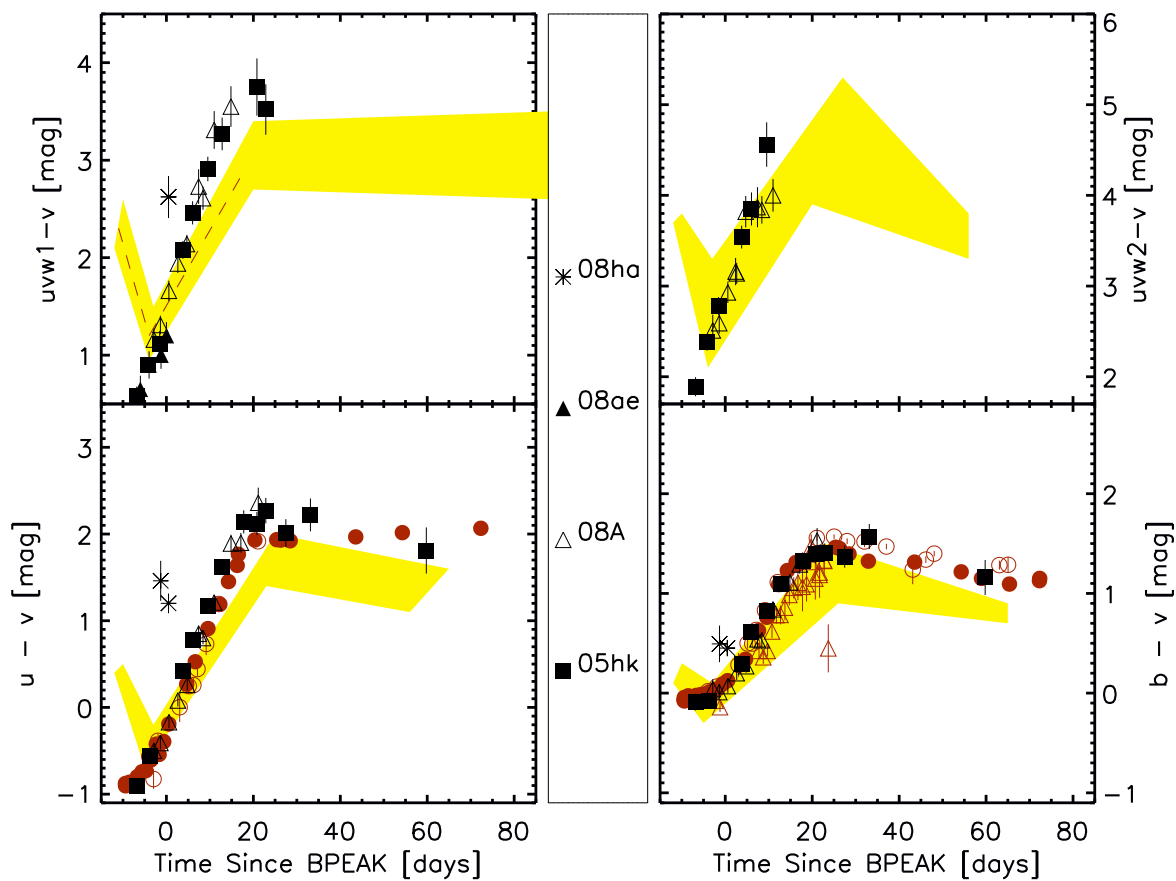}
\caption{Colors of four SN 2002cx-like SNe~Ia compared to the $v$ band.
In each panel, the 
range of colors of the normal SNe~Ia is shown as the shaded
region. SN 2008ha UVOT photometry is from Foley et al. (2009). 
Ground-based $UBV$ observations of SN 2002cx (Phillips et al. 2007, 
open triangles) and SN 2005hk (Sahu et al. 2008, open circles; 
Stanishev et al. 2008, filled circles) confirm the trends seen in the 
UVOT photometry. The colors of the SN 2002cx-like SNe~Ia are initially bluer 
than the normal SNe~Ia, but become redder with a steeper slope than seen 
for the normal SNe~Ia.}.
\label{col_N_02cx}
\end{figure}

\begin{figure}
\epsscale{1.0} \plotone{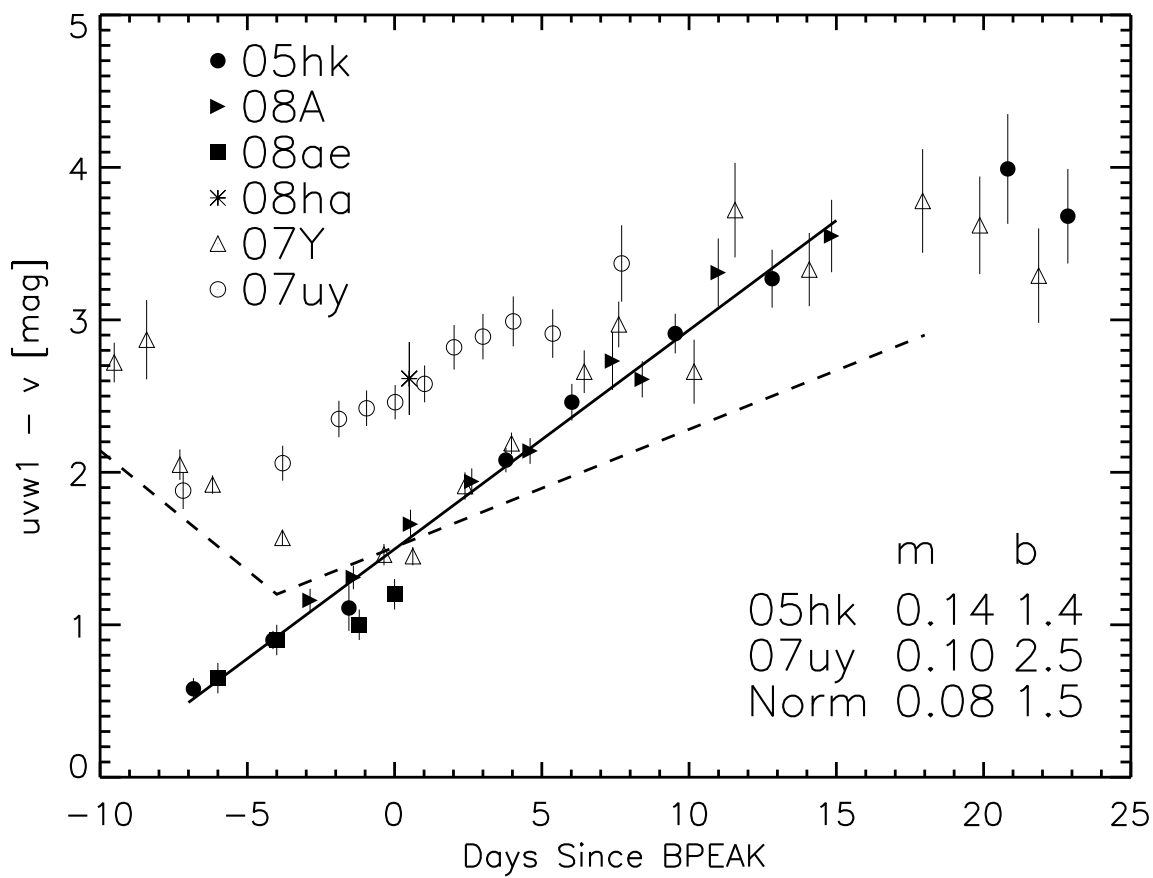}
\caption{$uvw1-v$ color curves for four SN 2002cx-like SNe~Ia 
compared with normal SNe~Ia and two SNe~Ib/Ic. The solid 
line is a linear fit to the SN 2005hk data, and the dashed line is a 
two-line fit to the twelve normal SNe~Ia shown in 
Fig. \ref{col_N_4panel}. Data for SN 2007Y (open triangles) and 
SN 2007uy (open circles) are from 
Brown et al. (2009). The three SN 2002cx-like SNe~Ia 
have color evolution consistent with a linear evolution out to +13 days, 
with a slope which is steeper than that seen in the normal SNe~Ia.} 
\label{slope_02cx}
\end{figure}

\begin{figure}
\epsscale{1.0} \plotone{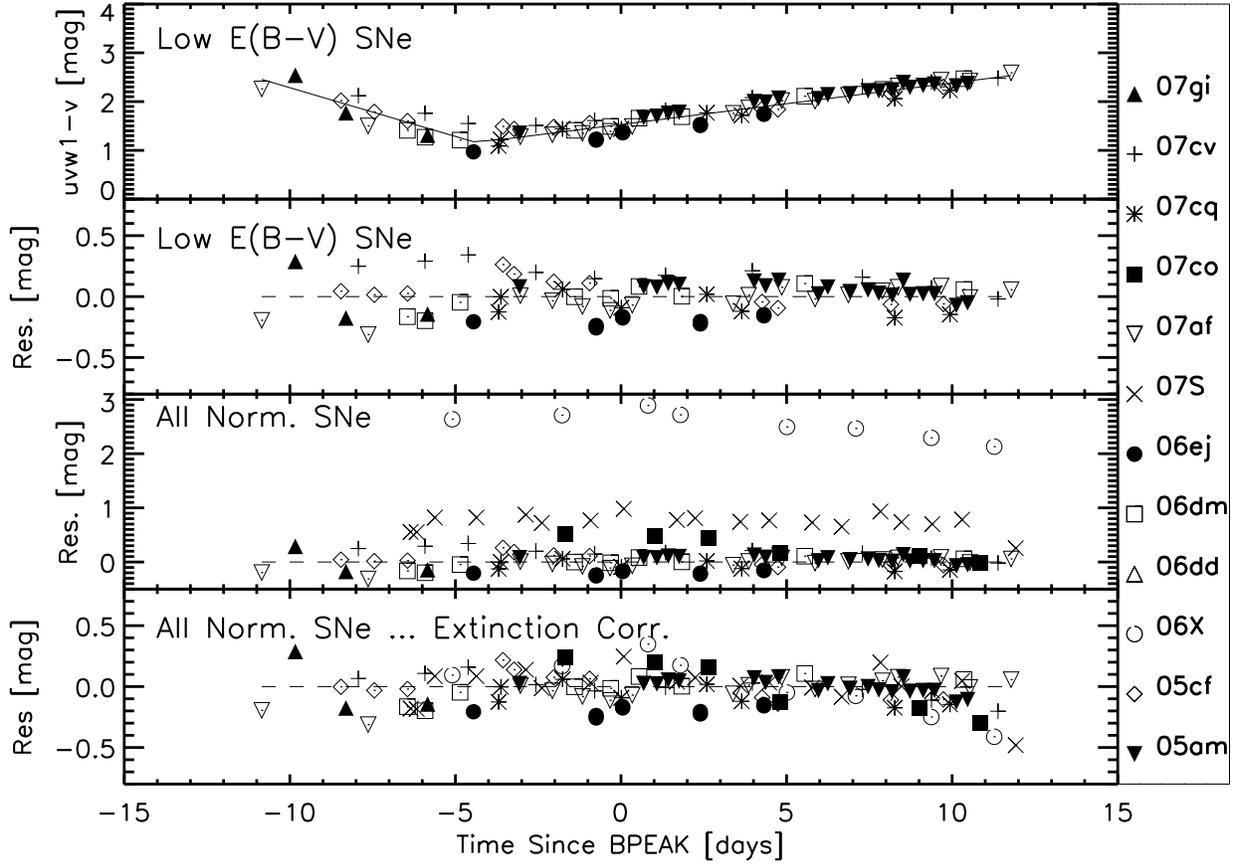}
\caption{$uvw1-v$ colors as a function of extinction. The $uvw1-v$ 
colors of nine low-extinction, normal SNe~Ia are shown 
fitted with a two-line function (upper panel), along with 
the resulting residuals (second panel). The residuals 
increase when higher-extinction SNe are added (third panel), 
but applying a reddening correction decreases the 
residuals (fourth panel). SN 2008Q was not included in 
this preliminary exploration of a reddening correction, as it 
appears to be an anomalously blue SN Ia.}
\label{ext_corr_w1}
\end{figure}

\begin{figure}
\epsscale{1.0} \plotone{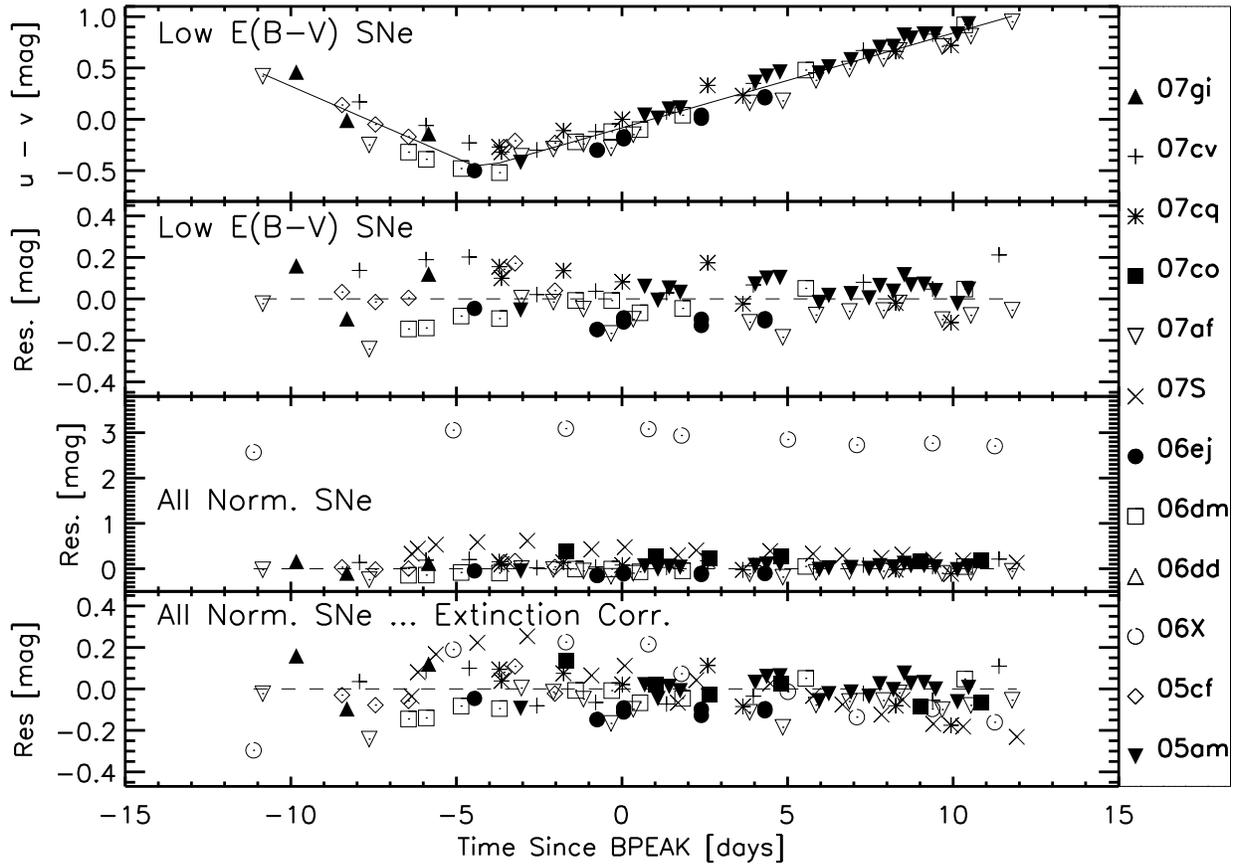}
\caption{$u-v$ colors as a function of extinction. 
The panels are the same as in Fig. \ref{ext_corr_w1}.}
\label{ext_corr_u}
\end{figure}

\begin{figure}
\epsscale{1.0} \plotone{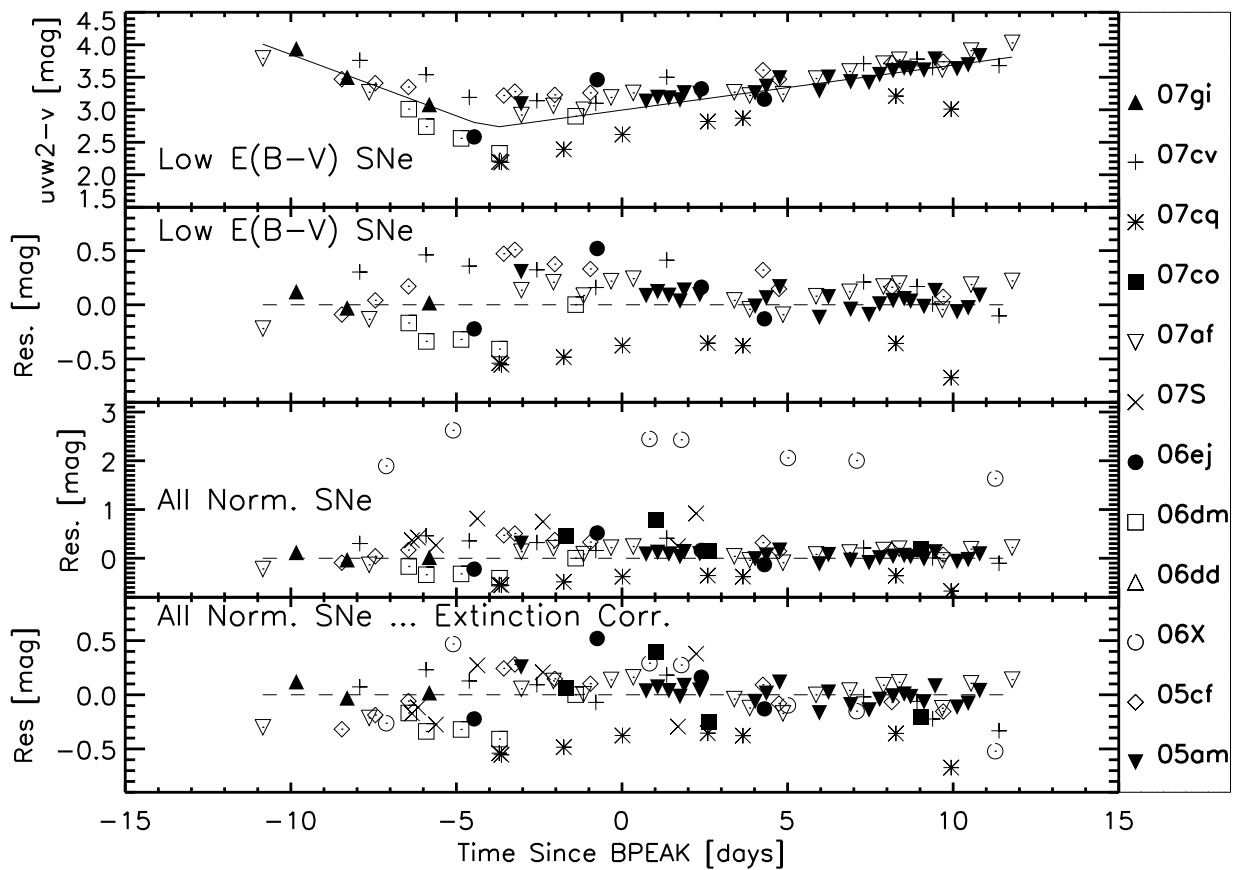}
\caption{$uvw2-v$ colors as a function of extinction. 
Symbols are the same as in Fig. \ref{ext_corr_w1}. 
The color curves appear to deviate from the two-line fit 
during the near-peak epoch, with some bluer than the 
lines and some redder than the lines.}
\label{ext_corr_w2}
\end{figure}

\begin{figure}
\epsscale{1.0} \plotone{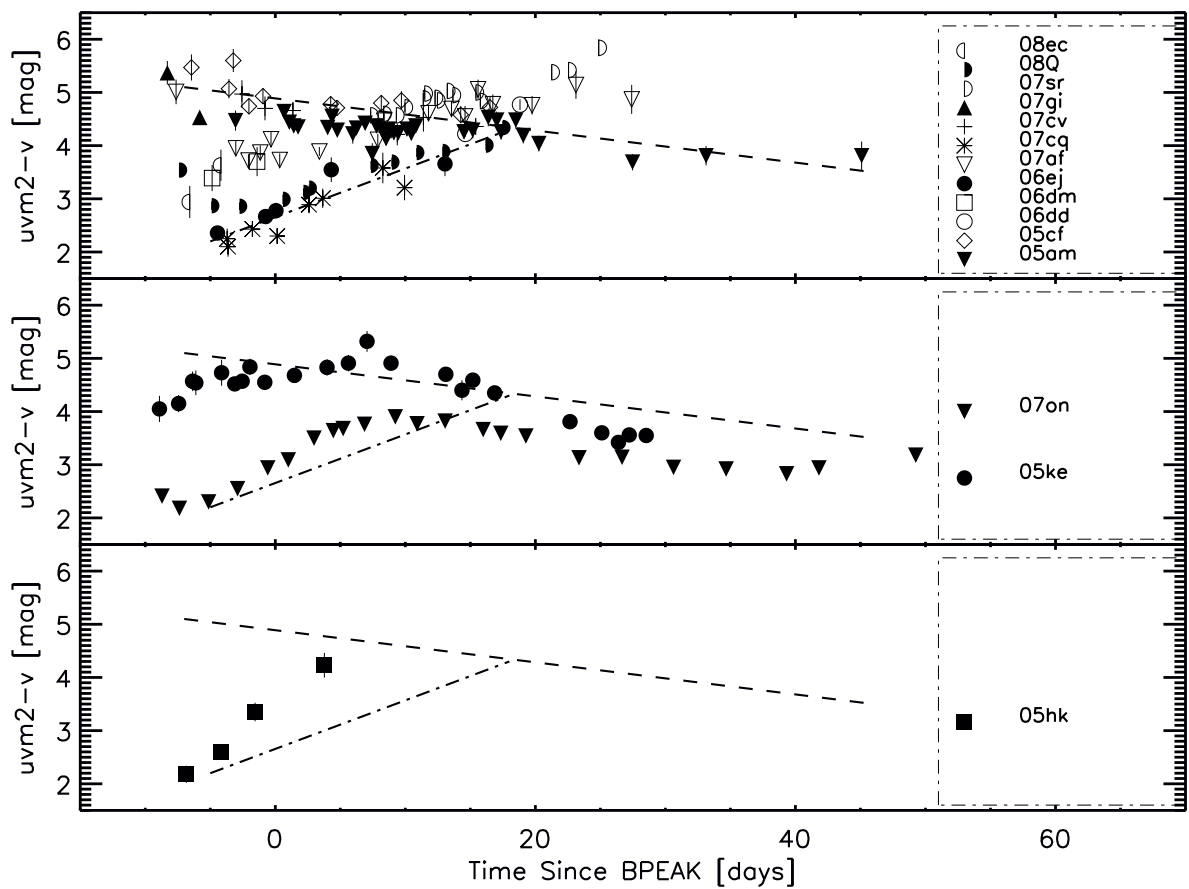}
\caption{$uvm2-v$ colors of 15 SNe~Ia. Twelve normal SNe~Ia 
are shown in the upper panel, the subluminous SNe 2005ke 
and 2007on in the middle panel, and the SN 2002cx-like SN 2005hk 
in the lower panel. The normal SNe~Ia exhibit two branches of 
color-curve shape, while SNe 2005ke, 2007on, and 2005hk appear to follow 
the general evolution as seen for those SNe in the $uvw1-v$ and $uvw2-v$ 
color curves.}
\label{col_m2_v}
\end{figure}

We investigate further the UV emission from normal SNe~Ia by 
comparing the $uvw2-v$ and $uvm2-v$ color curves.  
The near-peak $uvw2-v$ color curves (Fig. \ref{col_N_zoom}) show that  
a few SNe~Ia (SNe 2007cq and 2008Q) 
are significantly bluer from $-7$ to +5 days relative to the
time of $B$-band maximum than the larger collection of SNe~Ia. 
These same SNe are blue in the $uvm2-v$ color curves (Fig. \ref{col_m2_v})
and evolve toward redder emission, while the larger group has less 
color evolution, evolving toward bluer emission. 
Perhaps more interestingly, three SNe~Ia (SNe 2006dm, 2006ej, and 2008ec) 
are initially as blue as that first group in the $uvw2-v$ color curves, but 
abruptly shift to a redder color curve that matches that of the larger group. 
By contrast, the $uvm2-v$ color curve of SN 2006ej remains similar to that
of the bluer SNe at all epochs, exhibiting no jump toward redder emission, 
while SNe~2006dm and 2008ec have color curves somewhat intermediate to 
the blue and the red groups of SNe~Ia, evolving toward bluer emission. 

\begin{figure}
\epsscale{1.0} \plotone{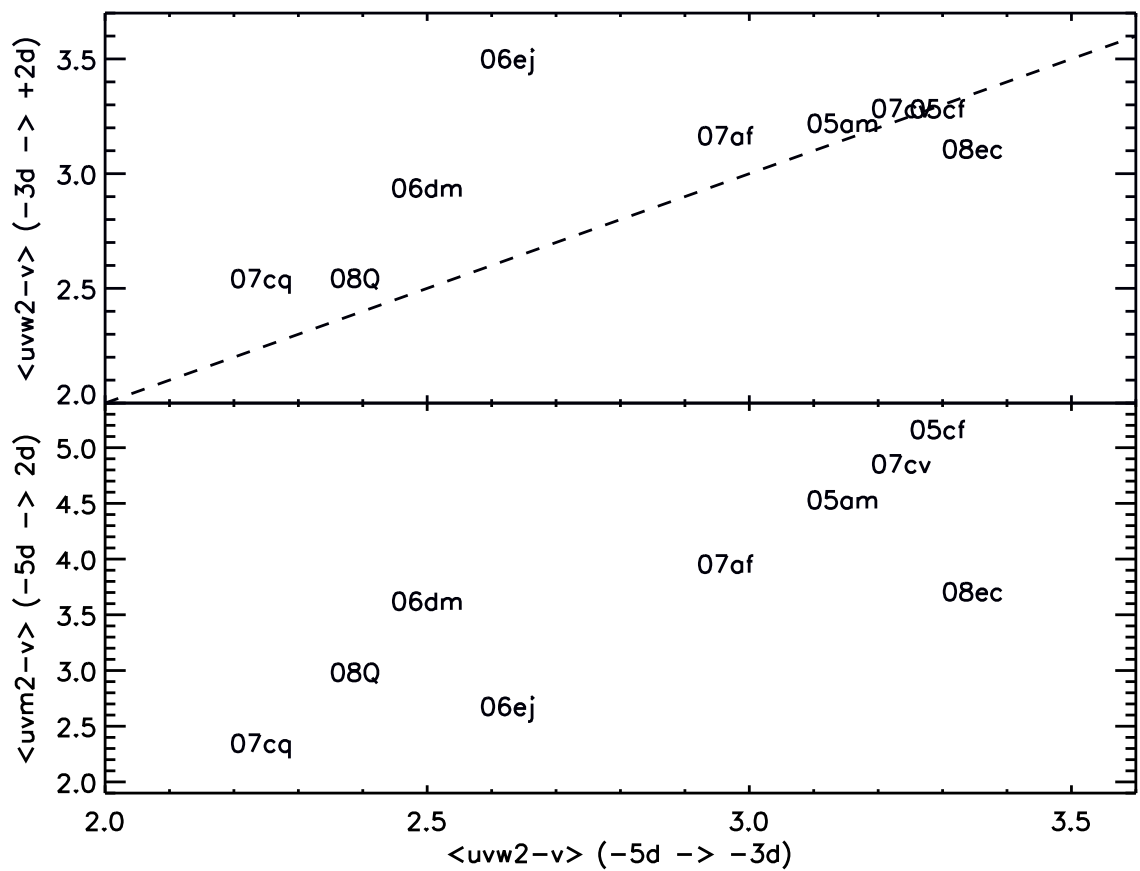}
\caption{Color-color plots for normal SNe~Ia. The mean $uvw2-v$ 
color during the epoch from $t = -5$ to $-3$ days relative to $B$ maximum
is compared to the mean $uvw2-v$ color during the epoch $t = -3$ to +2 days 
(upper panel). Similarly, the mean $uvw2-v$ color during the epoch
$t = -5$ to $-3$ days is compared to the mean $uvm2-v$ color during the 
epoch $t = -5$ to +2 days (lower panel). 
The general tendency for SNe~Ia with red $uvw2-v$ color to also 
have a red $uvm2-v$ color is apparent.} 
\label{color_color}
\end{figure}

We quantify these tendencies in Figure \ref{color_color}. The upper 
panel shows the mean $uvw2-v$ color from $-5$ to $-3$ days relative
to $B$ maximum (abcissa) versus the mean $uvw2-v$ color in the
interval $t = -3$ to +2 days (ordinate). 
Most SNe~Ia are above the equality line, meaning that there 
is some evolution toward redder emission, but SNe~2006ej and 2006dm are 
notable in the larger shift toward redder emission. SN 2008ec is 
below the equality line, but we note that that SN experiences the 
redward jump mentioned above at an earlier epoch, making this SN 
poorly represented in this figure. The lower panel  
again displays the mean $uvw2-v$ color from $-5$ to $-3$ days relative
to $B_{\rm peak}$ on the abcissa, but shows the mean $uvm2-v$ color 
in the interval $t = -5$ to +2 days (ordinate). The larger interval was 
chosen to accomodate the poorer sampling in the $uvm2$ filter. 
The general trend for the 
SNe~Ia with redder $uvw2-v$ color to also emit with redder $uvm2-v$ color 
is apparent, and this is in line with expectations based on the fact that 
the $uvw2$ and $uvm2$ filters have considerable overlap in their 
bandpasses. SN~2008ec appears to deviate from this general trend, 
but we note that the color curves shown in Figure \ref{col_N_zoom} 
appear to evolve on an earlier time-scale than the other 
SNe Ia. This might be evidence of an incorrect estimate of the 
B-band peak date, which was derived from UVOT data. Shifting the 
color curves by +2 days puts SN 2008ec in better agreement with 
the other SNe Ia, and would alter the values shown in 
Figure \ref{color_color}. Our attempt to quantify tendencies in UV color 
evolution is crude, and biased by the variations in the sampling 
of individual events, but the existence of basic trends beyond 
the scatter in the observations is apparent.  

\subsection{Other Color Curves}

\subsubsection{$uvw1-b$ Color Curves}

\begin{figure}
\epsscale{1.0} \plotone{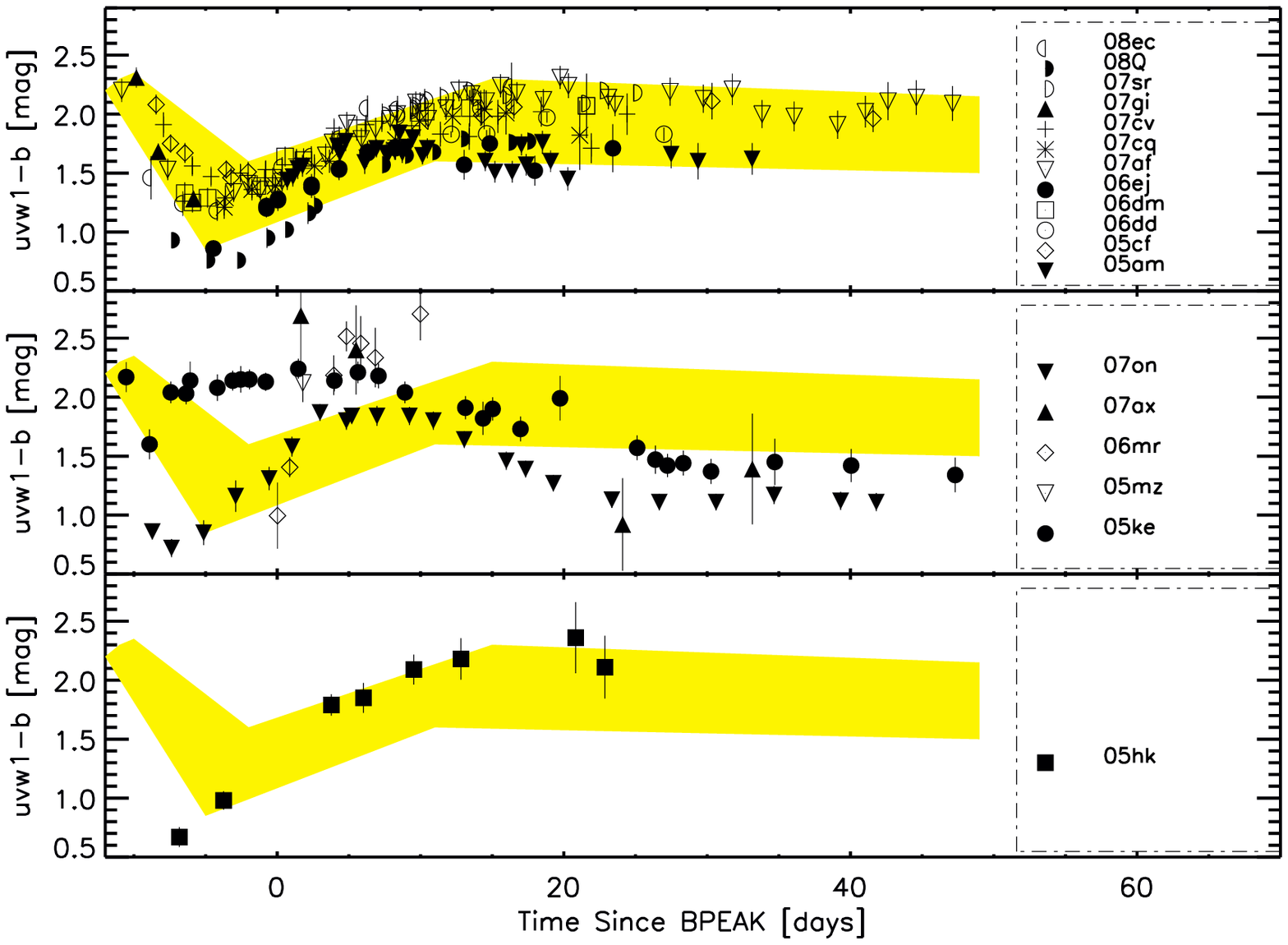}
\caption{$uvw1-b$ colors of 18 SNe~Ia. Twelve normal SNe~Ia are shown in 
the upper panel, five subluminous SNe~Ia in the 
middle panel, and the SN 2002cx-like SN 2005hk in the lower 
panel. The range of the normal SNe Ia is shown as a shaded region.} 
\label{col_w1_b}
\end{figure}

Figure \ref{col_w1_b} shows the $uvw1-b$ color curves with the 
SNe~Ia broken into the same three subclasses as for the $v$-band 
comparisons. The general curve features are similar to those of 
the $v$-band comparisons: the normal SNe~Ia exhibit a blue peak 
and a late flattening, the subluminous SNe~Ia are redder transitioning to 
bluer, and the SN~2002cx-like SNe~Ia are initially very blue, 
rapidly becoming redder. 
It is readily apparent that there is considerable 
divergence in the normal color curves after +5 days. SNe 2005am, 2006ej,
and 2007cq are bluer at the later epochs than SNe 2005cf, 2006dm, 2007af,
and 2007cv by $\sim$0.5 mag. SN~2008Q is a blue outlier. 
The transition to the later, flatter epoch 
is at approximately +10 days, whereas in $uvw1-v$ the inflection was 
at +15 to +20 days. The differences between the peak and +15 day 
color should be studied in tandem with Figure~\ref{decline_rates}, 
which shows clumping in the 
$\Delta m_{15}(uvw1)$ vs. $\Delta m_{15}(B)$ plane.

\subsubsection{$u-b$ Color Curves}

\begin{figure}
\epsscale{1.0} \plotone{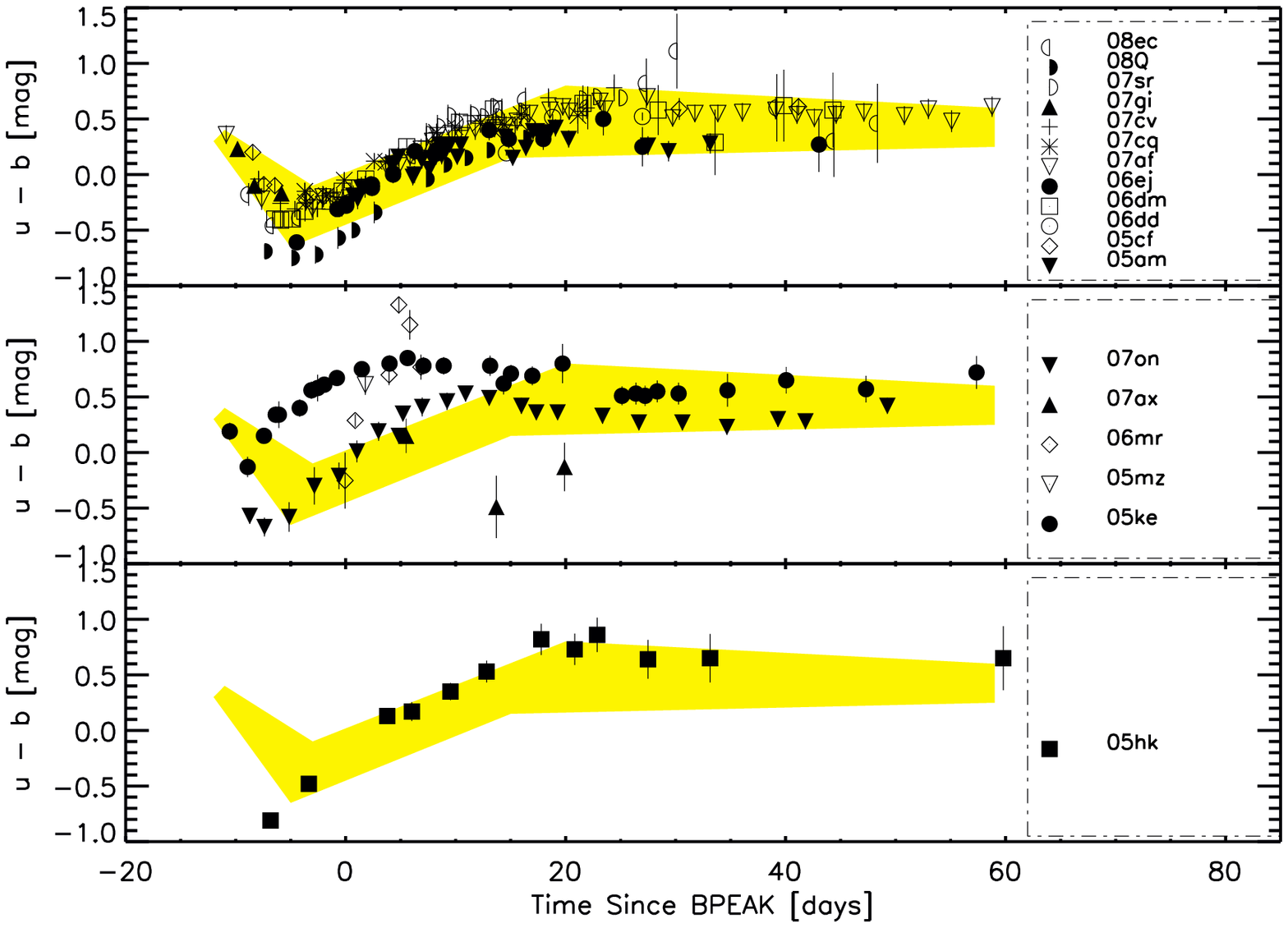}
\caption{$u-b$ colors of 18 UVOT-observed 
SNe~Ia. Twelve normal SNe~Ia are shown in
the upper panel, five subluminous SNe~Ia in the
middle panel, and the SN 2002cx-like SN 2005hk in the lower
panel. The range of the normal SNe Ia is shown as a shaded region.}
\label{col_u_b}
\end{figure}

Figure \ref{col_u_b} illustrates the $u-b$ colors with the SNe~Ia 
broken into the same three subclasses as for the $v$-band 
comparisons. The general curve features are similar to those of the
$v$-band and $uvw1-b$ comparisons: the normal SNe~Ia exhibit a blue peak
and a late flattening, the subluminous SNe~Ia are redder from 
$-10$ to +20 days, and the SN~2002cx-like SNe~Ia are initially very 
blue, rapidly becoming redder. 

\subsubsection{$uvw1-u$ Color Curves}

\begin{figure}
\epsscale{1.0} \plotone{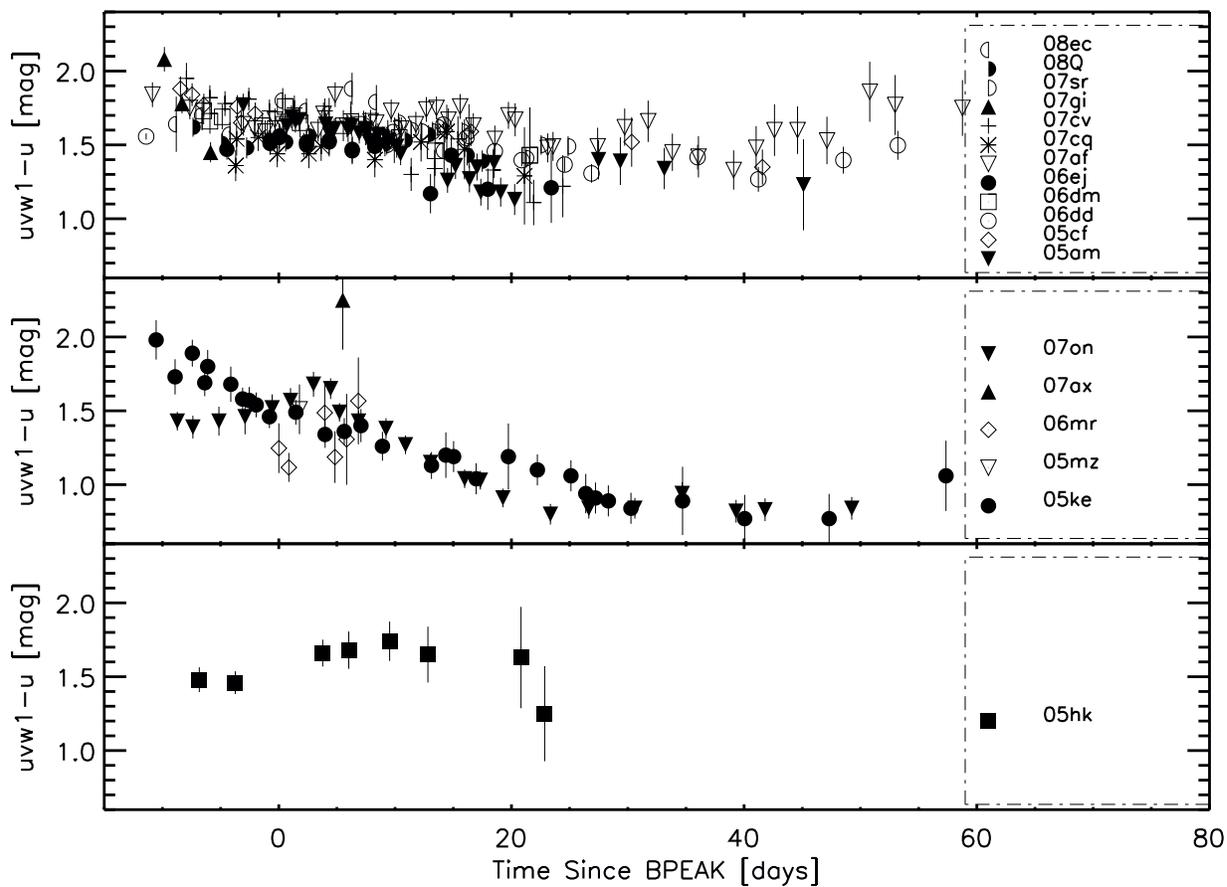}
\caption{$uvw1-u$ colors of 18 SNe~Ia. Twelve normal SNe~Ia are shown in
the upper panel, three subluminous SNe~Ia in the
middle panel, and the SN 2002cx-like SN 2005hk in the lower
panel.  The normal SNe~Ia gradually become bluer from 
the earliest epochs to +20 days. The subluminous SNe~Ia appear 
to rapidly become bluer for all observed epochs. There is no 
clear trend in the SN 2002cx-like SN~Ia.}  
\label{col_w1_u}
\end{figure}

The $uvw1-u$ colors permit a comparison of UVOT $uvw1$ photometry with 
ground-based $U$-band photometry. The color curves 
(Fig. \ref{col_w1_u}) have two interesting features. 
First, the normal SNe~Ia show a gradual shift toward 
bluer emission during the entire epoch of the $uvw1-v$ ``Z"-shaped 
evolution, though this gradual trend is moderated by appreciable 
scatter. Second, the subluminous SNe~Ia become bluer from 
the earliest epochs to the latest epochs. 
No clear trend is apparent for the SN~2002cx-like 
subclass, which would benefit from observations of additional events. 

\subsubsection{$uvw2-uvw1$ Color Curves}

\begin{figure}
\epsscale{1.0} \plotone{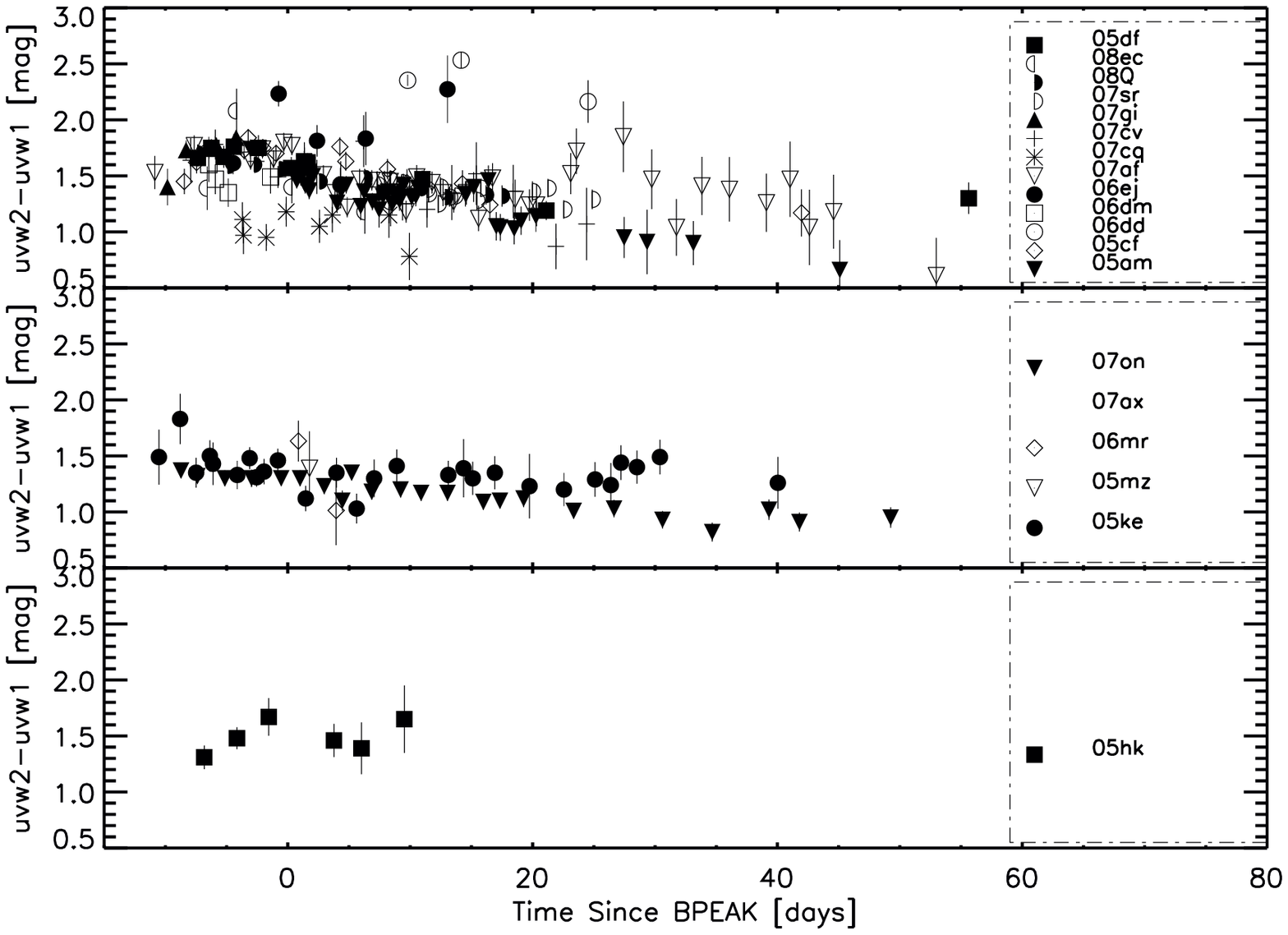}
\caption{$uvw2-uvw1$ colors of 16 SNe~Ia. Nine normal SNe~Ia are shown in
the upper panel, three subluminous SNe~Ia in the
middle panel, and the SN 2002cx-like SN 2005hk in the lower
panel.} 
\label{col_w2_w1}
\end{figure}

Finding purely rest-frame UV indicators of SN~Ia subclass is 
important for the study of SNe~Ia at very high redshifts. Foley et al. 
(2008b) suggest a $R_{\rm UV}$ ratio, comparing wavelength bands at 
2900~\AA\ and 2770~\AA; the ratio increases with increasing MLCS-$\Delta$ 
(i.e., the low-luminosity events have the largest $R_{\rm UV}$ ratio). 
The UVOT filters are not ideally suited to study such a ratio, 
but we can show the $uvw2-uvw1$ color curves. As seen in 
Figure \ref{col_w2_w1}, the UV emission becomes slightly bluer 
for the normal SNe~Ia, but with considerable scatter. The subluminous 
SNe~Ia appear to exhibit a shallower shift toward blue emission. 
The SN~2002cx-like SNe~Ia are not sampled sufficiently well 
to yield information regarding the temporal evolution of this
color index.

\section{Summary} 

We present a UVOT study of the UV and optical emission from 
26 SNe~Ia, collected during 2005--2008. The normal SNe~Ia 
exhibit definite similarities in the UV$-v$ colors, particularly 
in the $uvw1-v$ and $u-v$ colors. The colors rapidly become 
bluer from the earliest epochs until $\sim-5$ days relative to
$B$-band maximum, followed 
by a redward shift until $\sim$+20 days. At the later epochs, 
the color change is modest. By contrast, the subluminous 
subclass features redder emission initially, becoming still 
redder until $\sim$+20 days, at which time it becomes bluer, 
roughly matching the colors of the normal SNe~Ia. Members of the 
SN 2002cx-like subclass are initially very blue, but rapidly 
redden, with a slope steeper than the evolution seen among
the normal SNe~Ia. 

The $uvw1$ light curves of normal and subluminous SNe~Ia exhibit 
less variation in the peak shapes than seen in the $B$ or $b$ bands. 
This relative lack of variation makes discrimination between 
subclasses more difficult in the UV emission than in the 
$B$ or $b$-band emission. The SN 2002cx-like SNe do exhibit a 
steeper decline from peak than the normal or subluminous SNe~Ia, 
thus appearing different in peak shape than these latter subclasses. 
The SN 2002cx-like $uvw1-v$ color evolution is linear from $-8$ 
days to +30 days relative to $B$-band peak.

The subluminous SNe~Ia do differ from the normal SNe~Ia in the 
earlier transition to a late-time, flatter light curve. This difference 
could be utilized to discriminate this subclass from normal  
SNe~Ia when only rest-frame UV emission can be observed. 

The UV emission peaks at earlier epochs than the $B$-band emission, 
roughly two days earlier for normal events. This UV to 
$B$-band delay is consistent with the $U$-band study by Jha et al. (2006). 
The SN 2002cx-like SN 2005hk peaked anomalously early in the UV, and the  
other SN 2002cx-like object, SN~2008A, is consistent with 
a very large UV to $B$-band peak delay. 

Compared with a $U$-band template for normal SNe~Ia, the $uvw1$ mean 
template rises more steeply to the peak, but declines at a similar 
rate. The $uvw2$ mean template for normal SNe~Ia is very similar 
to the $U$-band template. 

These features require further study to determine how they constrain 
the physics of SN~Ia explosions. The nature of the $uvw1$ and $uvw2$ 
filters is such that they can  be contaminated to some extent by 
redder emission; this dictates that interpretation of these findings 
must include folding theoretical spectra through the UVOT filter 
response. Bearing that in mind, the data presented here introduce 
interesting challenges for SN~Ia theory.

The UVOT instrument has been observing, and will continue to observe, 
additional SNe~Ia. Based upon the findings presented in this work, as 
well as the new questions introduced in this work, the ongoing 
observational campaign will focus on the critical epochs described in 
the text to maximize the scientific yield of future UVOT observations. In 
particular, more  SN 2002cx-like, subluminous, and transitional subluminous 
SNe must be observed. Observations of all subclasses at very early epochs 
are critical, but such early detections of SNe~Ia are always a challenge.
The normal subclass exhibits dramatic color evolution during the week 
preceding $B$-band maximum. Dense sampling (daily, or even twice daily) 
during that week with high signal-to-noise ratio photometry in the four 
bluest filters would be 
important to further investigate that emission. Subluminous SNe~Ia evolve 
to a red peak, and back toward the blue during the first three weeks 
after $B_{\rm peak}$. Regular sampling (daily or every two days) with 
high-quality detections would optimize that study. For all 
subclasses, late-time observations every 3--5 days would permit measurement 
of the late-break time and slope.   

The results presented here invite additional investigations. 
Intermediate-redshift surveys have also observed rest-frame UV emission. 
Comparisons between those data and UVOT observations of nearby
SNe~Ia would be important to conduct. The generation of bolometric light 
curves of various subclasses of SNe~Ia will benefit through better 
estimation of the UV contributions to the optical and NIR emission. 
This will be particularly powerful for SNe~Ia simultaneously 
observed in all three wavelength ranges. 

Our study of the absolute magnitudes of these SNe~Ia is presented in 
a companion paper (Brown et al. 2010). The cosmological application 
of the combined studies is clear, as high-redshift surveys will study 
UV emission redshifted into the optical and NIR wavelength ranges. 
In addition, very nearby SNe~Ia will be observed with the UVOT 
UV grism, continuing spectral studies first presented by
Bufano et al. (2008). Multi-wavelength SN~Ia observing campaigns 
that include UVOT observations should continue for years to come. 


\bigskip
\medskip

We thank the Mission Operations team at Penn State University for
scheduling so many UVOT SN~Ia observations on short notice. We are
grateful to Stephane Blondin for his efforts on behalf of the CfA Supernova
Program, and X. Wang for categorization of NHV and NNN SNe~Ia. 
 P.A.M. would like to thank K. Krisciunas for assistance in
obtaining information on specific supernovae. The research of A.V.F.'s
supernova group at UC Berkeley is supported by National Science
Foundation (NSF) grants AST--0607485 and AST-0908886, the TABASGO Foundation, 
US Department of Energy SciDAC grant DE-FC02-06ER41453, US Department
of Energy grant DE-FG02-08ER41563, and {\it Swift} Guest Investigator 
grant NNX09AG54G.  The work at PSU is
sponsored NASA contract NAS5-00136. KAIT and its ongoing operation
were made possible by donations from Sun Microsystems, Inc., the
Hewlett-Packard Company, AutoScope Corporation, Lick Observatory, the
NSF, the University of California, the Sylvia \& Jim Katzman
Foundation, and the TABASGO Foundation.
Supernova research at the Harvard College Observatory is supported in
part by the NSF through grants AST--0606772 and AST--0907903.

\begin{table}
\scriptsize
\vspace{-9mm}
\caption{Optical Parameters for the SN Ia Sample: 2005 and 2006}
\begin{tabular}{l|cc|c|cccc|cccc}
\hline
\hline
SN &  Discovery & Host & Sub- & $\Delta m_{15}(B)$ & $t(B)_{\rm peak}$ & $B_{\rm peak}$ & 
Ref$^{a}$ & $E(B-V)$ & Ref.$^{ab}$ & $\Delta$ & Ref. \\
     & CBET      & Galaxy &  Class$^{c}$ &  [mag]$^{d}$  & [days]       &  [mag]$^{e}$  & 
          & [mag]$^{f}$ &             &          &      \\
\hline
2005am$^{g}$ & 113 & NGC 2811  & NNN & 1.51(04) & 3437.1 & 13.91    & 1 & 0.075(014)  & 1-M31 &  0.456(045)   &  1 \\
       &     &           &   & 1.73(05) & 3435.5 & 13.79(04) & 2 & 0.07(02)  & 2-M17 & 0.40(09)  & 2 \\
       &     &           &   & 1.48(02)& 3433.5 & 13.857(003)& 3 & 0.06& 3-P &  & \\
2005cf & 158 & MCG & NNN & 1.11  &   3534.0 & 13.54    & 4 & 0.097  & 4-Av  & --- &  \\
       &     &  -1-39-3   &    & 1.03(01)&  3533.9 & 13.65     & 1 & 0.143(019)  & 1-M31  & $-$0.127(017) & 1 \\   
       &     &           &    & 1.05(03) & 3533.7 & 13.63(02) & 5 & 0.09(03)  & 5-Av  & --- &  \\   
       &     &           &    & 1.06(08)&  3533.9 & 13.54(07)   & 2 & 0.22(04)& 2-M17  & $-$0.15(08) & 2 \\   
2005df$^{h}$ & 192 & NGC 1559 & N  & 1.2   & 3598.3  & 12.32 & 6 & 0.03  & 6-M31  & -0.12    & 6  \\
2005hk & 268,269 & UGC 272 & 02cx & 1.56  & 3685.1 & 15.91   & 7 & 0.09    & 8-NaD  & --- &  \\
       &     &           &    &    1.47(14) &     &15.91(05) & 2 & 0.48(03)& 2-M17 & $-$0.31(05)& 2\\
2005ke$^{i}$ & 287 & NGC 1371 & sub & 1.76(01) & 3699.2 & 14.882(005) & 3 & 0.06  & 3-P & ---  &   \\
             &     &          &    & 1.66(06) & 3699.2 & 14.95(05) & 2 & 0.06(02)  & 2-M17  & 1.55(03)  & 2  \\
             &     &          &    &          &        &           &   & 0.12  & T  &     &     \\
             &     &          &    &          &        &           &   & 0.27  & P  &     &     \\
2005mz & 347 & NGC 1275 & sub & 1.96(14)& 3745.9 & 16.88(11)& 2  & 0.32(05) & 2-M17  &  1.36(07)   & 2 \\
2006E$^{j}$ & 8658 & NGC 5338 & N & ---    & ---   & ---        & 14 & ---      & --- & ---      & --- \\
2006X  & 8667& NGC 4321 & NHV  & 1.10(12)& ---  &  14.05(02) & 2  & 1.50(02) & 2-M17  & $-$0.10(04) & 2 \\
       &     &          &  & 1.17(05)$^{k}$ & 3786.2 & 15.40(03) & 13 & 1.42(04) & 13-Av  &     &     \\
2006dd & 553,723 & NGC 1316 & NNN & 1.34  & 3918.8 & 12.18 & 9 & 0.08  & 9-P  & 0.01  & 9  \\ 
       &   &       &           &       &       &        &  &  0.26  & T  &     &     \\
2006dm & 569,570 & MCG & NNN & 1.28  & 3928.6 & 16.18 & T & 0.005  & T &     &     \\
       &         & -01-60-21  &    & 1.53(06)& 3929.0 & 16.13 & 1 & 0.060(014) & 1-M31 & 0.550(042) & 1 \\
       &         &             &    &       &        &       &   & 0.01 & P &      &     \\
2006ej & 603 & NGC 191 & NHV & 1.38(11) & 3976.9 & 15.88 & 1 & 0.050(009) & 1-M31 & 0.277(045) & 1 \\
       &   &        &      &      & 3973.6 & 15.90 & 12 & 0.03  & 12-M17 & 0.19(08) & 12 \\
       &   &        &      &      &        &       &   & 0.36  & T  &      &     \\
2006mr & 723 & NGC 1316 & sub & 1.82(02) & 4049.&       & 3  & 0.00 & 3-P  &      &    \\
\hline
\hline
\end{tabular}
\begin{tabular}{l}
$^{a}$REFERENCES. --- (1) KAIT, this work, (2) CfA3, Hicken et al. 2009, 
(3) CSP, Folatelli et al. 2010, \\
(4) Pastorello et al. 2007, (5) Wang et al. 2009a, (6) ANU, Appendix B, this work, 
(7) Phillips et al. 2007, \\
(8) Chornock et al. 2007, (9) Krisciunas 2009, private communication, 
(10) Simon et al. 2007, (11) Kasliwal et al. 2008, \\
(12) Blondin 2009, private communication, (13) Wang et al. 2008, 
(T) UVOT tail observations, this work, and \\ 
(P) UVOT peak observations, this work. \\
$^{b}$Extinction estimates are from MLCS2k2 with $R_{V} = 3.1$ (M31), 
MLCS2k2 with $R_{V} = 1.7$ (M17), Lira tail relation (L), \\
Phillips peak color relation (P), Na~I~D absorption (NaD), and
average of multiple methods (Av). \\
$^{c}$Utilizing the results of Wang et al. 2009b, the normal SNe are 
further divided into the following \\
categories. NHV = normal SN with high-velocity absorption features, 
NNN = normal SN without high-velocity features, \\
N = normal SN where existence of high-velocity features has not been 
determined, sub = subluminous SN~Ia, and \\
02cx-like = SN 2002cx-like SN~Ia (as explained in $\S$3.1). \\
$^{d}$Uncertainties are in parentheses, and are given in units of 0.01 mag. \\
$^{e}$Apparent magnitudes, no correction for Milky Way extinction. \\
$^{f}$Total of Milky Way extinction plus host-galaxy extinction. \\
$^{g}$Preliminary photometry was presented by Brown et al. 2005. \\
$^{h}$Only observed in UVOT UV filters ($uvw1$, $uvm2$, $uvw2$). \\
$^{i}$Preliminary photometry was presented by Immler et al. 2005. \\
$^{j}$SN 2006E was spectrally identified to be a SN~Ia. \\
$^{k}$The true $\Delta m_{15}(B) = 1.31 \pm 0.05$ mag differs from the 
observed value for this highly reddened SN. \\
\end{tabular}
\label{snparams_0506}
\end{table}

\begin{table}
\scriptsize
\vspace{-9mm}
\caption{Optical Parameters for the SN Ia Sample: 2007 and 2008}
\begin{tabular}{l|cc|c|cccc|cccc}
\hline
\hline
SN &  Discovery & Host & Sub- & $\Delta m_{15}(B)$ & $t(B)_{\rm peak}$ & $B_{\rm peak}$ & 
Ref$^{a}$ & $E(B-V)$ & Ref.$^{ab}$ & $\Delta$ & Ref. \\
     & CBET      & Galaxy &  Class$^{c}$ &  [mag]$^{d}$      & [days]       &  [mag]$^{e}$  & 
          & [mag]$^{f}$ &             &          &      \\
\hline

2007S  & 829 & UGC 5378 & N91T & 0.88(08) & --- & 15.82(03)& 2 & 0.53(03)  & 2-M17 &  $-$0.32(04)  &  2  \\
       &     &          &   & 0.91  & 4143.5 & 15.95    & T & 0.45 &Av&            &   \\
       &     &          &   & 0.80  & ---    & ---      & 3 &      &    &        &  \\
2007af & 865 & NGC 5584 & NNN & 1.11  & 4173.2 & 13.34     & T & 0.15  & Av  &  --- &  \\
       &   &     &          & 1.20(05) & --- & 13.13(03) & 2 & 0.17(03)  & 2-M17  & $-$0.04(05) & 2\\
       &     &          &   &       & 4174.3 & ---   & 10 & 0.13  & 10-Av  &  $-$0.04(02) &  10\\
2007ax & 904 & NGC 2577 & sub & 1.90 & 4187.3 & 16.3  & 11 & 0.05  & 11-Av  &  ---  &   \\
       &   &     &            &     & 4186.0 &       &   & 0.19(03) & 12-M17  & 1.33(07) & 12    \\
2007co & 977,978 & MCG & NHV & 1.06(02)  & 4264.9 & 16.94 & 1 & 0.161(025) & 1-M31 & $-$0.079(076) & 1 \\
       &         &+05-43-16 &  & 1.14(09)& --- & 16.43(08) & 2 & 0.34(04) & 2-M17 & $-$0.04(08)&2\\
       &          &              &   &        &     &           &   & 0.26     & P  &     &     \\
2007cq & 980 & Anon. Gal. & N91T & 1.02(03) & 4280.8 & 16.34 & 1 & 0.123(012) & 1-M31 & $-$0.024(027) & 1 \\
       &     &            &   &1.17(18) & 4281.& 15.82(07)& 2 & 0.17(03) & 2-M17 &  0.05(07)  &  2 \\
2007cv & 989 & IC 2587 & N & 1.31  & 4290.2 & 15.29 & ? & 0.19  & 2 & ---   &    \\
2007gi & 1017 & NGC 4036 & NHV & 1.31(09) & 4327.  & 13.25(04) & 15  & 0.17(04)   & 15-Av &    &  15  \\
2007on & 1121 & NGC 1404 & sub & 1.89 & ----   & ----  & 3 & ---  &   &    &  \\ 
2007sr & 1172 & NGC 4038 & NHV & 0.97(01) & 4448.8 &  & 13 & 0.18(02)  &  13   & ---  &   \\
2008A  & 1193 & NGC 634 & 02cx & --- & 4476. &  ---  & --- & ---   & 2 &    &     \\
2008Q  & 1228 & NGC 524 & NNN & 1.39 & 4505.5   & 13.81  & P & 0.07     & P &    &  \\
2008ec & 1437 & NGC 7469 & NNN & 1.06 & 4672.4   & 15.84  & P & 0.22   & PT-ave   &    &  \\
2008ha & 1567 & UGC 12682 & 02cx & 2.17(02) & 4783.2 & 18.23(01) & 14 & 0.076 & 14 & --- & \\  
\hline
\hline
\end{tabular}
\begin{tabular}{l}
$^{a}$REFERENCES. --- (1) KAIT, this work, (2) CfA3, Hicken et al. 2009, 
(3) CSP, Folatelli et al. 2010, \\
(4) Pastorello et al. 2007, (5) Wang et al. 2009a, (6) ANU, private communication, 
(7) Phillips et al. 2007, \\
(8) Chornock et al. 2007, (9) Krisciunas 2009, private communication, 
(10) Simon et al. 2007, (11)  Kasliwal et al. 2008, \\
(12) Blondin 2009, private communication, (13) Schweizer et al. 2008, 
(14) Foley et al. 2009b, (15) Zhang et al. 2009, \\
(T) UVOT tail observations, this work, and (P) UVOT peak observations, this work. \\
$^{b}$Extinction estimates are from MLCS2k2 with $R_{V} = 3.1$ (M31), MLCS2k2 with 
$R_{V} = 1.7$ (M17), Lira tail relation (L), \\
Phillips peak color relation (P), Na~I~D absorption (NaD), and average 
of multiple methods (Av). \\
$^{c}$Utilizing the results of Wang et al. 2009b, the normal SNe are 
further divided into the following \\
categories. NHV = normal SN with high-velocity absorption features, 
NNN = normal SN without high-velocity features, \\
N = normal SN where existence of high-velocity features has 
not been determined, sub = subluminous SN~Ia, and \\
02cx-like = SN 2002cx-like SN~Ia (as explained in $\S$3.1). \\
$^{d}$Uncertainties are in parentheses, and are given in units of 0.01 mag. \\
$^{e}$Apparent magnitudes, no correction for Milky Way extinction. \\
$^{f}$Total of Galactic extinction plus host-galaxy extinction. \\
\end{tabular}
\label{snparams_0708}
\end{table}

\begin{table}
\scriptsize
\caption{Parameters Derived from UVOT UV Observations: Early Epochs}
\vspace{4mm}
\begin{tabular}{l|lc|cccccc}
\hline
\hline
 & \multicolumn{8}{c}{Function-fitted $uvw1$ Parameters}  \\
\hline
SN & Model$^{a}$ & $\chi^{2}/DOF$ & $t$(pk)$_{uvw1}$ & $uvw1_{\rm max}^{b}$ & $R_{\rm 5d}(uvw1)^{c}$ & 
$\Delta m_{5}(uvw1)^{d}$ & $\Delta m_{15}(uvw1)^{d}$ & $t$(LB)$_{uvw1}^{f}$  \\
 & (1) & (2) & (3) & (4) & (5) & (6) & (7) &  (8) \\ 
\hline
2005am &FXPKLT& 0.42 & 3435.3 &  15.34(--) & ---  &  --- & ---  & 3465.0  \\
2005cf & CMPLT& 0.24 & 3531.0 &  15.11(07) & 1.20 & 0.10 & 1.25 & 3557.7 \\
2005df & CMPLT& 0.54 & 3597.4 &  13.92(07) & 0.49 & 0.16 & 1.32 & 3626.4 \\
2005hk & MID  & 0.81 & 3678.4 &  16.90(10) & ---  & 0.15 & 1.69 & 3696.6 \\
2005ke & MID  & 1.11 & 3696.7 &  17.09(08) & 0.73 & 0.24 & 1.49 & 3707.4 \\
2006dm & MID  & 0.54 & 3926.2 &  17.63(09) & 0.71 & 0.29 & 1.62 &  ---   \\
2006ej & FXPK & 0.45 & 3974.5 &  16.91(--) & ---  & ---  & ---  &  ---   \\
2007S  & MID  & 0.66 & 4141.1 &  17.83(13) & 0.42 & 0.13 & 1.11 &  ---   \\
2007af & CMPLT& 1.05 & 4172.0 &  14.74(07) & 0.62 & 0.49 & 1.64 & 4203.9 \\
2007co & MID  & 0.75 & 4264.1 &  18.81(15) & 0.06 & 0.13 & 1.31 &   ---  \\
2007cv & MID  & 0.69 & 4288.5 &  16.83(08) & 0.83 & 0.37 & 1.56 &   ---  \\
2007on & CMPLT& 2.11 & 4414.73 & 14.34(05) & 1.97 & 0.23 & 1.80 & 4439.9 \\ 
2008Q  & MID  & 0.29 & 4501.99 & 14.84(05) & 1.13 & 0.23 & 1.63 & --- \\
2008ec & MID  & 0.55 & 4669.28 & 17.31(11) & 1.01 & 0.20 & 1.31 & --- \\ 
\hline
\end{tabular}
\begin{tabular}{l|lccc|lcccc}
\hline
\hline
 & \multicolumn{4}{c}{Function-fitted $uvw2$ Parameters}  & \multicolumn{5}{c}{Function-fitted $u$ Parameters}\\
\hline
SN & Model$^{a}$ & $\chi^{2}/DOF$ &  $t$(pk)$_{uvw2}$ & $uvw2_{\rm max}^{b}$ & 
Model$^{a}$ & $\chi^{2}/DOF$ & $t$(peak)$_{U}$ & $U_{\rm max}^{a}$ & $\Delta m_{15}(U)^{d}$ \\
& (9) & (10) & (11) & (12) & (13) & (14) & (15) & (16) & (17)  \\ 
\hline
2005am &  --- &  --- &    --- &  ---     & FXPKLT & 0.79 & 3435.8 & 13.55(05) & ---  \\ 
2005cf & MID  & 0.79 & 3533.0 & 16.86(08)& CMPLT  & 0.39 & 3530.9 & 13.41(05) & 1.17 \\
2005df & CMPLT& 0.87 & 3598.9 & 15.61(07)& ---    &  --- & ---   & ---      &  ---   \\
2005hk & ---  & ---  &  ---   & ---      & CMPLT  & 6.87 & 3680.3 & 15.49(08) & 1.90 \\
2005ke & MID  & 1.13 & 3697.7 & 18.41(11)& MID    & 1.08 & 3695.3 & 15.52(07) & 1.79 \\
2006dm & ---  & ---  & ---    & ---      & CMPLT  & 0.11 & 3926.0 & 15.99(07) & 1.71 \\
2006ej & ---  & ---  & ---    & ---      & FXPKLT & 1.31 & 3974.1 & 15.47(06) & ---  \\
2007S  & ---  & ---  & ---    & ---      & MID    & 1.01 & 4143.0 & 15.86(07) & 1.38 \\
2007af & CMPLT& 0.82 & 4170.0 & 16.53(09)& CMPLT  & 0.96 & 4170.1 & 13.16(07) & 1.29 \\
2007co & ---  & ---  &  ---   & ---      & MID    & 0.91 & 4263.9 & 16.99(07) & 1.56 \\
2007cv & MID  & 2.22 & 4287.8 & 18.60(12)& MID    & 3.20 & 4288.4 & 15.08(06) & 1.88 \\
2007on & CMPLT& 2.03 & 4414.7 & 15.65(05)& CMPLT  & 1.39 & 4415.0 & 12.93(05) & 2.08 \\
2008Q  & MID  & 1.04 & 4502.3 & 16.43(05)& MID    & 3.15 & 4501.5 & 13.39(05) & 1.54  \\
2008ec & ---  & ---  & ---    & ---      & CMPLT  & 1.42 & 4669.9 & 15.62(07) & 1.53 \\
\hline
\end{tabular}
\begin{tabular}{l}
$^{a}$Light-curve model used in fitting. \\
$^{b}$Uncertainties are in parentheses, and are given in units of 0.01 mag. \\
$^{c}$Change in magnitude in the 5 days before the peak to the peak date. \\
$^{d}$Change in magnitude from the peak date to 5 days after peak. \\
$^{e}$Change in magnitude from the peak date to 15 days after peak. \\
$^{f}$The date of the late break, JD$-$2,450,000. \\
\end{tabular}
\label{fit_params_uv}
\end{table}

\begin{table}
\scriptsize
\caption{Parameters Derived from UVOT Optical Observations: Early Epochs}
\vspace{4mm}
\begin{tabular}{l|lcccc|lcccc}
\multicolumn{11}{l}{ } \\
\hline
 & \multicolumn{5}{c}{Function-Fitted $b$ Parameters}  & \multicolumn{5}{c}{Function-Fitted $v$ Parameters}\\
\hline
SN & Model & $\chi^{2}/DOF$ & $t$(peak)$_{B}$ & $B_{\rm max}$ & $\Delta m_{15}(B)^{d}$  & 
Model & $\chi^{2}/DOF$ & $t$(peak)$_{V}$ & $V_{\rm max}$ & $\Delta m_{15}(V)^{d}$   \\
& (18) & (19) &  (20) & (21) & (22) &  (23)  & (24)  & (25) &  (26) & (27)  \\
\hline
2005am & ---   &      & [3437.5]&  ---      &  --- & MID   & 5.01 & 3439.0 & 13.83(05) & 0.49 \\  
2005cf & ---   &      & [3534.0]&  ---      &  --- & CMPLT & 0.95 & 3532.0 & 13.62(05) & 0.35 \\
2005hk & CMPLT & 2.31 &  3684.3 & 15.90(07) & 1.52 & MID   & 0.99 & 3688.1 & 15.72(05) & 0.75 \\
2005ke & CMPLT & 0.86 &  3698.0 & 14.93(07) & 1.74 & CMPLT & 1.00 & 3699.8 & 14.21(05) & 1.07 \\
2006dm & CMPLT & 0.57 &  3928.7 & 16.17(07) & 1.28 & CMPLT & 0.53 & 3930.6 & 16.13(07) & 0.86 \\
2006ej & FXPKLT& 1.87 &  3977.2 & 15.89(07) & ---  & ---   & ---  & ---    & ---       & --- \\
2007S  & CMPLT & 0.34 &  4143.7 & 15.94(07) & 0.77 & CMPLT & 0.58 & 4145.6 & 15.53(07) & 0.61 \\
2007af & MID   & 1.41 &  4172.3 & 13.39(07) & 1.05 & CMPLT & 0.53 & 4175.9 & 13.25(05) & 0.58 \\
2007co & MID   & 0.48 &  4263.1 & 16.86(07) & 0.96 & MID   & 0.20 & 4265.2 & 16.69(07) & 0.63 \\
2007cv & CMPLT & 0.20 &  4290.0 & 15.30(07) & 1.39 & MID   & 0.29 & 4291.9 & 15.15(05) & 0.76 \\
2007on & CMPLT & 1.04 &  4418.6 & 13.14(07) & 1.70 & CMPLT & 0.65 & 4420.4 & 13.06(05) & 0.92 \\  
2008Q  & MID   & 0.27 &  4504.7 & 13.85(07) & 1.39 & MID   & 0.45 & 4505.1 & 13.80(05) & 0.66 \\
2008ec & CMPLT & 0.88 &  4672.5 & 15.83(07) & 1.06 & CMPLT & 0.61 & 4673.6 & 15.70(07) & 0.58 \\
\hline
\hline
\end{tabular}
\begin{tabular}{l}
$^{a}$Light-curve model used in fitting. \\
$^{b}$Uncertainties are in parentheses, and are given in units of 0.01 mag. \\
$^{c}$Change in magnitude in the 5 days before the peak to the peak date. \\
$^{d}$Change in magnitude from the peak date to 5 days after peak. \\
$^{e}$Change in magnitude from the peak date to 15 days after peak. \\
$^{f}$The date of the late break, JD$-$2,450,000. \\
\end{tabular}
\label{fit_params_opt}
\end{table}

\begin{table}
\caption{Parameters Derived from UVOT Observations: Late Epochs} 
\begin{tabular}{l|ccc|ccc}
\hline
\hline
 & \multicolumn{3}{c}{$uvw1$-Band  Parameters} & 
\multicolumn{3}{c}{$u$-Band Parameters}  \\
SN & $\beta_{uvw1}^{a}$ & $t$(LB)$_{uvw1}^{b}$ &  $\gamma(uvw1)^{c}$ & 
$\beta_{u}^{a}$ & $t$(LB)$_{u}^{b}$ &  $\gamma(u)^{c}$ \\
\hline
2005am & 12.04(0.07) & 3461.5(0.5)  & 3.88(0.39) & --- & --- & ---  \\
2005cf & 13.15(0.36)  & 3559.0(0.9)  & 3.38(0.51) & --- & --- & --- \\
2005df & 8.55(0.96) & 3626.5(0.8)  & 2.52(0.05) & --- & --- & --- \\
2005hk &    ---      &  ---    & ---    &   17.66(0.65) & 3706.7(1.0) & 1.44(0.36) \\
2006E  &  ---    & ---          & 2.12(0.07) & --- & --- & ---  \\
2006dd & 14.52(0.27) & 3944.1(0.7)  & 3.62(0.13) & --- & --- & ---   \\
2006dm & 12.68(3.27) & --- & --- &  13.56(0.97) & 3952.8(0.9) & 3.86(0.65) \\
2006ej & 11.92(1.76) & ---  & --- & --- & --- & --- \\
2007S  & 10.85(0.67) & --- & --- & 10.60(0.23) & 4172.2(1.5) &  3.40(0.65) \\
2007af & 12.73(0.14)& 4199.8(0.4)  & 3.41(0.11) & 13.05(0.06) & 4200.9(0.1) & 2.91(0.03) \\
2007co & 9.35(1.98) & --- & --- & --- & --- & --- \\
2007cv & 11.87(0.07) & --- & --- & --- & --- & --- \\
2007on & 6.88(0.58) & 4439.5(1.4) &  3.52(0.11) & 10.77(0.35) & 4440.7(0.3) & 3.42(0.07) \\
2007sr & 11.68(0.11) & --- & --- & --- & --- & --- \\
2008ec &  ---      &  ---    & ---    & 13.13(0.71) & 4702.1(2.4) & -0.12(1.93) \\
\hline
\hline
\hline
\hline
& \multicolumn{3}{c}{$b$-Band  Parameters} & \multicolumn{3}{c}{$v$-Band Parameters} \\
SN & $\beta_{b}^{a}$ & $t$(LB)$_{b}^{b}$ &  $\gamma$($b$)$^{c}$ & 
$\beta_{v}^{a}$ & $t$(LB)$_{v}^{b}$ &  $\gamma$($v$)$^{c}$ \\
\hline
2005hk & 11.54(0.78) & 3709.1(0.6) & 1.59(0.16) & --- & --- & --- \\
2005ke & 4.67(0.12) & 3730.9(1.0) & 2.03(0.11) & 6.69(0.44) & 3728.6(1.4) & 3.23(0.10) \\
2006dm & 10.89(0.15) & 3960.0(0.6) & 1.76(0.40) & 7.56(0.20) & 3962.8(1.8) & 3.43(0.62) \\
2007S & 9.47(0.21) & 4177.9(0.3) & 1.52(0.41) & 5.38(0.05) & 4174.6(1.8) & 4.57(0.12) \\
2007af & 9.91(0.06) &  4208.0(0.2) & 1.61(0.02) & 6.10(0.03) & 4215.9(0.3) & 2.61(0.02) \\
2007cv & 12.53(1.15) & 4312.1(1.1) & 4.65(0.41) & --- & --- & --- \\
2007on & 11.84(1.41) & 4441.9(1.6) & 3.00(0.15) & 8.91(0.34) & 4445.5(1.0) & 3.71(0.17) \\
2008ec & 11.01(1.28) & 4703.2(1.1) & 2.53(0.65) & 6.21(0.32) & 4703.6(0.3) & 4.67(0.46) \\
\hline
\end{tabular}
\begin{tabular}{l}
$^{a}$The initial linear slope in mag per 100 days. \\
$^{b}$The date of the late break, JD$-$2,450,000. \\
$^{c}$The late linear slope in mag per 100 days. \\
\end{tabular}
\label{fit_params_late}
\end{table}

\begin{table}
\caption{Parameters Obtained Fitting with Mean Template}
\vspace{4mm}
\begin{tabular}{l|cccc}
\hline
\hline
 & \multicolumn{2}{c}{Mean Template-Fitted $uvw1$ Parameters} & 
\multicolumn{2}{c}{$t_{B}-t_{uvw1}$} \\
\hline
SN & $t_{uvw1}^{a}$ & $uvw1_{\rm max}^{b}$ & Function Fit$^{c}$ & Template Fit$^{c}$ \\
(20) & (21)   &  (22)  &  (23)     &  (24)  \\
\hline
2005am & 3433.8 & 15.32(06) & 2.2 & 3.7 \\ 
2005cf & 3532.4 & 15.08(07) & 3.0 & 1.8 \\
2005df & 3597.7 & 13.83(07) & 0.7 & 0.4 \\
2005hk & ---    &    ---    & 5.6 & --- \\
2005ke & ---    &    ---    & 1.0 & --- \\
2006dm & 3926.3 & 17.70(09) & 2.1 & 2.0 \\
2006ej & 3972.9 & 17.05(08) & --- & --- \\
2007S  & ---    &    ---    & 2.3 & 2.5 \\
2007af & 4171.7 & 14.84(07) & 1.2 & 1.5 \\
2007co & 4264.7 & 18.75(15) & 1.0 & 0.4 \\
2007cv & 4289.1 & 16.95(08) & 1.7 & 1.1 \\ 
2007on & ---    &    ---    & 4.1 & --- \\
2008Q  & 4502.9 & 14.92(05) & 3.5 & 2.6 \\
2008ec & 4671.6 & 17.32(11) & 3.1 & 0.8 \\
\hline
\multicolumn{3}{r}{Mean $t_{B}-t_{uvw1}$ = } & 2.22$\pm$1.10 & 1.68$\pm$1.06 \\
\hline
\hline
\end{tabular}
\begin{tabular}{l}
$^{a}$The date of the $uvw1$-band peak magnitude, JD$-$2,450,000. \\
$^{b}$Uncertainties in units of 0.01 mag. \\
$^{c}$Time delay between peak magnitude and late break in days. \\
\end{tabular}
\label{mean_temp_table}
\end{table}

\begin{table}
\caption{$R_{UV} - R_{v}$ Values from Extinction Corrections}
\vspace{4mm}
\label{ext_coeffs}
\begin{tabular}{l|c|ccc}
\hline
\hline
SN & $E(B-V)$ & $R_{uvw1} - R_{v}$ & $R_{u} - R_{v}$ & $R_{uvw2} - R_{v}$ \\
   & [mag]$^{a}$  & [mag]  & [mag] & [mag]  \\
\hline
2006X & 1.32 & 1.92 & 2.17 & 1.63 \\
2007S$^{b}$ & 0.45 & 1.63 & 0.81 & 1.20 \\
2007co$^{b}$ & 0.26 & 1.09 & 0.95 & 1.53 \\
\hline
\hline
\end{tabular}
\linebreak
\begin{tabular}{l}
$^{a}$See Tables 1 and 2 for reddening references. \\
$^{b}$Broader-peaked optical light curves than low-extinction sample. \\ 
\end{tabular}
\end{table}

\clearpage

\clearpage

\clearpage

\clearpage

\clearpage

\clearpage

\clearpage

\clearpage

\clearpage

\clearpage

\clearpage

\clearpage

\clearpage

\clearpage

\clearpage

\clearpage

\clearpage

\clearpage

\clearpage

\clearpage

\clearpage

\clearpage

\clearpage

\clearpage

\clearpage

\clearpage

\clearpage

\clearpage

\clearpage

\appendix
\renewcommand{\thetable}{A-\arabic{table}}
\setcounter{table}{0}
\section*{APPENDIX A: UVOT Photometry for 6 SNe~Ia}

Here we present UVOT photometry for six SNe~Ia that were not 
included in the Brown et al. (2009) survey. 
The underlying galaxy light has been subtracted 
(Brown et al. 2009) and the magnitudes 
calibrated to the UVOT Vega system (Poole et al. 2008).

\begin{table}
\scriptsize
\caption{UVOT Photometry of Six SNe Ia}
\begin{tabular}{lcccccc}
\hline
\multicolumn{7}{c}{SN 2005mz} \\
\hline
\hline
JD$-$2,450,000 & $uvw2$ & $uvm2$ & $uvw1$ & $u$ & $b$ & $v$ \\
$[$days$]$ & [mag]$^{a}$ & [mag]$^{a}$ & [mag]$^{a}$ & 
              [mag]$^{a}$ & [mag]$^{a}$ & [mag]$^{a}$ \\
\hline
3742.43 & 20.18(19) & ... & ... & ... & ... & ... \\
3747.74 &  20.67(30) & ... & 19.28(17) & 17.77(07) & 17.16(06) & 16.57(06) \\
\hline
\\
\multicolumn{7}{c}{SN 2007on} \\
\hline
\hline
JD$-$2,450,000 & $uvw2$ & $uvm2$ & $uvw1$ & $u$ & $b$ & $v$ \\
$[$days$]$ & [mag]$^{a}$ & [mag]$^{a}$ & [mag]$^{a}$ & [mag]$^{a}$ & [mag]$^{a}$ & [mag]$^{a}$ \\
\hline
4411.28 & 16.64(05) & 16.91(04) & 15.27(04) & 13.84(05) & 14.41(05) & 14.50(04) \\
4412.62 & 16.05(04) & 16.24(04)  & 14.73(05) & 13.34(06) & 14.01(06) &  14.06(04) \\
4414.87 & 15.64(05) & 15.82(04) & 14.34(05)  & 12.91(09) & 13.49(10)    & 13.52(05)\\
4417.04 & 15.62(05) & 15.82(04)  & 14.32(06) & 12.86(11) & 13.16(13)   & 13.27(06)\\
4419.46 & 15.79(04) & 16.04(04) & 14.49(05) &  12.97(08) & 13.18(09) & 13.10(06)  \\
4421.00 & 16.02(05) & 16.15(05) & 14.72(05) & 13.15(07) &  13.14(07) & 13.06(05)  \\
4422.98 & 16.40(05) & 16.60(05) & 15.17(05) & 13.49(07) & 13.30(05) &  13.10(05)  \\
4424.45 & 16.42(07) & 16.76(011) & 15.32(06) & 13.67(04) &    ---   & 13.12(04) \\
4425.22 & 16.71(05) & 16.89(05) & 15.36(04) & 13.87(05) & 13.52(05) & 13.21(05) \\
4426.86 & 16.80(05) & 17.03(05) & 15.62(05) & 14.19(05) & 13.78(07) & 13.27(05) \\
4429.22 & 17.27(05) & 17.45(06) & 16.07(05) & 14.69(05) & 14.23(06) & 13.55(05) \\
4430.90 & 17.37(05) & 17.38(06) & 16.20(05) & 14.93(04) & 14.40(06) & 13.61(05) \\
4433.05 & 17.58(05) & 17.60(07) & 16.41(05) & 15.26(04) & 14.77(05) & 13.78(04) \\
4435.98 & 17.69(05) & 17.70(06) & 16.60(05) & 15.56(04) & 15.14(06) & 14.04(05) \\
4437.33 & 17.77(05) & 17.76(07) & 16.67(05) & 15.64(04) & 15.28(05) & 14.17(05) \\
4439.27 & 17.85(06) & 17.90(08) & 16.73(05) & 15.82(04) & 15.46(05) & 14.36(04) \\
4443.36 & 17.91(05) & 17.86(07) & 16.90(05) & 16.10(05) & 15.77(05) & 14.73(04) \\
4446.66 & 18.08(06) & 18.08(08) & 17.05(05) & 16.21(05) & 15.94(05) & 14.94(04) \\
4450.62 & 18.13(06) & 18.08(08) & 17.20(05) & 16.36(05) & 16.09(05) & 15.13(04) \\
4454.67 & 18.23(06) & 18.20(09) & 17.41(06) & 16.47(05) & 16.24(05) & 15.28(04) \\
4459.32 & 18.45(07) & 18.30(09) & 17.43(06) & 16.61(05) & 16.31(05) & 15.47(05) \\
4461.80 & 18.50(06) & 18.45(09) & 17.59(06) & 16.76(05) & 16.48(05) & 15.51(04) \\
4469.25 & 18.75(07) & 18.92(11) & 17.80(06) & 16.96(05) & 16.54(05) & 15.74(05) \\
\hline
\end{tabular}
\begin{tabular}{l}
$^{a}$Uncertainties are in units of 0.01 mag. \\
\end{tabular}
\end{table}

\pagebreak
\begin{table}
\scriptsize
\caption{UVOT Photometry of 6 SNe Ia (continued)}
\begin{tabular}{lcccccc}
\hline
\multicolumn{7}{c}{SN 2007sr} \\
\hline
\hline
JD$-$2,450,000 & $uvw2$ & $uvm2$ & $uvw1$ & $u$ & $b$ & $v$ \\
$[$days$]$ & [mag]$^{a}$ & [mag]$^{a}$ & [mag]$^{a}$ & [mag]$^{a}$ & [mag]$^{a}$ & [mag]$^{a}$ \\
\hline
4455.33 & 16.54(04) & 17.41(07) & 15.12(04) & 13.49(05) & 13.13(09) & 12.84(05) \\
4456.33 & 16.68(04) & 17.46(07) & 15.21(04) & 13.64(05) & 13.21(09) & 12.86(05) \\
4456.34 & --- & --- & 15.19(04) & --- & --- & --- \\
4457.27 & 16.80(04) & 17.53(07) & 15.36(04) & 13.84(04) & 13.31(09) & 12.96(05) \\
4457.30 & --- & --- & 15.34(04) & --- & --- & --- \\
4457.96 & 16.83(04) & 17.69(07) & 15.38(04) & 13.82(05) & 13.35(08) & 12.97(05) \\
4457.98 & --- & --- & 15.44(04) & --- & --- & --- \\
4458.61 & --- & --- & 15.47(04) & --- & --- & --- \\
4459.38 & 16.97(04) & 17.90(08) & 15.59(04) & 13.99(04) & 13.45(08) &  13.01(05) \\
4459.62 & 16.93(06) & 17.99(09) & --- & --- & --- & --- \\
4460.29 & --- & 18.01(13) & 15.70(05) & 14.11(04) & 13.59(05) & 13.11(04) \\
4460.66 & 17.10(07) & 17.96(08) &  --- & --- & --- & --- \\
4461.23 & 17.13(06) & 18.14(14) & 15.82(05) & 14.23(04) & 13.62(05) & 13.11(04) \\
4461.66 & 17.20(05) & 18.15(10) & 15.88(04) & 14.26(04) & 13.74(06) & 13.20(04) \\
4463.09 & --- & 18.35(11) &   --- & --- & --- & --- \\
4463.77 & 17.53(06) & 18.19(11) & 16.14(05) & 14.55(04) & 13.98(05) & 13.35(04) \\
4467.99 & 17.92(07) & 18.71(17) & 16.56(05) & --- & --- & --- \\
4469.26 & 18.15(06) & 19.07(14) & 16.76(05) & 15.35(04) & 14.67(04) & 13.69(04) \\
4470.55 & 18.12(06) & 19.08(014) & 16.92(05) & 15.42(04) & 14.72(04) & 13.66(04) \\
4472.87 & 18.41(07) & 19.61(018) & 17.12(05) & 15.63(04) & 14.94(04) & 13.77(04) \\
\hline
\\
\hline
\multicolumn{7}{c}{SN 2008A} \\
\hline
\hline
JD$-$2,450,000 & $uvw2$ & $uvm2$ & $uvw1$ & $u$ & $b$ & $v$ \\
$[$days$]$ & [mag]$^{a}$ & [mag]$^{a}$ & [mag]$^{a}$ & [mag]$^{a}$ & [mag]$^{a}$ & [mag]$^{a}$ \\
\hline
4473.11 & 19.26(16) & 19.25(21) & 17.91(07) & 16.25(04) & 16.79(05) & 16.75(08) \\
4474.60 & 19.24(10) & 19.24(13) & 17.96(07) & 16.24(04) & 16.66(04) & 16.65(05) \\
4476.58 & 19.39(11) & 19.78(17) & 18.18(08) & 16.30(04) & 16.54(04) & 16.47(05) \\
4478.12 & 19.45(16) & --- & --- & --- & --- & --- \\
4478.62 & 19.43(11) & 19.66(17) & 18.23(08) & 16.37(04) & 16.49(04) & 16.29(05) \\
4480.60 & 20.09(19) & 20.04(21) & 18.41(08) & 16.53(04) & 16.54(04) & 16.27(05) \\
4483.40 & 20.06(25) & ---       & 18.92(18) & 17.04(07) & 16.73(05) & 16.19(06) \\
4484.42 & 20.07(16) & 20.81(35) & 18.84(11) & 17.03(05) & 16.76(04) & 16.23(05) \\
4486.99 & 20.26(20) & ---       & 19.57(22) & 17.47(06) & 17.10(05) & 16.26(05)  \\
4490.84 & ---       & 21.00(34) & 20.02(24) & 18.36(08) & 17.58(06) & 16.47(06) \\
4493.12 & ---       & ---       & ---       & 18.64(10) & 18.03(07) & 16.74(06) \\ 
4494.86 & ---       & ---       & ---       & 18.71(19) & 18.07(13) & ---   \\
4497.13 & ---       & ---       & ---       & 19.35(17) & 18.52(10) & 16.99(08) \\
\hline
\end{tabular}
\begin{tabular}{l}
$^{a}$Uncertainties are in units of 0.01 mag. \\
\end{tabular}
\end{table}

\pagebreak
\begin{table}
\scriptsize
\caption{UVOT Photometry of 6 SNe Ia (continued)}
\begin{tabular}{lcccccc}
\hline
\multicolumn{7}{c}{SN 2008Q} \\
\hline
\hline
JD$-$2,450,000 & $uvw2$ & $uvm2$ & $uvw1$ & $u$ & $b$ & $v$ \\
$[$days$]$ & [mag]$^{a}$ & [mag]$^{a}$ & [mag]$^{a}$ & [mag]$^{a}$ & [mag]$^{a}$ & [mag]$^{a}$ \\
\hline
4496.63 & 17.85(06) & 18.76(14) & 16.15(04) & 14.53(04) & 15.22(04) & 15.22(04) \\
4499.11 & 16.81(04) & 17.33(07) & 15.22(04) & 13.71(05) & 14.46(05) & 14.46(04) \\
4501.25 & 16.49(04) & 16.98(06) & 14.89(04) & 13.41(05) & 14.13(05) & 14.12(04) \\
4503.29 & 16.39(04) & ---       & 14.83(04) & 13.31(06) & 13.88(07) & 13.89(04) \\
4504.59 & 16.45(05) & 16.79(08) & 14.88(04) & 13.36(04) & 13.86(05) & 13.80(04) \\
4505.74 & 16.58(05) & 16.92(06) & 15.00(05) & --- & --- & --- \\
4506.60 & 16.51(04) & 17.00(05) & 15.06(04) & 13.50(06) & 13.84(07) & 13.80(04) \\
4510.29 & 16.97(07) &       --- & 15.49(04) & --- & --- & --- \\
4511.35 & 17.00(04) & 17.51(07) & 15.65(04) & 14.04(04) & 14.08(05) & 13.88(04) \\
4512.96 & 17.22(05) & 17.64(07) & 15.86(04) & 14.30(04) & 14.21(05) & 13.95(04) \\
4514.87 & 17.51(05) & 17.95(07) & 16.12(04) & 14.59(04) & 14.44(05) & 14.08(04) \\
4516.88 & 17.75(05) & 18.15(08) & 16.44(04) & 14.87(04) & 14.65(05) & 14.26(04) \\
4520.20 & 18.25(06) & 18.48(09) & 16.92(05) & 15.49(04) & 15.16(04) & 14.47(04) \\
4521.50 & 18.43(07) & 17.11(05) & --- & 15.72(04) & 15.34(04) & 14.54(04) \\
\hline

\\
\multicolumn{7}{c}{SN 2008ec} \\
\hline
\hline
JD$-$2,450,000 & $uvw2$ & $uvm2$ & $uvw1$ & $u$ & $b$ & $v$ \\
$[$days$]$ & [mag]$^{a}$ & [mag]$^{a}$ & [mag]$^{a}$ & [mag]$^{a}$ & [mag]$^{a}$ & [mag]$^{a}$ \\
\hline
4663.19 & ---       & ---       & 18.82(19) & 17.18(07) & 17.36(07) & 17.33(12) \\  
4665.41 & 19.30(18) & 19.52(36) & 17.91(10) & 16.21(05) & 16.67(05) & 16.58(07) \\
4667.81 & 19.49(21) & 19.82(36) & 17.41(07) & 15.84(05) & 16.23(05) & 16.19(06) \\
4672.34 & 18.81(13) & ---       & 17.41(07) & 15.61(05) & 15.86(05) & 15.74(05) \\
4674.38 & 18.97(14) & ---       & 17.48(08) & 15.75(05) & 15.84(05) & 15.67(05) \\
4676.33 & 19.11(20) & ---       & 17.69(10) & 15.98(05) & 15.90(05) & 15.71(06) \\
4678.30 & 19.23(18) & ---       & 18.05(11) & 16.17(05) & 16.00(05) & 15.72(05) \\
4680.40 & 19.46(20) & ---       & 18.19(11) & 16.40(05) & 16.15(05) & 15.74(05) \\
4681.81 & ---       & ---       & 18.43(13) & ---       & ---       & ---       \\
4683.91 & ---       & ---       & 18.41(17) & ---       & ---       & ---       \\
4686.21 & ---       & ---       & 18.72(16) & 17.26(07) & 16.67(06) & 16.16(11) \\
4688.35 & ---       & ---       & 19.23(22) & 17.67(09) & 17.00(07) & 16.29(06) \\
4694.00 & ---       & ---       & ---       & 18.37(17) & 17.77(11) & 16.65(12) \\
4699.33 & ---       & ---       & ---       & 18.92(21) & 18.10(11) & 16.87(08) \\
4702.12 & ---       & ---       & ---       & 19.61(35) & 18.50(13) & 17.22(10) \\
4711.20 & ---       & ---       & ---       & 19.33(27) & 18.73(16) & 17.55(12) \\
4716.39 & ---       & ---       & ---       & 19.28(26) & 18.98(20) & 17.76(14) \\
4720.34 & ---       & ---       & ---       & 19.42(31) & 18.96(20) & 18.01(17) \\
\hline
\end{tabular}
\begin{tabular}{l}
$^{a}$Uncertainties are in units of 0.01 mag. \\
\end{tabular}
\end{table}

\appendix
\renewcommand{\thetable}{B-\arabic{table}}
\setcounter{table}{0}
\renewcommand{\thefigure}{B-\arabic{figure}}
\setcounter{figure}{0}
\section*{APPENDIX B: ANU Photometry of SN 2005df}

We present photometry and light curve fitting results for SN 2005df, 
as obtained with the Australian National University 2.3~m and 1~m telescopes at Siding 
Springs.

\begin{figure}
\epsscale{1.1} \plotone{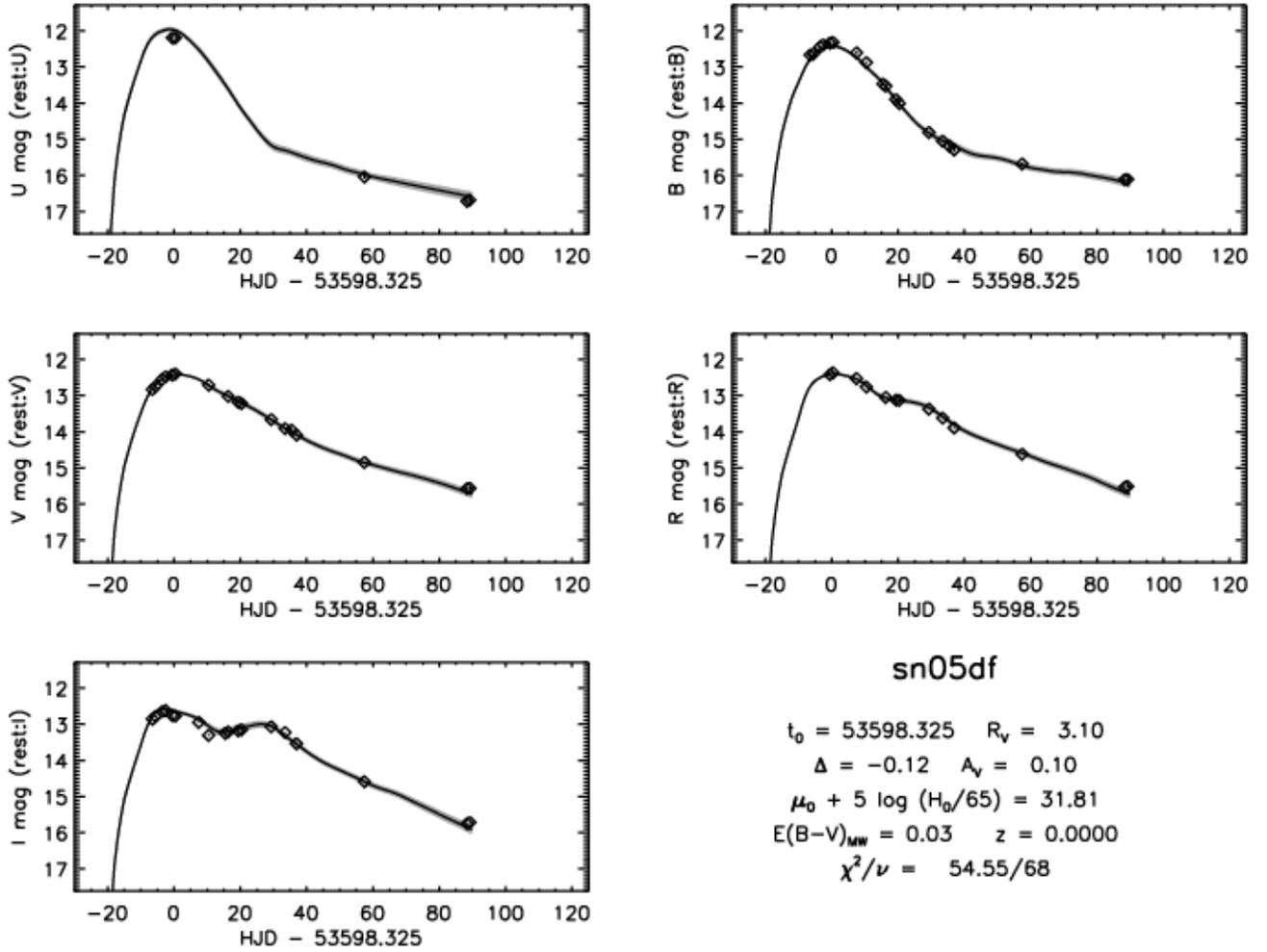}
\caption{$UBVRI$ light curves of SN~2005df. The light curves were fitted 
with the MLCS2k2 routine to estimate the date of maximum brightness 
peak magnitude ($B_{\rm peak}$), extinction, and $\Delta$ value, as shown.}
\label{sn2005df}
\end{figure}

\begin{table}
\scriptsize
\caption{ANU Photometry of SN 2005df}
\begin{tabular}{lccccccc}
\hline
\hline
MJD & Phase$^{a}$ & $U$ &  $B$ &  $V$ &  $R$ &  $I$ &  Inst. \\
$[$days$]$ & $[$days$]$ & [mag] & [mag] & [mag] & [mag] & [mag] &  \\
\hline
53591.79 & -7 & --- & 12.68(0.04) & 12.82(0.02) & --- & 12.82(0.01) & 1m+SITE  \\
53592.76 & -6 & --- & 12.65(0.02) & 12.73(0.01) & --- & 12.75(0.01) & 1m+SITE  \\
53594.75 & -4 & --- & 12.45(0.02) & 12.54(0.01) & --- & 12.63(0.01) & 1m+SITE  \\
53595.75 & -3 & --- & 12.39(0.01) & 12.47(0.01) & --- & 12.59(0.01) & 1m+SITE  \\
53597.79 & -1 & 12.24(0.02) & 12.33(0.03) & 12.42(0.02) & 12.41(0.02) & 12.72(0.02) & 2.3m+Im  \\
53598.70 & 0 & 12.24(0.03) & 12.32(0.02) & 12.39(0.01) & 12.36(0.01) & 12.73(0.01) & 2.3m+Im  \\
53605.79 & 7 & --- & 12.60(0.04) & --- & 12.53(0.01) & 12.93(0.01) & 1m+WFI  \\
53608.76 & 10 & --- & 12.89(0.02) & 12.70(0.02) & 12.75(0.02) & 13.30(0.02) & 2.3m+Im  \\
53613.72 & 15 & --- & 13.46(0.01) & --- & --- & 13.27(0.01) & 1m+WFI  \\
53614.63 & 16 & --- & 13.52(0.02) & 13.02(0.01) & 13.05(0.01) & 13.24(0.01) & 1m+WFI  \\
53617.68 & 19 & --- & 13.88(0.02) & 13.20(0.01) & 13.11(0.01) & 13.20(0.01) & 1m+WFI  \\
53618.66 & 20 & --- & 14.00(0.05) & 13.25(0.01) & 13.12(0.01) & 13.17(0.01) & 1m+WFI  \\
53627.62 & 29 & --- & 14.79(0.01) & 13.71(0.01) & 13.34(0.00) & 13.09(0.01) & 1m+WFI  \\
53631.78 & 33 & --- & 15.04(0.03) & 13.96(0.01) & 13.59(0.01) & 13.26(0.01) & 1m+WFI  \\
53633.76 & 35 & --- & 15.11(0.04) & 14.06(0.05) & --- & --- & 1m+SITE  \\
53635.25 & 36 & --- & 15.22(0.05) & 14.19(0.02) & 13.88(0.01) & 13.59(0.03) & 1m+SITE  \\
53655.72 & 57 & 16.05(0.05) & 15.72(0.04) & 14.86(0.01) & 14.62(0.01) & 14.58(0.01) & 2.3m+Im  \\
53686.68 & 88 & 16.72(0.04) & 16.14(0.03) & 15.60(0.04) & 15.56(0.03) & 15.78(0.02) & 2.3m+Im  \\
53687.51 & 89 & 16.70(0.05) & 16.14(0.05) & 15.59(0.03) & 15.54(0.03) & 15.75(0.01) & 2.3m+Im  \\
53742.60 & 144 & 18.07(0.05) & --- & 16.72(0.07) & 17.08(0.04) & 17.27(0.05) & 2.3m+Im  \\
53773.44 & 174 & --- & 15.72(0.04) & 17.46(0.03) & --- & 17.64(0.03) & 2.3m+Im  \\
53831.86 & 233 & --- & --- & 18.08(0.04) & 18.45(0.06) & --- & 2.3m+Im  \\
\hline
\end{tabular}
\begin{tabular}{l}
$^{a}$Phase relative to the fitted $B_{\rm peak}$. \\
\end{tabular}
\end{table}

\end{document}